\documentclass[preprint]{aastex63}

\received{February 22, 2019}
\revised{December 15, 2019}
\accepted{December 15, 2019}
\submitjournal{ICARUS}

\shorttitle{Dynamical evidence for an early giant planet instability}
\shortauthors{Ribeiro et al.}
\begin{document}

\title{Dynamical evidence for an early giant planet instability \footnote{Article accepted for publication on December, 15th, 2019, \url{https://authors.elsevier.com/tracking/article/details.do?aid=113605&jid=YICAR&surname=Ribeiro}}}

 \correspondingauthor{Rafael Ribeiro de Sousa}
 \email{rafanw72@gmail.com, rribeiro@oac.eu}

\author{Rafael Ribeiro de Sousa}
\affiliation{S\~ao Paulo State University, UNESP,Campus of Guaratinguet\'a, Av. Dr. Ariberto Pereira da Cunha, 333 - Pedregulho, Guaratinguet\'a - SP, 12516-410, Brazil}
\affil{ Laboratoire Lagrange, UMR7293, Universit\'e C\^ote d'Azur, CNRS, Observatoire de la C\^ote d'Azur, Boulevard de l'Observatoire, 06304 Nice Cedex 4, France}

\author{Alessandro Morbidelli}
\affil{ Laboratoire Lagrange, UMR7293, Universit\'e C\^ote d'Azur, CNRS, Observatoire de la C\^ote d'Azur, Boulevard de l'Observatoire, 06304 Nice Cedex 4, France}

\author{Sean N. Raymond}
\affil{Laboratoire dAstrophysique de Bordeaux, Univ.  Bordeaux, CNRS, B18N, alle Geoffroy Saint-Hilaire, 33615 Pessac, France}

\author{Andre Izidoro}
\affiliation{S\~ao Paulo State University, UNESP,Campus of Guaratinguet\'a, Av. Dr. Ariberto Pereira da Cunha, 333 - Pedregulho, Guaratinguet\'a - SP, 12516-410, Brazil}

\author{Rodney Gomes}
\affil{Observat\'orio Nacional, Rua General Jos\'e Cristino 77, CEP 20921-400, Rio de Janeiro, RJ,  Brazil}
% \email{rodney@on.br}

\author{Ernesto Vieira Neto}
\affiliation{S\~ao Paulo State University, UNESP,Campus of Guaratinguet\'a, Av. Dr. Ariberto Pereira da Cunha, 333 - Pedregulho, Guaratinguet\'a - SP, 12516-410, Brazil}



\begin{abstract}

The dynamical structure of the Solar System can be explained by a period of orbital instability
experienced by the giant planets. While a late instability was originally proposed to explain the Late 
Heavy Bombardment, recent work favors an early instability. 
Here we model the early dynamical evolution of the outer Solar System to self-consistently constrain the 
most likely timing of the instability. 
We first simulate the dynamical sculpting of the primordial outer 
planetesimal disk during the accretion of Uranus and Neptune from migrating planetary embryos during the gas 
disk phase, and determine the separation between Neptune and the inner edge of the planetesimal disk. 
We performed simulations with a range of (inward and outward) migration histories for Jupiter. 
We find that, unless Jupiter migrated inwards by 10 AU or more, the instability almost certainly happened within 100 Myr 
of the start of Solar System formation. There are two distinct possible instability triggers.
The first is an instability that is triggered by the planets themselves, with no appreciable influence 
from the planetesimal disk. About half of the planetary systems that we consider have a self-triggered instability.
Of those, the median instability time is $\sim4$Myr. 
Among self-stable systems -- where the planets are locked in a resonant chain that remains 
stable in the absence of a planetesimal's disk-- our self-consistently sculpted planetesimal 
disks nonetheless trigger a giant planet instability with a median instability time of 
37-62 Myr for a reasonable range of migration histories of Jupiter.
The simulations that give the latest instability times are those that invoked long-range inward 
migration of Jupiter from 15 AU or beyond; however these simulations over-excited the inclinations of Kuiper belt objects and are 
inconsistent with the present-day Solar System. We conclude on dynamical grounds that the giant planet 
instability is likely to have occurred early in Solar System history.     

\end{abstract}



\section{Introduction} \label{sec:intro}
The orbital evolution of the giant planets played a central role in shaping the dynamical structure of the present-day Solar System. The so-called \textit{Nice model} invokes a dynamical instability among the giant planets that was triggered by interactions with the primordial outer planetesimal disk \citep{Tsiganisetal2005,Levison2011,NesvornyMorby2012}. 
There are a number of lines of circumstantial evidence that support this model.  A giant planet instability can explain the capture of the Trojans of Jupiter and Neptune \citep{Morbidellietal2005,Nesvornyetal2013,GomesNesvorny2016}, the irregular satellites of the giant planets \citep{Nesvorny2007,Nesvornyetal2014} and the Kuiper belt's orbital structure  \citep{Nesvorny2015a,Nesvorny2015b,Nesvorny2016a,Gomesetl2018} see also \citet{Nesvorny2018} for a review.

The giant planet instability caused a cometary bombardment in the inner Solar System \citep{Gomesetal2005}. According to the \textit{Nice model} scenario, after the gas dissipation Jupiter, Saturn, Uranus, and Neptune's gravitational interactions with a massive outer 
planetesimal disk eventually drove them into an unstable configuration. During the giant planet instability, the outer planetesimal disk was mostly destabilized. The asteroid belt was also strongly perturbed, removing $\sim 90\%$ of the original asteroids \citep{Nesvornyetal2017,deiennoet2016,Deiennoetal2018,clement2019}. As a consequence, a large number of planetesimals (both asteroids and comets) was delivered into the inner Solar System, causing a massive bombardment on the terrestrial planets and the Moon assuming that the Moon and the planets existed at this time \citep{Gomesetal2005,bottke2012}. If this happened late, 
the instability would have had a high probability of disrupting the orbits of the terrestrial 
planets~\citep{KaibandChambers2016} or at least of over-exciting their orbits~\citep{roig16}. 
However, an instability {\em during terrestrial planet formation} would have excited the terrestrial planets' 
building blocks -- for instance by removing mass from Mars' feeding zone but not Earth's, 
simulations of terrestrial planet formation with an early giant planet instability provide a good match to the Solar System \citep{Clementetal2018,clement2019,Clement2019b}.

%The timing of the giant planet instability is hard to constrain empirically.

The Lunar crater record suggests an epoch of intense bombardment around $3.9$ Gy ago, when impacts created the 
youngest basins of the Moon. These impacts took place late in the Solar System formation time line, roughly $500-700$ My after the planets formed; this period is known as the Late Heavy Bombardment (LHB). Two hypotheses have been suggested to explain the origin of the LHB. The first claims that the LHB was due to a surge on 
the bombardment rate  \citep{Tera1974,Ryder1990,Ryder2002}. This may be explained by the destabilization of the asteroid belt and trans-Neptunian disk due to a sufficiently late giant planet instability, as described above \citep{Gomesetal2005}. In this line of thinking the instability would have 
happened  $ \sim 500-700$ My after the planets formed. The second hypothesis is that the LHB was the tail-end of the terrestrial planet accretion, presumably 
due to planetesimals left-over from the main phase of planet formation \citep{Hartmann1975,NeukumHarmann2001,Hartmann2003}.
\citet{Morbietal2012} showed that in this hypothesis the Moon would have accreted a mass that is an order of magnitude larger than that deduced from the abundance 
of highly siderophile elements (HSE) in its mantle \citep{Walker2009,Dayetal2007,Dayetal2015}. But \citet{Morbidellietlal2018} showed that the lunar HSEs could have been sequestered into the lunar core during the crystallization of the lunar magma ocean (LMO). The HSE budget of the lunar mantle 
would then trace only the amount of material accreted after LMO crystallization. If the latter occurred late, as argued in \citet{Elkins-Tantonetal2011}, 
the small amount of lunar mantle HSEs can be explained also in the accretion-tail hypothesis for the LHB.  
Therefore, both late or early instabilities may be consistent with the lunar crater record and geochemical properties. 

To date it has been assumed that the giant planet instability was triggered by gravitational interactions between the giant planets and an outer planetesimal disk \citep{Tsiganisetal2005,Morbidellietal2007,Levison2011,Deiennoetal2017}, (see \citet{tom_2002} for an exception).
However, this was not necessarily the case. Below we demonstrate that the instability may instead have been self-triggered by the giant planets themselves. Upon dissipation of their natal planet-forming disks, giant planets emerge on marginally stable (and often quickly unstable) configurations~\citep[e.g.,][]{moeckel08,matsumura10}. A large number of studies have simulated self-triggered instabilities, mainly in the context of explaining the eccentricity distribution of giant exoplanets \citep[e.g., review by ][]{davies14}. Instabilities arising as gravitational jostling between planets cause their orbits to intersect, leading to a phase of close encounters (often called `planet-planet scattering'), which typically leads to the ejection of one or more planets into interstellar space~\citep{rasio96,weidenschilling96,lin97,adams03}. Simulations of self-triggered instabilities find that 1) more closely-packed sets of planets generally become unstable on shorter timescales \citep{chambers96,marzari02,chatterjee08} and 2) marginally-stable systems with outer planetesimal disks that do become unstable tend to do so relatively quickly, with a median instability time of ~$10^5$ years but a tail extending out to hundreds of Myr~\citep{raymond10,raymond11}. 

%There are possible evolutionary histories in which the outer planetesimal disk triggered the instability but these are only relevant if the giant planets emerged from the gaseous disk on extremely stable orbits, which we only produce by artificially increasing the disk's dissipative effects (see Section 3.2).

Interactions with the planetesimal disk remains a possible instability trigger if the giant planets emerged from the gaseous disk on long-term stable orbits. The timing of such an instability is set by the distance between the
outermost planet (presumably Neptune) and the inner edge of the primordial trans-Neptunian planetesimal disk \citep{Gomesetal2005,Levison2011}. \citet{Gomesetal2005} argued that the planetesimal disk should only contain particles with dynamical lifetimes (times required to encounter a planet) longer than the lifetime of the gaseous disk~\citep[a few My e.g.,][]{haisch01,hillenbrand08,pascucci09}. Planetesimals with short dynamical lifetimes should have been removed with little effect on the dynamical evolution of the planets because planet-gas interactions dominate over the planet-planetesimal interactions \citep{Capobianco2011}.

\citet{Gomesetal2005} thus argued for a roughly 1 AU-wide gap between Neptune's early orbit and the inner edge of the planetesimal disk and found that the giant planet instability occurred in this case between 200 Myr and 
1 Gyr, roughly at the time expected for the LHB. \citet{Levison2011} considered 4 resonant giant planets where the inner edge of the planetesimal disk was several AUs beyond the orbit of Neptune (the outermost planet). 
They showed that viscous stirring (due to the disk's self-gravity, assuming the presence of a thousand 
Pluto-mass objects) leads to an exchange of energy between the planet and disk particles even in the absence of close encounters. 
This energy exchange proceeds at a very slow pace and the system crosses many weak secular resonances 
that lead to instability on a timescale consistent with the Late Heavy Bombardment. 
\citet{Deiennoetal2017} simulated the outer Solar System's evolution 
assuming the presence of an additional primordial ice giant \citep{Nesvorny2011}
and testing a variety of orbital configurations of the 5 giant planets. 
They found that a late instability is possible for a specific minimum distance
between the inner edge of the disk and Neptune. However, in some cases, the timescale of instability was dependent on the numerical resolution of the 
planetesimal disk (mass and number of the planetesimals). Of course, the real primordial planetesimal disk was sculpted by the growth and migration of the giant planets.
If the giant planets or their precursors migrated inward from farther out in the gaseous disk (e.g, \citet{Bitshetal2015}), 
this could plausibly have created a significant gap between the outermost ice giant and 
the primordial planetesimal disk. On the other hand, planetesimals excited onto eccentric orbits by the forming/migrating planets may have their orbits re-circularized at perihelion distance by gas drag~\citep[e.g.][]{raymond17}, 
narrowing the gap between the planets and planetesimal disk.

The goal of this paper is to constrain the timing of the giant planet instability by simulating interactions between the giant planets and a sculpted outer planetesimal disk. To do this, we need a reliable model of the growth and early evolution of the giant planets' orbits, in particular for the ice giants. Indeed, understanding the accretion of Uranus and Neptune is a long-standing problem in Solar System formation. Early studies showed that their growth by planetesimal accretion takes longer than the gas disk lifetime  \citep{Safronov69,Levisonsteward2001,Thommesetal2003}, even at $10-15$ AU \citep{Levisonetal2010}. The accretion of mm- to cm-sized ``pebbles'' aerodynamically drifting through the disk has been shown to drastically accelerate core growth \citep{LambrechtsJohansen2012,LambrechtsJohansen2014}. For reasonable pebble fluxes cores of $10-20$ Earth masses can form within the gas disk lifetime \citep{Lambrechtsetal2014} even when gravitational interactions among the growing protoplanets are accounted for \citet{Levisonetal2015n}. Yet pebble accretion should produce planets with zero obliquity \citep{DonesTremaine1993,JohansenLacerda2010}. 
The $3$ degree obliquity of Jupiter and the $26$-degrees obliquity of Saturn can be explained by spin-orbit resonances with Uranus and Neptune, respectively \citep{WardHamilton2004,HamiltonWard2004,Boueetal2009,VokrouhlickNesvorn2015,BrasserandLee2015}.
However, Uranus and Neptune's obliquities of 98 and 30 degrees, respectively, are thought to be the result
of giant collisions during their formation \citep{Slatteryetal1992,Boueaskar2010,Morbidellietal2012,Jakub2012,Kegerreisetal2018}.
%If this is the case, Uranus and Neptune cannot have formed solely by pebble accretion. 

\citet{Izidoroetal2015} proposed that the ice giants formed in two steps. By the time Jupiter and Saturn had undergone rapid gas accretion to become giants, pebble accretion had produced a system of protoplanets of $\sim5$ Earth masses, comparable in mass to most super-Earths \citep[e.g.,][]{mayor11,batalha13,marcy14,wolfgang16}. 
These protoplanets migrated inward in the Type-I regime  due to tidal torques from the gas disk \citep{GoldreichTremaine80,Ward86,Tanakaetal2002}. While it is unclear whether Jupiter and Saturn at this point would have been migrating inward, outward or have roughly stationary orbits \citep{MassetSnellgrove2001,MorbidelliCrida2007,PierensNelson2008,PierensRaymond2011,RaymondMorbidelli2014,Pierensetal2014}, 
the gas giants' migration was certainly slower than that of the protoplanets for typical disk viscosities. In addition, the co-presence 
of Jupiter and Saturn has the effect to slow-down, stop or reverse, the resonant migration of these 
two planets, depending on the aspect ratio of the disk \citep{CridaMorbidelli2007}. 
% 
% 
% \textbf{In the classical type II migration, with speed proportional to the viscosity of the disk, occurs only if the gap is deep and clean. If there is a significant amount of gas in the gap, the migration of the planet is generally slower than the classical type II migration rate (Crida and Morbidelli, 2007). In the intermediate regime between type I and type II migration, if the gap is not clean, the gas in the gap exerts a positive viscous torque on the outer disk, so the outer disk does not push the planet inward as efficiently as when the gap is clear. A planet with Jupiter mass at 5 AU in an accreting disk does not open a very deep clean gap, in under some conditions on the disk parameters ( an isothermal disks in a  range of viscosities) the type II migration may be avoided for a Jupiter mass planet. Thus, type-II migration of gas giants is significantly slower than the type-I migration of protoplanets with a mass of few super-Earths or for a set of larger parameters of the traditional disk.}  

The protoplanets cannot generally migrate past Jupiter and Saturn's orbits; rather, the gas giants act as a migration barrier \citep{Izidoroetal2015b}. Protoplanets become trapped in mean motion resonances with Saturn, and successive protoplanets form chains of mean motion resonances. As more protoplanets migrate inward, the resonant chain is destabilized, leading to giant (obliquity-generating) collisions and another phase of migration into resonant chains.  \cite{Izidoroetal2015} showed that this process typically produces 2-3 ice giants with masses comparable to those of Uranus and Neptune in resonant chains that include Jupiter and Saturn.  

% \textbf{Deienno et al. (2017, AJ 153) studied the early and late instability scenario to find out whether or not a late instability would still be possible considering Neptune’s migration proposed by Nesvorny (2015,a,b). Deienno et al. (2017) claimed in a favor of an early instability given the high complexity of the mechanism applied for the late scenario.  More than that, Deienno et al. (2017) presented a very detailed description of the relationship between the distance of the inner edge of the planetesimal disc and Neptune’s orbit that would or would not lead towards an instability within 400 Myr. Although our approach is different of Deienno et al. (2017) not only by the proceeding to build the planetesimal disk structure but we also here we address whether the distance between the inner edge of the planetesimal disk and Neptune can be large enough or not.}

We constrain the timing of the giant planet instability in two ways.  First, we simulate the dynamical evolution of the successful ice giant formation systems produced by \citet{Izidoroetal2015}. Second, we generate planetesimal disks that are dynamically consistent with the growth and migration of the giant planets, again using the framework of \citet{Izidoroetal2015}. Our paper is laid out as follows.  In \S \ref{Methods} we discuss how we simulated the evolution of the primordial planetesimal disk during the growth of Uranus and Neptune during the gas disk phase. We also tested different migration histories for Jupiter. In \S \ref{instime} we simulate the giant planet instability under two different assumptions. 
We first show that the giant planets in successful runs from \citet{Izidoroetal2015} are often self-unstable after the dissipation of the gaseous disk. 
Next, we show that even planetary systems that would be long-term stable without any influence of the planetesimal disk
do become quickly unstable under the influence of planetesimal disks consistent with the growth and dynamical evolution
of the giant planets. This result holds for all migration histories of Jupiter in which the planet migrated inwards by less than 10 AU. In \S \ref{discussion} we discuss caveats to our study. We present our conclusions in \S \ref{conclusion}.

 \section{Sculpting of the planetesimal disk during the ice giants' growth}
 
 \subsection{ Model of Izidoro et al. (2015)}
 \label{izidoro}
 
\begin{figure*}
\gridline{\fig{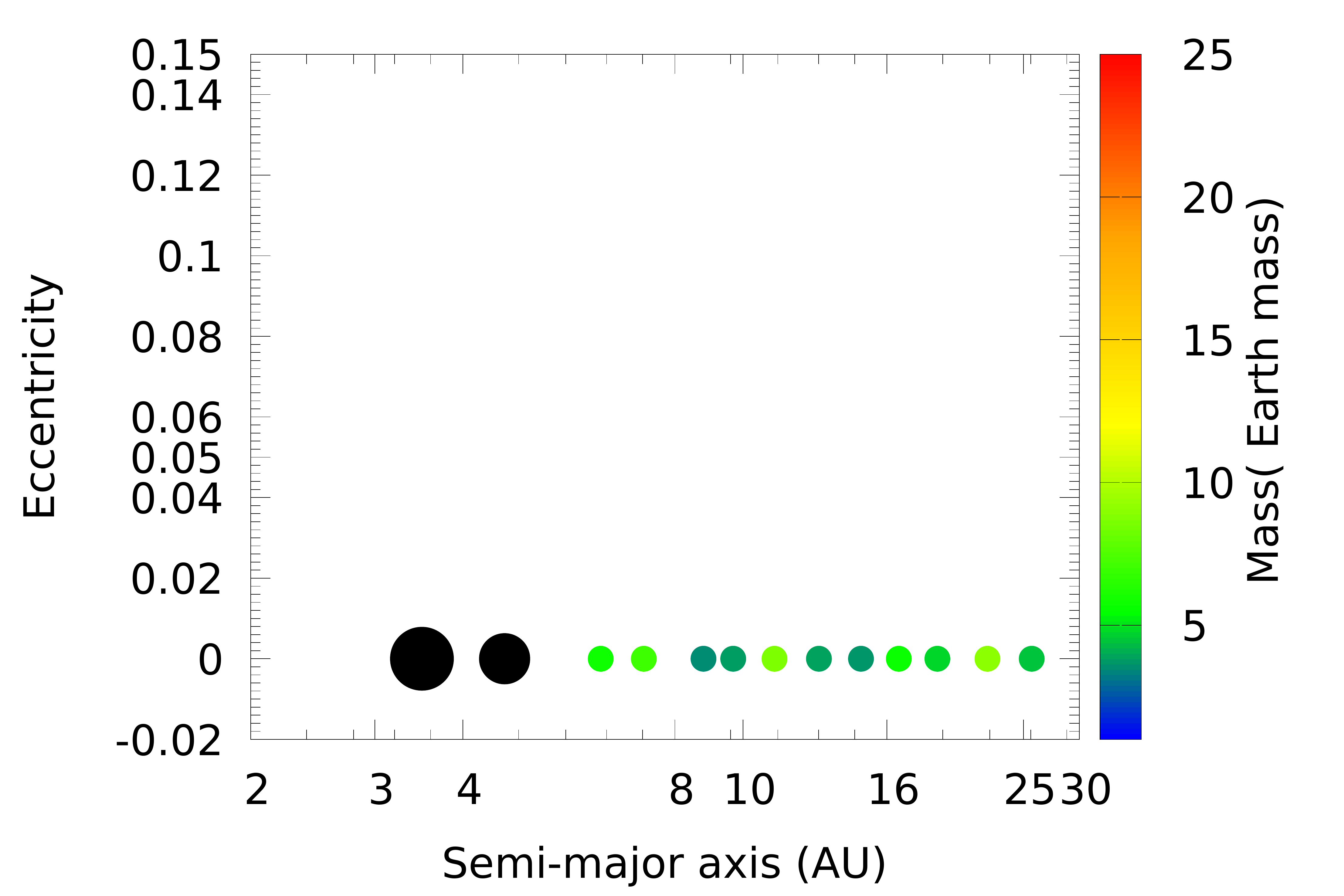} {0.45\textwidth}{(a)}
          \fig{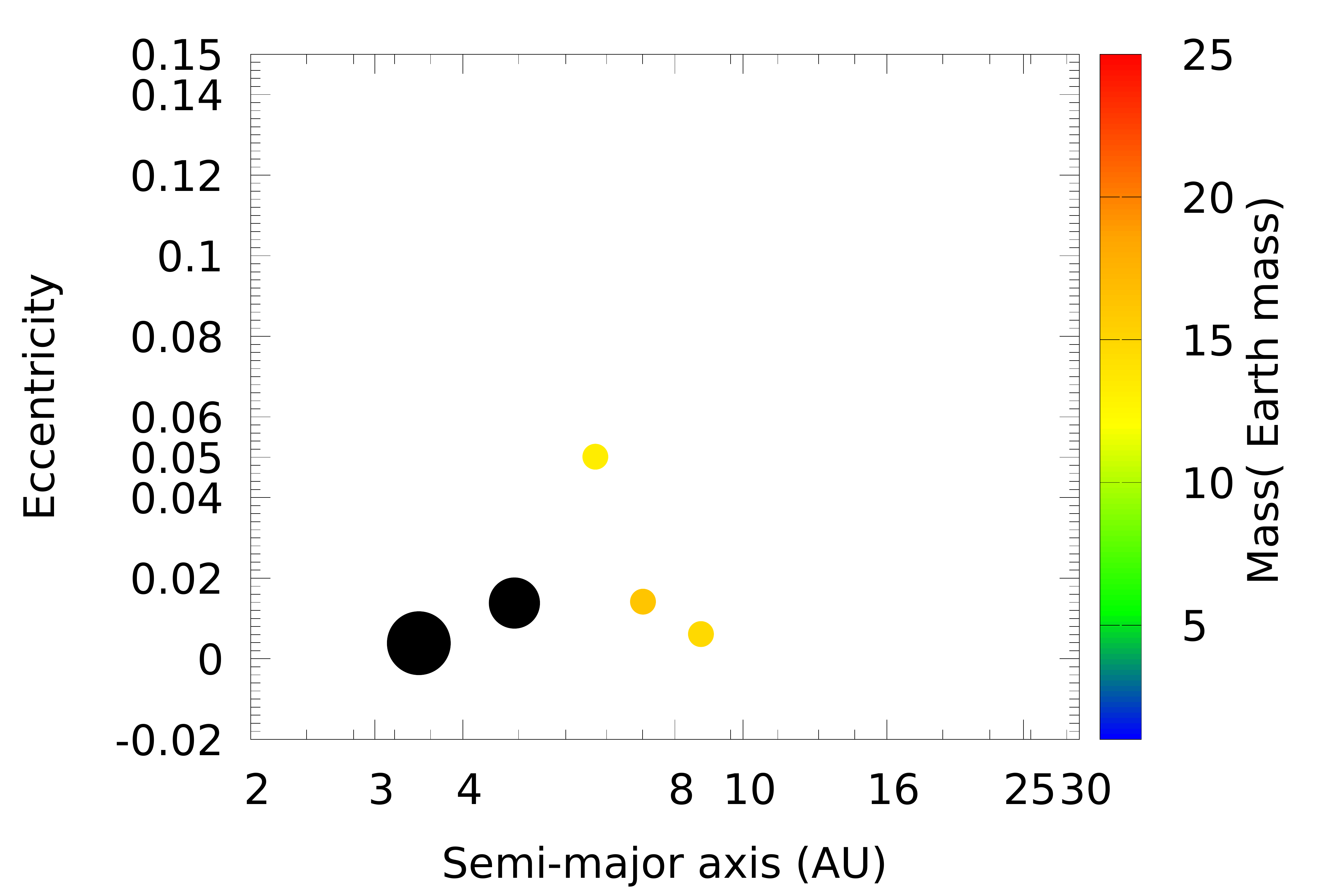}{0.45\textwidth}{(b)}
          }

\gridline{\fig{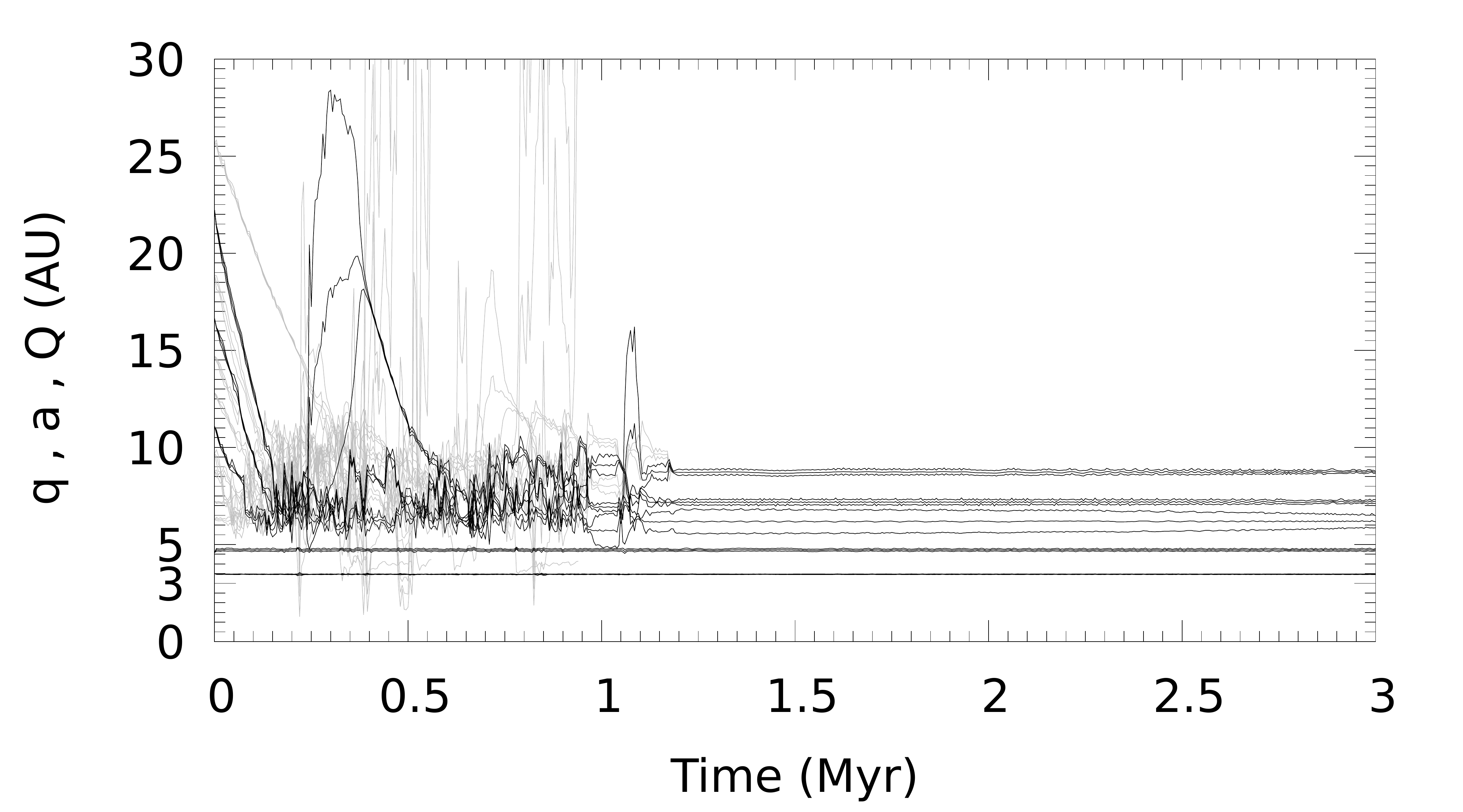} {0.5\textwidth}{(c)}
          }   
\caption{Panels (a) and (b) are two snapshots which represent the eccentricity as a function of semimajor axis of the system with Jupiter, Saturn and a collection of planetary embryos (initial (a) and final (b)) of the
considered \citet{Izidoroetal2015} simulation. Note that 3 planets are produced beyond Saturn, but one will be ejected during the planet instability \citep{NesvornyMorby2012}. Panel (c) represents a complete evolution in pericenter (q), semimajor axis (a), and apocenter (Q) of the same simulation. 
Black lines correspond to the planets and gray lines correspond to the planetary embryos.
\label{fig:interpolation1}}
\end{figure*}

 In this section, we focus on the best simulation of \citet{Izidoroetal2015} in which three Neptune-mass
planets formed and no initial protoplanets were left behind in the Solar System. This final outcome is consistent with the most modern 
version of the Nice model \citep{NesvornyMorby2012}, which drastically increases the success rate in reproducing 
the outer Solar System by invoking the existence of an additional primordial Neptune-mass planet. 
The \citet{Izidoroetal2015} simulation is illustrated in Figure \ref{fig:interpolation1}. In panel (a), we show the initial semi major axes and eccentricities of the planetary system which contains Jupiter (at $3.5$ AU), 
Saturn (at $4.58$ AU) and a collection of $11$ protoplanets distributed with semi major axis in a range from $6$ to $26$ AU with masses between $3$ to $9$ Earth's mass.
All bodies started in circular orbits. In panel (b), we show the final semi major axis and eccentricity of Jupiter, Saturn, Uranus, Neptune and a fifth comparable-mass planet at the end of the $3$ Mys lifetime of the protoplanetary disk 
of gas. This simulation reproduces the real masses of Uranus and Neptune, around of $15$ Earth's masses.
The dynamical evolution of this planetary system is showed in panel (c).  
The planetary embryos migrated inward because of the presence of the gas and were trapped in mean-motion resonances with the giant planets.  
The mutual gravitational interaction among the planetary embryos eventually broke the resonant chains and the system became dynamically unstable.
During this phase, the planetary embryos were scattered by mutual close encounters and encounters with the giant planets. Some objects were ejected from the system while other objects merged,
building the three final massive planets.

For the gas disk, \citet{Izidoroetal2015} used the 1 D radial density distribution obtained from hydrodynamical simulations that made use of the prescription of 
\citet{SSunyaev1973} using the value of alpha of 0.002 (more details about the disk are given in the section \ref{gas}).  To include the effects of type I migration for the planetary embryos,
they followed \citet{Paardekooper2008} invoking the locally isothermal approximation to describe the thermodynamics of the disk. In this case, the Lindblad and the coorbital torques contributions can be 
written in terms of the negative of the local gas surface density and temperature gradients.  To mimic the effects of eccentricity and inclination damping in their simulations, they followed 
\citet{papaloizelarwod2000} and \citet{CNelson2006,CNelson2008}. 
 Note that in Fig. \ref{fig:interpolation1} the ice giants show moderate orbital eccentricity. This is due to the way that the planets are assembled together into mean motion ressonances and about the disk’s properties (eccentricity damping, aspect ratio of the disk, to cite a few).

 \subsection{Methods}
  \label{Methods}
We used N-body simulations conducted in the framework of the study on the formation of Uranus and Neptune of 
\citet{Izidoroetal2015}. We interpolated the orbital elements of all massive bodies (Jupiter, Saturn and the proto-planets) from the best simulation of \citet{Izidoroetal2015} using spline functions. We then used this forced interpolated evolution of the massive bodies (perturbers) in our simulations of planetesimal evolution. In addition to the perturbers, we included also the effects of a gaseous protoplanetary disk on the planetesimals (gas drag and dynamical friction) as detailed in section~\ref{gas}. For consistency, we adopted the surface density for the gas used by \citet{Izidoroetal2015} to compute the migration and tidal damping rates on the proto-planets. 

The simulations of \citet{Izidoroetal2015} assumed Jupiter and Saturn on fixed orbits at $3.5$ AU and $4.58$ AU, respectively, motivated by the initial conditions of the Grand Tack model
\citep{Walshetal2011}. However, the actual migration history of Jupiter and Saturn is not well known, \citet{Bitshetal2015}, for example, using a disc evolution 
and pebble accretion, claimed that a giant planet ending at 5 AU can only form from a seed initially located beyond 20 AU. We expect that, by controlling the migration of the proto-planets precursor of Uranus and Neptune, 
the migration of the Jupiter and Saturn plays an important role in the evolution of the planetesimals disk and its final structure. Thus, we scale the evolution of Saturn and the proto-planets in \citet{Izidoroetal2015} 
simulation relatively to the semi-major axis of Jupiter and we adopt different migration histories for Jupiter in different simulations of the planetesimal disk evolution (See in Sect.~\ref{interp} 
for a complete explanation). Obviously, the time is also rescaled with the orbital radius, so that the orbital periods of all massive bodies obey the Kepler law.

We consider three cases of inward migration of Jupiter from $20$ AU, $15$ AU, and $10$ AU to $5$ AU, one case of outward migration from $2$ AU to $5$ AU and 
one case where the orbit of Jupiter is fixed at $5$ AU.  The interpolation of Saturn and the proto-planets evolution in rescaled coordinates and its implementation with an imposed Jupiter's migration are detailed in Sect.~\ref{interp}, as well as the 
rescaling of the gas properties. This procedure has the merit that each simulation implements the exact same evolution of the embryos (which would not have been the case if we had simulated the initial conditions of 
\citet{Izidoroetal2015} for each Jupiter's migration history), which highlights how the final structure of the planetesimal disk depends on the giant planets migration range, removing stochastic effects. 

We re-run 4 times each simulation with a different planetesimal size of  $1$ km, $10$ km, $100$ km and $1000$ km, each a bulk density of $3.0$ $g/cm^3$ in a total of  $4000$ planetesimals. 
The planetesimals are initially distributed beyond the orbits of the giant planets, but throughout the region occupied by the proto-planets and up to 60 AU.  Thus, the planetesimals initial semi-major axis range depends of the 
initial position of Jupiter. The eccentricities and inclinations of the planetesimal disk are initially chosen as $10^{-3}$. Their argument of pericenter and longitude of ascending node are randomly selected between $0$ and $360$ 
degrees.

We used the REBOUND code \citep{Rein2012,ReinSpiegel2015,ReinTamayo2015} to perform the N-body simulations of the  planetesimals interacting with the Sun, Jupiter, Saturn and planetary embryos interpolated from the output of 
\citet{Izidoroetal2015} simulation, and with the gas of the protoplanetary disk. In our simulations, during the gas disk phase planetesimals are assumed to be non-interacting test particles.

\subsubsection{Interactions of planetesimals with the gaseous protoplanetary disk}
\label{gas}

As explained in \citet{Izidoroetal2015}, although real hydrodynamical simulations would be ideal to study the problem, they would be impracticable given the multi-Myrs timescales and the number of bodies involved. 
Thus, the gas density is used to compute synthetic forces acting on the planetesimals, and then integrating the evolution of the system with a N-body code. 

\begin{figure}[]
\plotone{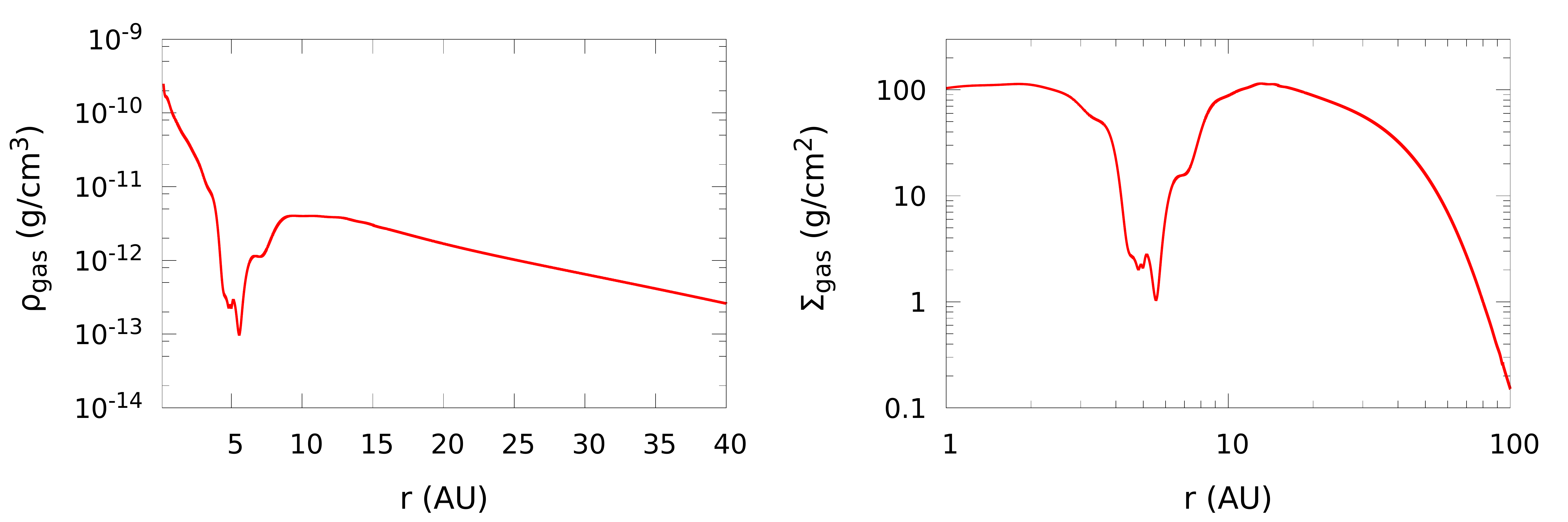}
\caption{The protoplanetary disk profile at $t=0$ from hydrodynamical simulations in \citet{Izidoroetal2015}. The left panel shows the gas volumetric density and the right panel shows the surface density.\label{fig:diskprofile22}}
 \end{figure}

From the simulation of \citet{Izidoroetal2015} we have the surface density of the disk $\Sigma(r,t)$, which accounts for the gaps opened by Jupiter and Saturn in a disk as well as the partial depletion of the inner disk. 
This profile had been calculated from a hydrodynamical simulation, starting from a disk with initial surface density $\Sigma(r)=1,000 g/cm^2 (AU/r)$, (see right panel of Fig. \ref{fig:diskprofile22}) and assuming a uniform decay 
over time with $\exp(-t/\tau_{gas})$. We also adopt from \citet{Izidoroetal2015} the aspect ratio of the disk   
\begin{equation}
h=\frac{H(r)}{r} =0.033r^{0.25}\ ,
\end{equation}
where $H(r)$ is the pressure scale height at radius $r$. Thus, the density of the gas has a $z$-distribution given by 
\begin{equation}
\rho(r,z,t)=\frac{\Sigma(r,t)}{\sqrt{2\pi}H(r)}\exp{\left(-{\frac{z^2}{2H^2}}\right)}\ , \end{equation}
which is shown in the left panel of Figure \ref{fig:diskprofile22}.  
In the hydrodynamical simulations, the disk viscous stress was modeled using the standard ``alpha'' prescription \citep{SSunyaev1973}. 
The disk viscosity is given by: 
\begin{equation}
\nu=  \alpha c_{s} H\,
\end{equation}
where, $c_s$ is the isothermal sound speed and the value of $\alpha$ is 0.002 in Izidoro et al. (2015)'s simulations.

The sub-keplerian velocity of the gas in the mid-plane was also read from the hydrodynamical simulations and includes the pressure gradient effect, consistent with the profiles of $\Sigma(r)$ and $H(r)$ reported above. 
Section~\ref{interp} will discuss how all these quantities are rescaled when Jupiter is assumed to be on an orbit with a different semi-major axis. 

With this information, we have all quantities needed to compute the gas drag effects on our planetesimals. 
The aerodynamic gas drag force on
a particle moving in a gas disk environment is expressed in function of its shape, size, velocity and gas conditions. In the particular case of spherical body with radius $R$, the drag force is in a direction 
opposite to the particle velocity and could be expressed by:
\begin{equation}
\vec{F}_D= - \frac{1}{2} C_D \pi R^2 \rho_g v_{rel} \vec{v}_{rel} , 
\end{equation}
where $C_D$ is the drag coefficient and $\vec{v}_{rel}$ is the vector relative velocity between the gas and the planetesimal.
The drag coefficient for a spherical object is a function of the Reynolds number ($R_e$) which is a measure of the turbulence 
of the gas in the wake of a planetesimal, the Mach Number ($M$) and Knudsen number ($K$). 
To evaluate the drag coefficient we used the approach of \citet{Brasseretal2007} where they estimated 
the values of $M$, $K$ and $R_e$ as a function of planetesimal's size and velocity.

\citet{Grishin2015} showed that for planetesimals within the mass range of $m\sim {10}^{21}-{10}^{25}\;{\rm{g}}$ there is another planetesimal-gas interaction, possibly dominating over gas-drag. 
This force is known as gas dynamical friction (GDF) and considering the same approach of \citet{Grishin2015}, 
the GDF force is given by: 
\begin{equation}
\vec{F}_{GDF}= - \frac{4 \pi G^2m^2 \rho_g}{v_{rel}^3} \vec{v}_{rel} I(M) , 
\end{equation}
$I(M)$ is a dimensionless factor depending on the Mach number (M) and it is given by:
\begin{itemize}
 \item If $M<1$ then:
 
\begin{equation}
I(M)=\frac{1}{2}\ln{\left(\frac{1+M}{1-M}\right)} -M
\end{equation}
 
 \item If $M>1$ then: 
 
\begin{equation}
 I(M)=\frac{1}{2}\ln{\left(1- \frac{1}{M^2}\right)+ \ln{\left(\frac{v_{rel}t}{R}\right)}} 
\end{equation}
 
\end{itemize}

\begin{figure*}
\gridline{\fig{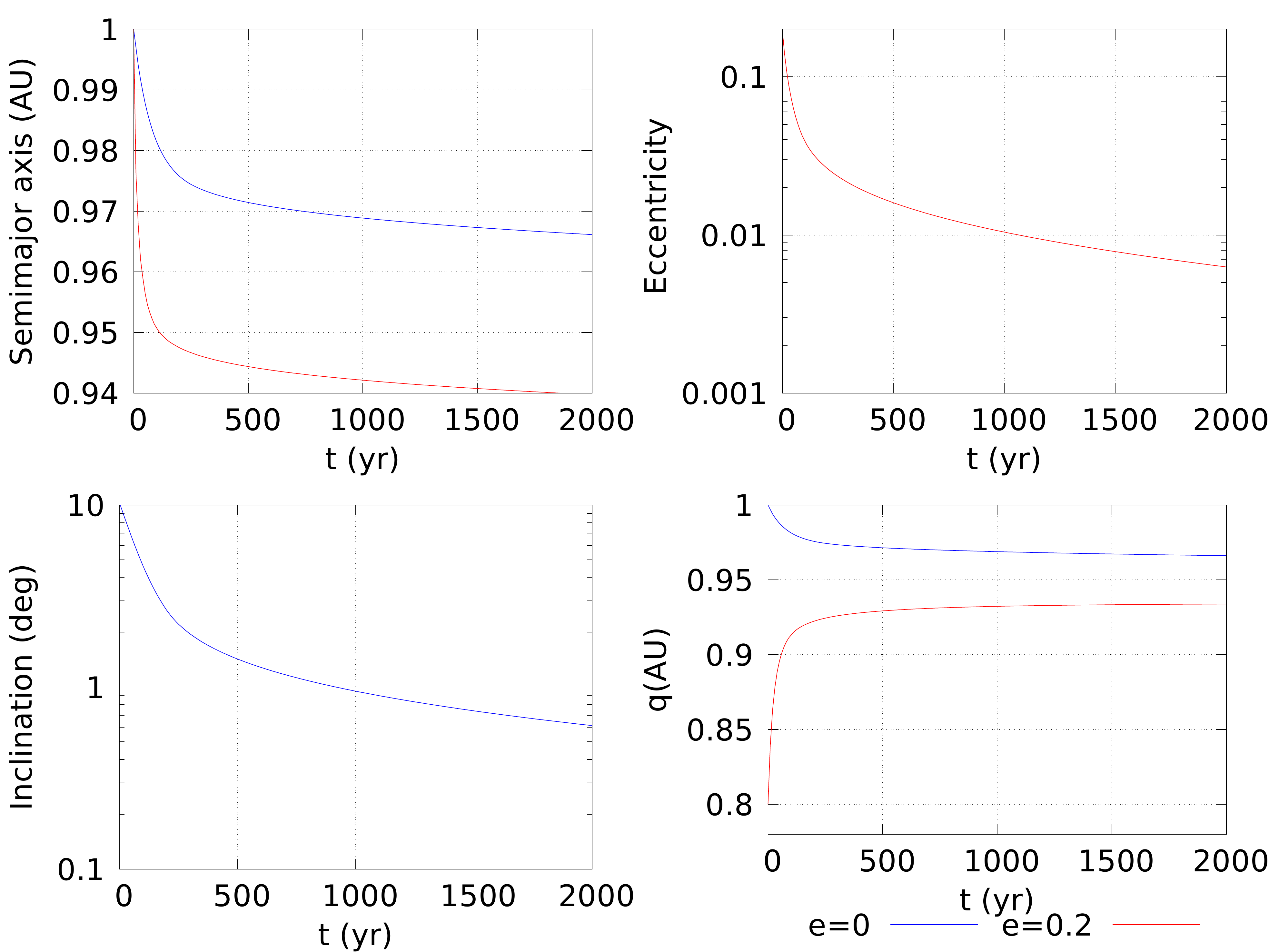} {0.6\textwidth}{}}
\caption{A summary of the action of the gas drag on a planetesimal's orbital elements. The blue solid line shows a first particle with initial eccentricity and inclination equal to 0 and 10 degrees, respectively.
The red solid line shows a second particle with initial eccentricity equal to 0.2 and a planar orbit. Going clockwise from the top-left the orbital elements are the semi major axis, eccentricity, perihelion and inclination ploted in function of the time. \label{fig:gaseffect}
}
\end{figure*}

We show in Figure \ref{fig:gaseffect} the results of the effects of the aerodynamic gas drag in the evolution of two test particles of a radius of 2 km using the gas profile of \citet{Grishin2015}, 
for comparison with that work. The first particle (blue) has 10 degrees of inclination and eccentricity equal to 0. The second particle has eccentricity equal to 0.2 and it has a planar orbit (red). 
The eccentric particle moves faster than the local gas when it is at pericenter. As a consequence, the drag decreases the apocenter distance of the particle. At apocenter, the particle moves slower than the local gas so that the drag 
increases the pericenter distance. The orbit will circularize at a distance between the two extremes, which depends on the radial profile of the gas since the latter governs how much damping occurs near perihelion vs. aphelion. 
We can see that $q$ decreases over time for the orbit initially circular and inclined, while it increases for the orbit initially eccentric and planar (see also in \citet{Brasseretal2007}).

\begin{figure*}
          \gridline{ \fig{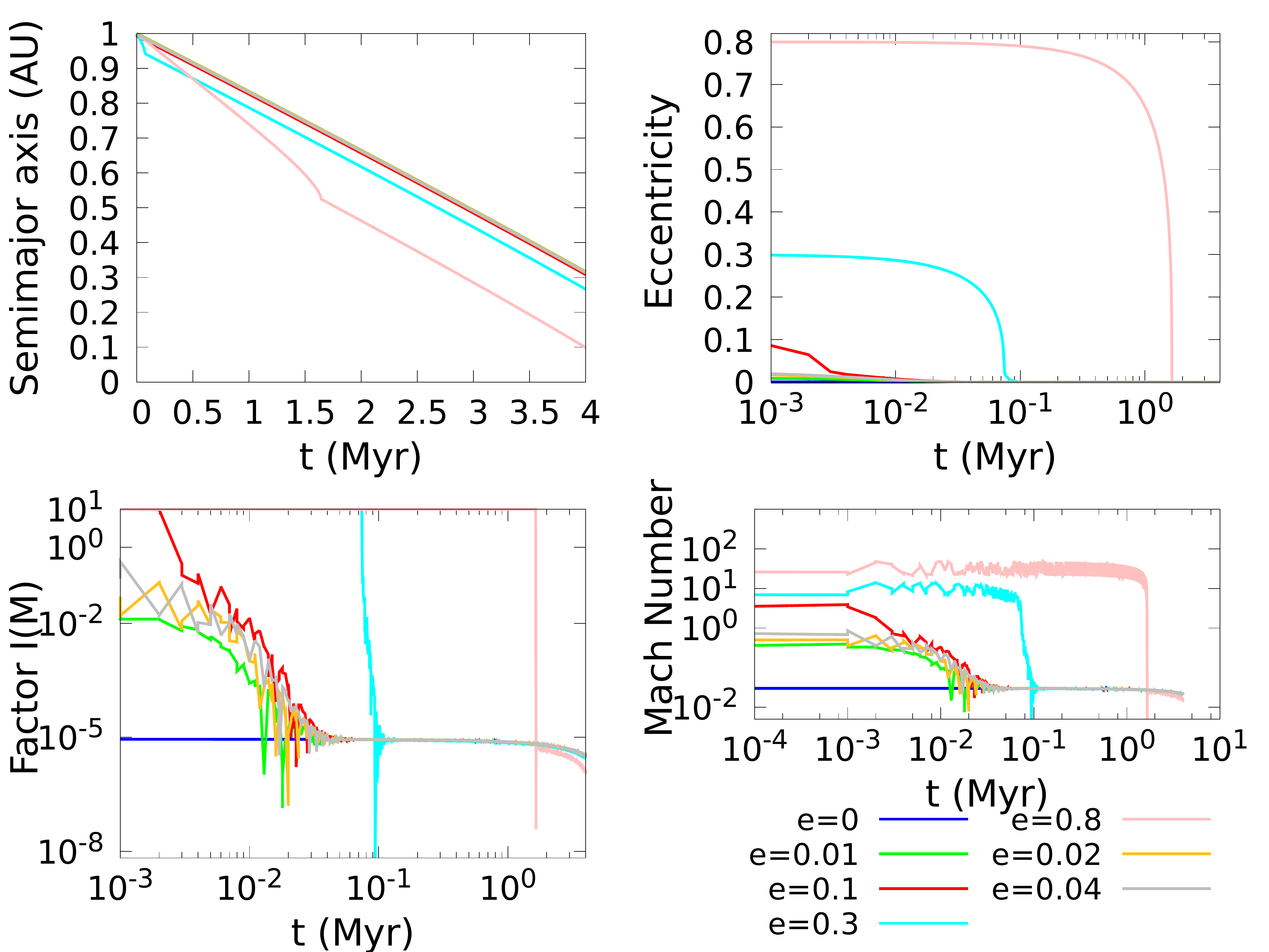}{0.65\textwidth}{}
          }
\caption{A summary of the effect of the gas dynamical friction force on particle with a mass of $10^{22}$ kg with different initial eccentricities. The orbits starts at $1$ AU
and they are planar orbits. Going clockwise from the top-left the orbital elements are the semi major axis, eccentricity, the Mach number ($M$) and the factor $I(M)$ ploted in function of the time.
\label{fig:gaseffect2}}
\end{figure*}

We show in Figure \ref{fig:gaseffect2} the effects of the gas dynamical friction on particles with mass of $10^{22}$ kg and with different eccentricities. 
We found the same results of \citet{Grishin2015} (aerodynamic gas drag is neglected in these tests): particles with initial low eccentricities circularized much faster than high eccentricity orbits. 
There are two regimes of radial drift that depend on the eccentricities: an exponential decay in semi major axis before the circularization of the orbit and a linear decay when the orbits of the particles are circular 
\footnote{Although \citet{Grishin2015} have found an exponential decay when the orbits of the particles are circular,
this seems to be due to a mistake in their calculation of 
the sound speed of the gas (Grishin and Perets, private communication).}.  

\begin{figure*}
          \gridline{ \fig{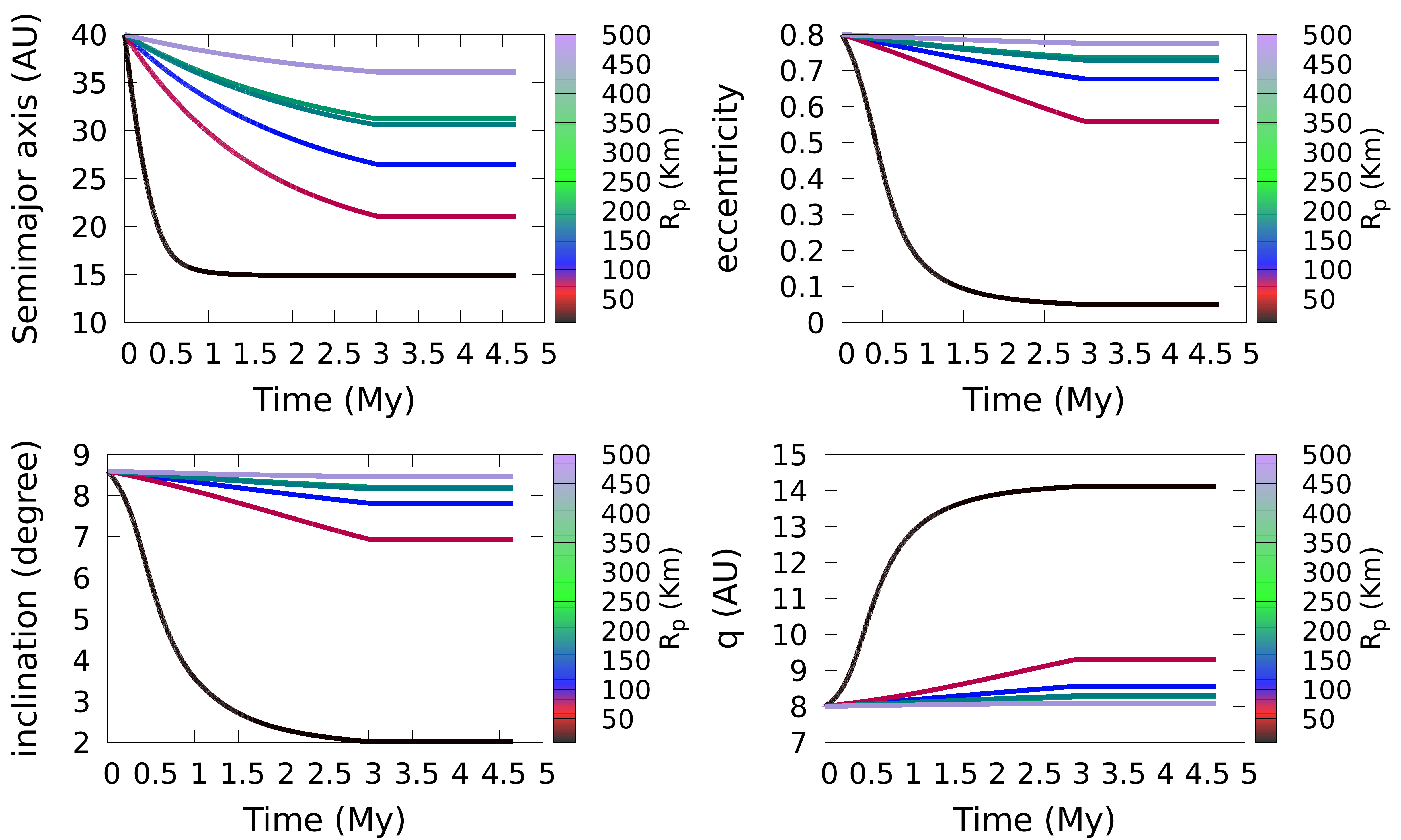}{0.7\textwidth}{}
          }
\caption{A summary of the action of the aerodynamic gas drag and dynamical friction gas drag on a planetesimal's orbital elements. In this simulation, we used the surface density presented in Fig. \ref{fig:diskprofile22} 
   assuming a uniform decay over time of 3 My, the gas is removed at 3 My. Going clockwise from the top-left the orbital elements are the semi major axis, eccentricity, perihelion and inclination ploted in function of the time.}   
   \label{fig:traject} 
\end{figure*}

In Fig. \ref{fig:traject} we show a simulation of the set of particles with different sizes under effects of the aerodynamic gas drag and dynamical friction gas drag. 
For this simulation, we used the surface density of the disk presented in Fig. \ref{fig:diskprofile22}. The density of the particles is 3 $g/cm^{3}$ and minimum particle radius ($R_p$) is 1 km (black lines) whereas maximum particle radius is 500 km (purple lines). 
The orbits of the particles have initial eccentricity of 0.8, initial semi-major axis of 40 AU and initial inclination of 8.6 degrees. 
We observed that km-size planetesimals present a more significative radial drift, eccentricity and inclination damping than 100 km-size planetesimals. 
It shows that the effects of the aerodynamic gas drag are more pronounced than dynamical friction effects. We will show in 
section \ref{results}, that the final dynamical state of the primordial planetesimal disk under the 
influence of the Jupiter, Saturn, planetary embryos and the gas drags is planetesimal size dependent. 
With a dominant aerodynamic drag, there is strong eccentricity and inclination damping in favor of km-sized planetesimals, 
which makes them a dynamically cold configuration.

% 
%     \begin{figure}[H]
%     \begin{center}
%     \includegraphics[scale=0.15]{/media/rafanw72/hda/Pesquisa/Morby/project_Morby/gas_interaction/Izidoro_Uranus_Neptune_Gas/test1/interpolation_gas/code_validation/both_effects/size_distribution/version1/size_distribution/ClassB/b1/graph_figure_new/traject.pdf} 
%    \caption{A summary of the action of the aerodynamic gas drag and dynamical friction gas drag on a planetesimal's orbital elements. In this simulation, we used the surface density presented in Fig. \ref{fig:diskprofile22} 
%    assuming a uniform decay over time of 3 My, the gas is removed at 3 My. Going clockwise from the top-left the orbital elements are the semi major axis, eccentricity, perihelion and inclination ploted in function of the time.}   
%    \label{fig:traject} 
%     \end{center} 
%     \end{figure}
%   
%   

Note that we do not need to implement the tidal forces of the disk of gas on the proto-planets, because we take their evolution from the output of the \citet{Izidoroetal2015} simulation, which already implemented these forces.

\subsubsection{Interpolation of Uranus and Neptune accretion}
\label{interp} 

In this section, we present how we interpoled one successful simulation of \citet{Izidoroetal2015}, which we take as reference.
For the interpolation we used cubic splines. We interpolate all orbital elements between two successive outputs. The time-resolution of the output in \citet{Izidoroetal2015} simulation is $5,000$ y. 
Because this exceeds the orbital period of the bodies, we calculate the number of orbits between two successive outputs using the information of the orbital frequency (mean motion, $n$) 
to obtain an averaged orbital period, which is then slightly adjusted so that the position of the bodies at the end of the $5,000$ y timesep matches that recorded in the original simulation. 
The mean orbital period is then used to calculate the fast variation of the mean longitude within the timestep. Our interpolation deals with events of collisions (merge) and ejection of the proto-planets at 
the exact moment that these events happen. We eliminate the proto-planets when they are ejected from the Solar System ($e>1.0$). 
When a collision occurs, the remaining proto-planet gets a new mass equal to the sum of the masses of the two bodies involved in 
the event and we put to zero the mass of the proto-planet eliminated during the merge, so that it won't have any further dynamical influence on the planetesimals.
The masses of proto-planets and giant planets do not increase due to planetesimal accretion in the simulations. 
% The interpolated evolutions of the semi major axis, eccentricity, inclination, longitude of the perihelion ($\varpi$), node ($\Omega$) and mean ($\lambda$) of Jupiter are 
% showed in Figure \ref{fig:interpolation2}, the curves represent the results of our interpolation whereas the points come from the original simulations of \citet{Izidoroetal2015}, showing that our interpolation procedure 
% reproduces with fidelity the original evolution of the system.
There are other works using interpolation in order to reassess evolution from previous simulations (\citet{Deienno2011,Roigenesvorny2015,roig16,Deiennoetal2018,RIBEIRODESOUSA2018} to cite a few).
 
% 
% 
% \begin{figure*}
%  \gridline{\fig{fig/interpolation/fig1a.pdf}{0.5\textwidth}{(a)}
%           \fig{fig/interpolation/fig2a.pdf}{0.5\textwidth}{(b)}} 
%           
%             \gridline{\fig{fig/interpolation/fig3a.pdf}{0.5\textwidth}{(c)}
%           \fig{fig/interpolation/fig4a.pdf}{0.5\textwidth}{(d)}} 
%          
%             \gridline{\fig{fig/interpolation/fig5a.pdf}{0.5\textwidth}{(e)}
%           \fig{fig/interpolation/fig6a.pdf}{0.5\textwidth}{(f)}} 
%           
% \caption{The orbital evolution of the semi major axis (a), eccentricity (b), inclination (c), longitude of the perihelion, $\varpi$, (d), node, $\Omega$ (e)
% of Jupiter. The solid red curves show the results of our interpolation using cubic splines whereas the solid curves come from the \citet{Izidoroetal2015} simulations. Panel (f) shows
% the evolution of the mean longitude of Jupiter, $\lambda$, where the points represent the values of $\lambda$ which come from \citet{Izidoroetal2015} simulations.\label{fig:interpolation2}}
% \end{figure*}

\begin{figure}[]
\plotone{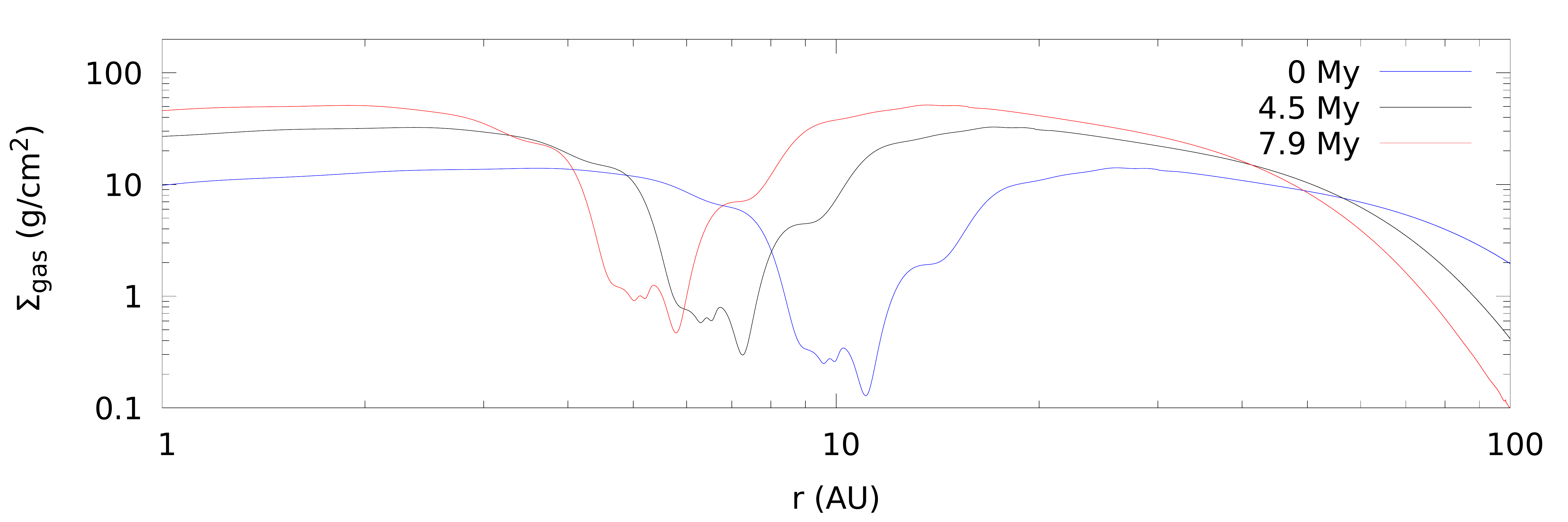}
\caption{The surface density profiles $\Sigma(r/R)/R^2$. Different colours lines correspond to different times (0 My, 4.5 My and 7.9 My). It corresponds the case with Jupiter migrating from 10 to 5 AU.\label{fig:diskprofilesigma}}
 \end{figure}

When we interpolate the evolution of the massive bodies, we rescale their semi-major axes according to the desired location of Jupiter at time t, given an imposed migration pattern $a_{jup}(t)$. 
In other words, if $a^I(t)$ is the semi-major axis of a body $I$ at time $t$ in \citet{Izidoroetal2015} output (or the interpolated value from the output) and  $a^I_{jup}$ is the semi major axis of Jupiter, 
we convert $a^I$ into $a(t)=a^I(t) R(t)$ where $R(t)=a_{jup}(t)/a^I_{jup}(t)$. All the other orbital elements are kept unchanged.
Due to the fact that the direct perturbation of the protoplanets embedded in the planetesimal disc is larger and happens in a shorter 
timescale than any perturbation by a possible secular or mean motion resonance, this should not place a problem nor change our main results/conclusions.
Then, the orbital elements are converted into positions, which are 
used in the N-body code to compute the forces that the massive bodies exert on the planetesimals. In order to preserve the orbital periods, the length of the timestep $dt^I$ of Izidoro's simulation is 
rescaled as $dt=dt^I [a_{jup}(t)/a^I_{jup}(t)]^{3/2}$. The simulation time is incremented by $dt$ (and not $dt^I$). By doing so, the total duration of the simulation increases, if Jupiter is farther 
than the original $3.5$ AU (e.g. the simulation illustrated in Fig. \ref{fig:nonmigrat1jc} where Jupiter is kept at $5$ AU will be $5.12$ My instead of the original $3$ My). 
Concerning the gas, given the rescale factor $R(t)$ on semi-major axes, the surface density at the heliocentric distance $r$ is computed as $\Sigma(r/R(t)) /( R(t) )^{2}$, where $\Sigma$ is the surface gas density in \citet{Izidoroetal2015}'s simulation. 
As for the gas velocity in the azimuthal direction, it is rescaled as $v_{\theta}(r/R(t)) /\sqrt{R(t)}$ . The rescaled gas quantities are used to compute forces due to gas drag and gas dynamical friction on the planetesimals. The aspect ratio 
of the disk is $h(r/R(t))$. In hydrodynamical simulations, the migration of the planets 
temporarily affects the surface density distribution of the gas, for instance by pushing gas inside its orbit and leaving 
a slightly depleted outer disc behind (acting like a snowplough as shown by \citet{Crida_Bit2017}). However, 
on longer timescales, the situation stabilises and migration rate of the planets is proportional to the viscosity
of the disk and the shape of the gap is not affected when the planet moves \citep{C.Robert2018}. We 
show in Fig. \ref{fig:diskprofilesigma} how the location of the surface density of the gas disk changes overtime during the Jupiter migration.

%Of course, Jupiter's growth would already have left a strong imprint on the local planetesimal distribution~\citep{raymond17}.

\subsection{Results}
\label{results}
In this section, we present the results of our numerical simulations considering a planetesimal disk interacting gravitationally with the Sun, with the giant planets and the proto-planets 
(described in section~\ref{interp}), and with the gas using the protoplanetary disk described in section~\ref{gas}. The goal is to determine the dynamical state of the 
planetesimal disk after the gas phase, in particular to assess the distance between Neptune and the inner edge of the planetesimal disk.

We performed five simulations using the package REBOUND code \citep{Rein2012} with the integrator IAS15 \citep{ReinSpiegel2015}.
Each simulation represents a different imposed orbital migration history of Jupiter, with all other massive bodies rescaled accordingly from the \citet{Izidoroetal2015} simulation, as explained before. 
The semimajor axis of Jupiter evolved as a function 
\begin{equation}
a_{J}(t)= a'_J+ a'\exp(-t/\lambda),
\end{equation}
where $a'_J$ is the final semimajor axis of Jupiter and $a'$ is set so that $a'_J+a'$ is the initial position of Jupiter.
% 
% 
% Because Jupiter is massive enough to open a gap in the disk, and it migrates in the type-II regime, the migration time 
% scale of the planet is the accretion time scale of the disk which depends of the disk viscosity. 
% In our simulations, we set the viscosity parameter $\alpha=0.002$ as a consequence both the growth and migration timescale 
% are shorter (Pierens and Raymond 2011). 
We adopted an e-folding timescale $\lambda=1$ My for all the simulations.

Table \ref{tab:mathmode} shows our set of the simulations. We performed a simulation considering the giant planets on non-migrating orbits (section {\ref{JC}}), with Jupiter and Saturn at $5$ and $6.8$ AU, and a simulation 
considering Jupiter and Saturn migrating outward (section {\ref{JO}}) in agreement with Grand Track Scenario \citep{Walshetal2011,Brasseretal2016}.
% \textbf{Bitsch et al. (2015) reproduced the masses and planetary orbits of the four giant planets in the solar system using a disc evolution 
% and pebble accretion model. They showed that Jupiter can be formed from distances around of 20 AU. Motived by their work,  
However, because the migration history of Jupiter and Saturn in their natal gas disk is poorly constrained \citep{Bitshetal2015}, we also performed simulations in which Jupiter migrated inward from $10$ to $5$ AU, from $15$ to $5$  AU and from $20$ to $5$ AU (in section {\ref{JI}}). 
In all these scenarios, Saturn migrated inward as well because we rescaled Saturn's semi-major axis using Jupiter's semimajor axis (as we have shown in Section 2.2).

For each imposed migration history, we used four independent simulations using different planetesimals sizes, with diameters of $1$ km, $10$ km, $100$ km and $1000$ km.
The planetesimal size matters because of the size-dependent effects of gas drag and gas dynamical friction.
All the simulations start with fully formed Jupiter and Saturn, multiple planetary embryos
and a planetesimal disk with 4,000 planetesimals with eccentricities and inclination of $0.001$ and extended from $10$ up to $60$ AU.

\begin{deluxetable*}{ccCrlc}[b!]
\tablecaption{Set of simulations\label{tab:mathmode}}
\tablecolumns{3}
\tablenum{1}
\tablewidth{0pt}
\tablehead{
\colhead{Name of the simulation set} &
\colhead{Criteria for Jupiter's migration} & 
\colhead{Time of the dispersal of the gas} &
}
\startdata
\textbf{Jup\_static} & Jupiter on non-migrating orbits at 5 AU & 5.2  Myr\\
\textbf{Jup\_outward} & Jupiter migrating outward from 2 to 5 AU & 3.7  Myr\\
\textbf{Jup\_10AU\_in} & Jupiter migrating inward from 10 to 5 AU & 7.9  Myr\\
\textbf{Jup\_15AU\_in} & Jupiter migrating inward from 15 to 5 AU & 11  Myr\\
\textbf{Jup\_20AU\_in} & Jupiter migrating inward from 20 to 5 AU & 14.9  Myr\\
\enddata
\end{deluxetable*}

\clearpage

\subsubsection{Jupiter on non-migrating orbits}
\label{JC}
    \begin{figure}
     \gridline{ \fig{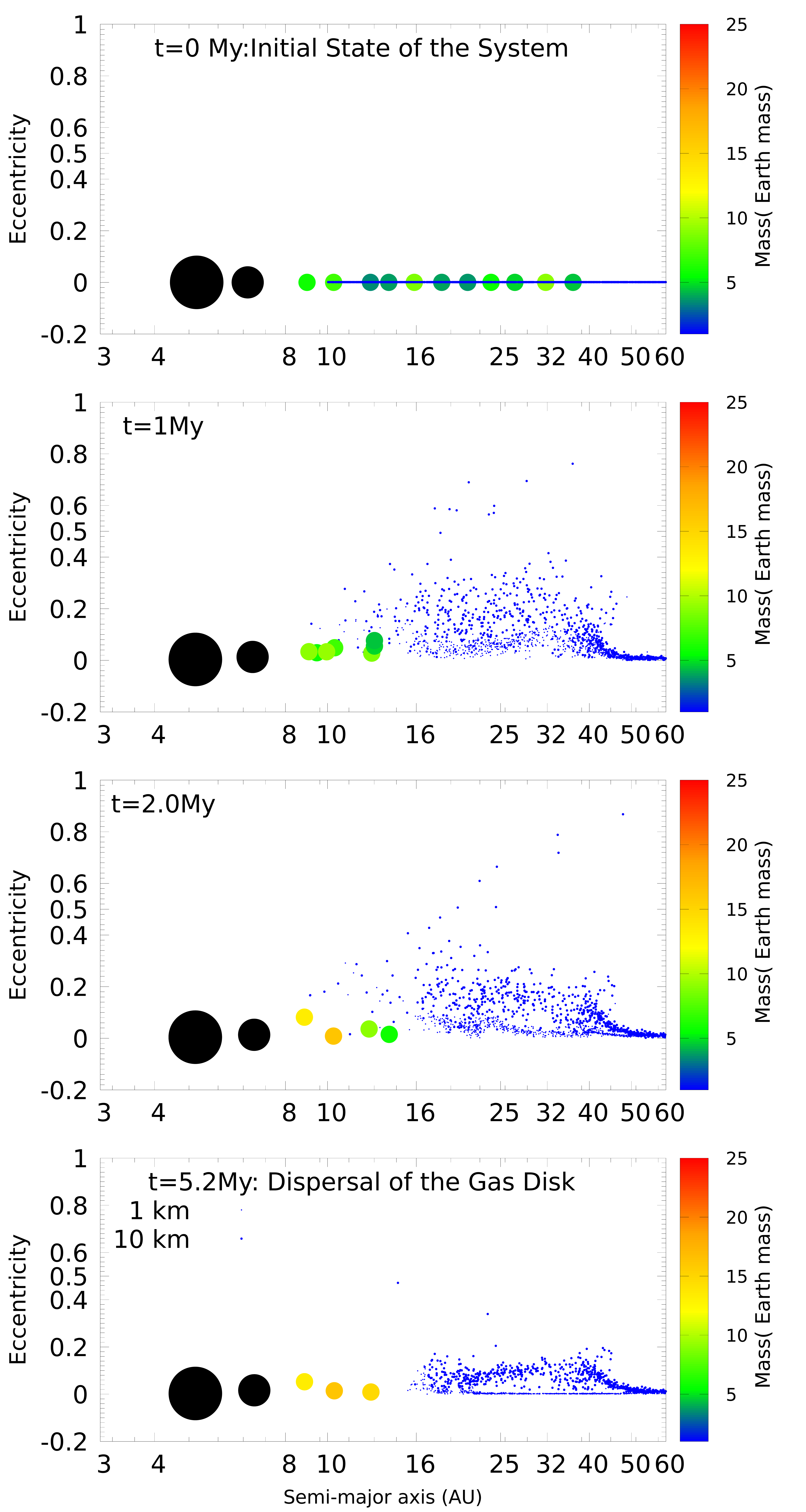}{0.5\textwidth}{(a)}
          \fig{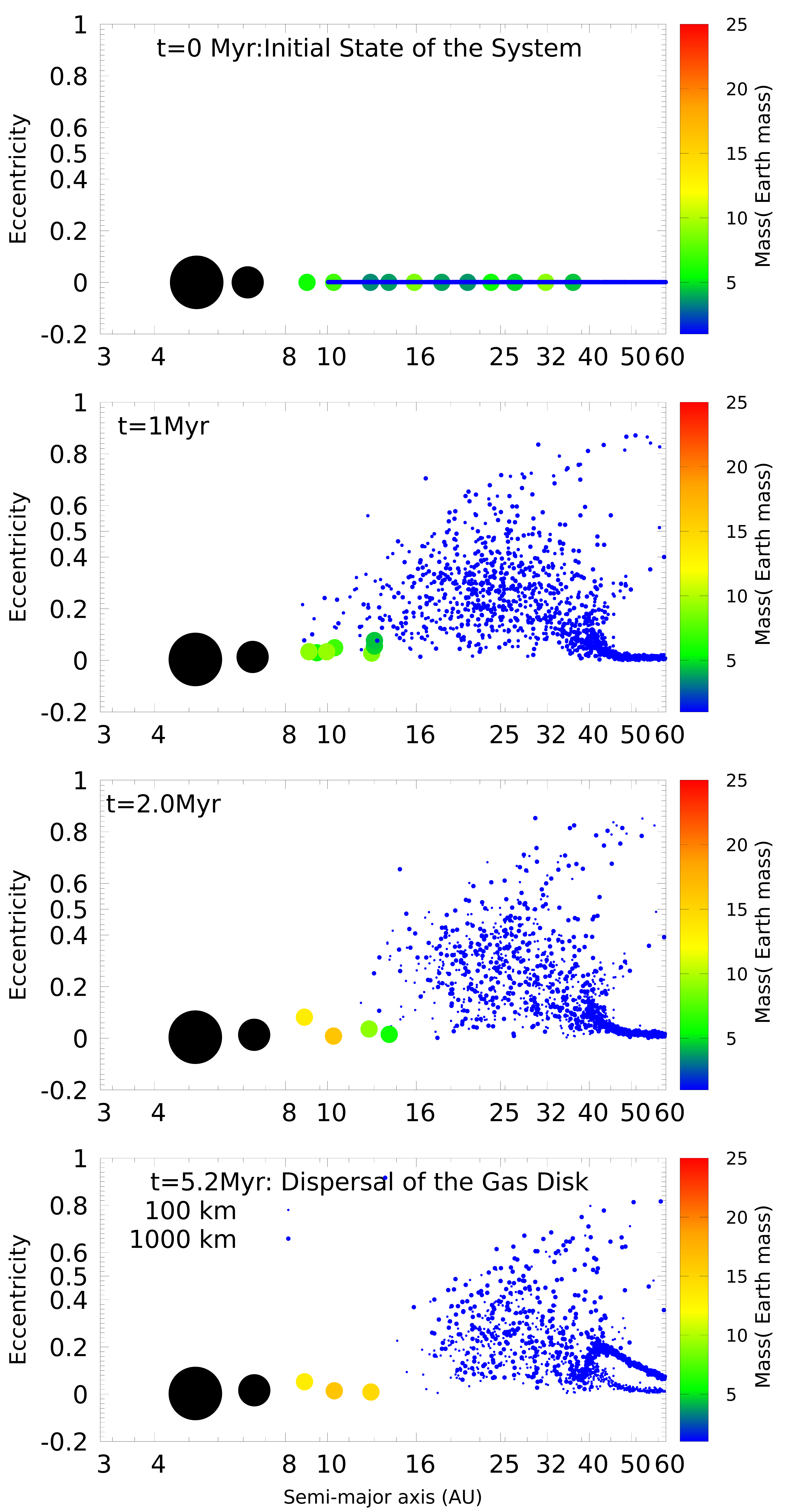}{0.5\textwidth}{(b)}}
%           \gridline{
%           \fig{fig/Jup_static/fig2.pdf}{0.52\textwidth}{(b)}
%            \fig{fig/Jup_static/fig3.pdf}{0.52\textwidth}{(c)}
%           }
\caption{Eccentricity/Semi-major axis plot portraying the dynamical evolution of the case \textbf{Jup\_static} (table \ref{tab:mathmode}). Panel (a)
represents the dynamical evolution for a co-addition of planetesimals with sizes of 1km and 10km . Panel (b) shows the dynamical evolution but for a co-addition of planetesimals sizes of 100km and 1000km.  
The color box represents the mass of the particles, except for Jupiter and Saturn (we use the size of each point to represent the mass of Jupiter and Saturn and the planetesimals sizes).}
\label{fig:nonmigrat1jc}
\end{figure}

    \begin{figure}
     \gridline{ \fig{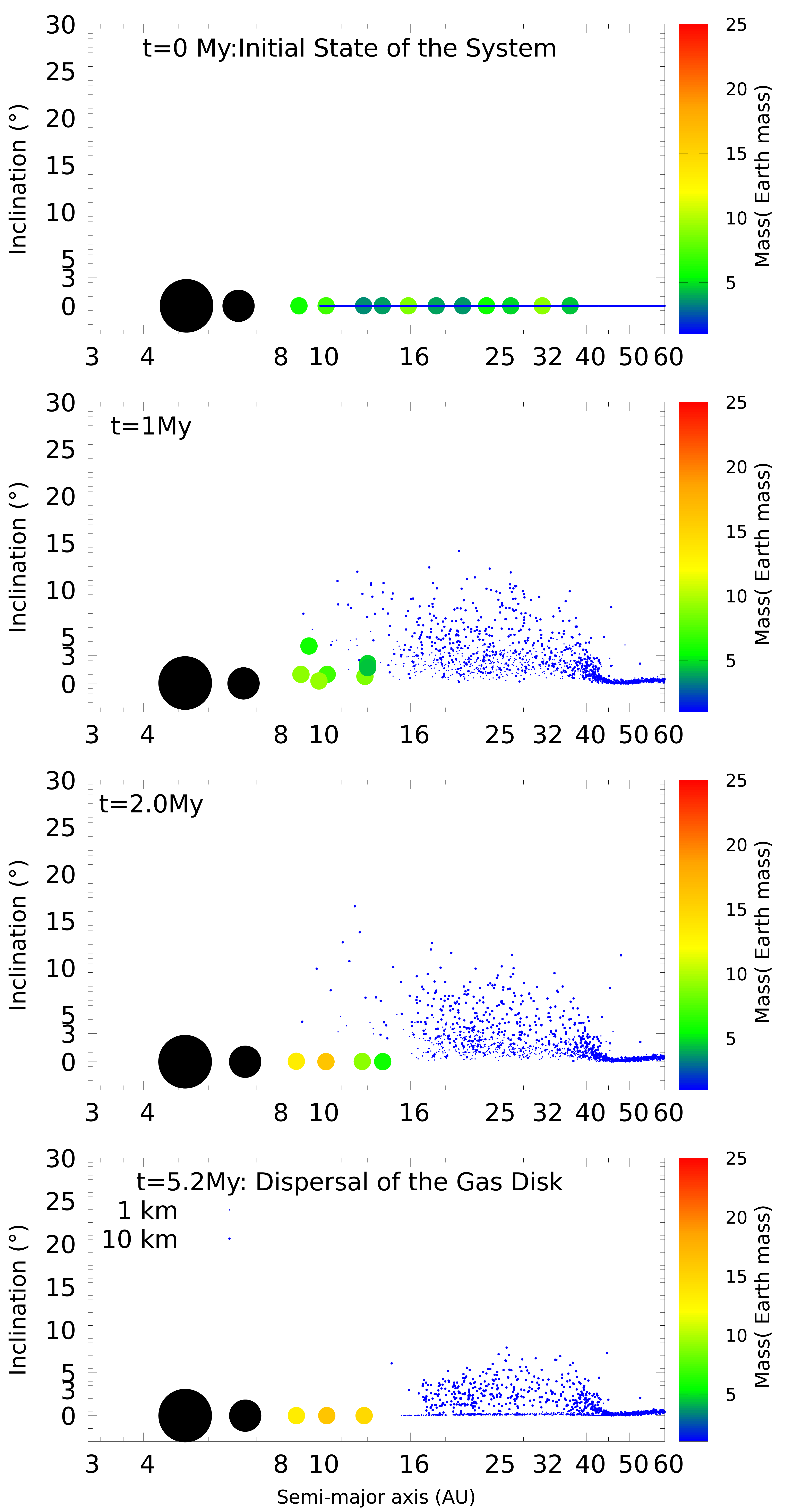}{0.5\textwidth}{(a)}
          \fig{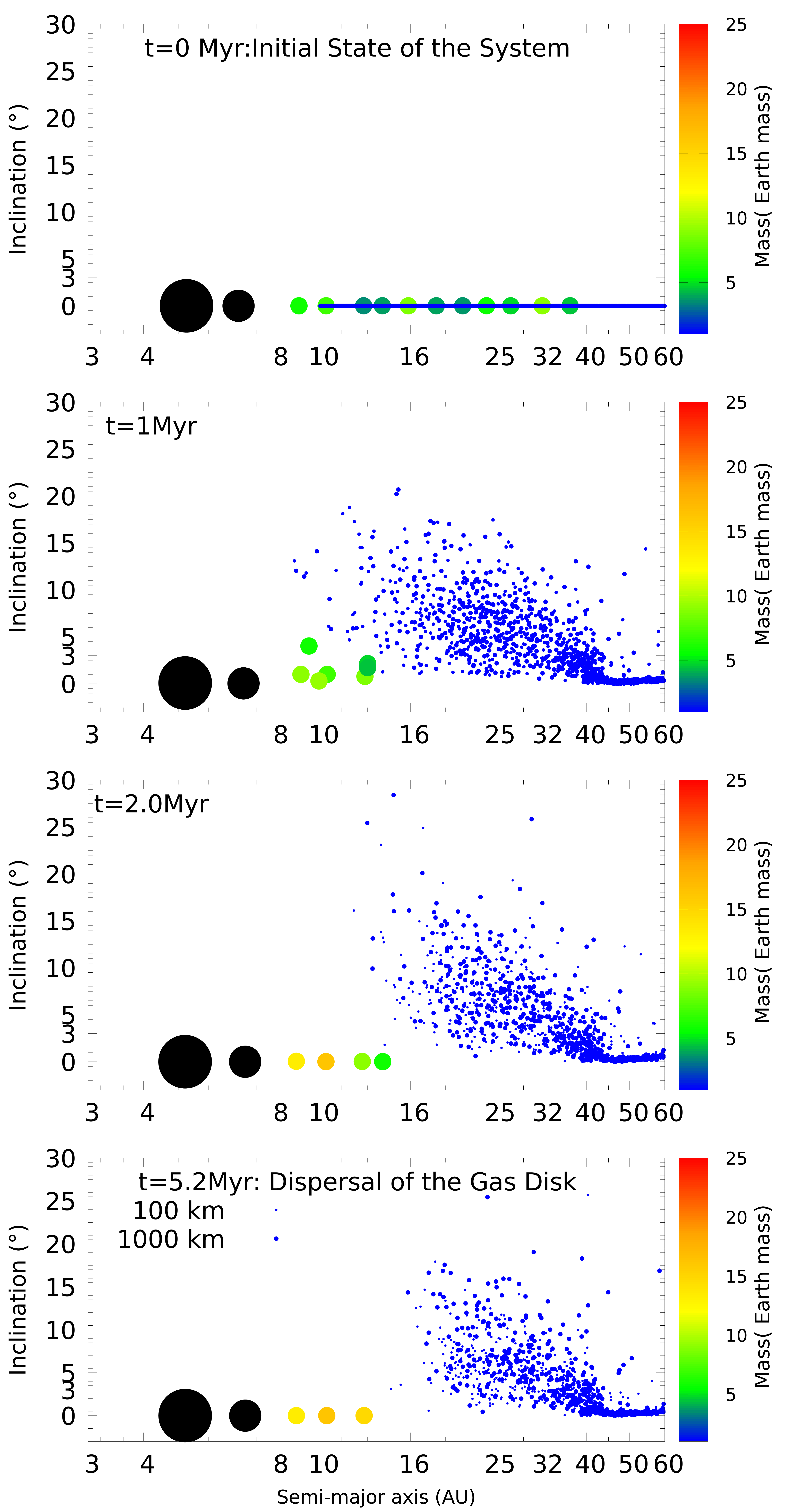}{0.5\textwidth}{(b)}}
%           \gridline{
%           \fig{fig/Jup_static/fig2.pdf}{0.52\textwidth}{(b)}
%            \fig{fig/Jup_static/fig3.pdf}{0.52\textwidth}{(c)}
%           }
\caption{Inclination/Semi-major axis plot portraying the dynamical evolution of the case \textbf{Jup\_static} (table \ref{tab:mathmode}). Panel (a)
represents the dynamical evolution for a co-addition of planetesimals with sizes of 1km and 10km . Panel (b) shows the dynamical evolution but for a co-addition of planetesimals sizes of 100km and 1000km.  
The color box represents the mass of the particles, except for Jupiter and Saturn (we use the size of each point to represent the mass of Jupiter and Saturn and the planetesimals sizes).}
\label{fig:nonmigrat1jcincli}
\end{figure}

Figures \ref{fig:nonmigrat1jc} and \ref{fig:nonmigrat1jcincli} show the eccentricity, inclination and semi-major axis evolution for the simulation set 
\textbf{Jup\_static}. The gas lifetime (hence the duration of the simulation) in this simulations is set to $5.12$ Mys.
% 
% Figure \ref{fig:nonmigrat1jc} (b) and (c) show four snapshots of the final eccentricities and inclinations of the giant planets at $5.12$ My, and the planetesimal disk for the different planetesimals sizes. 
Notice that the smaller are the planetesimals, the colder is the final disk. 
This is because of the stronger effect of the aerodynamic gas drag and a negligible gas dynamical friction (see in Fig. \ref{fig:traject}). 
During the dynamical evolution of the planetary embryos and the giant planets showed in Fig. \ref{fig:interpolation1} and described in the section \ref{izidoro}, the planetesimal disk is depleted
by close encounters and collisions with the planetary embryos or Jupiter and Saturn.
% 
% \textbf{Planetary embryos collide and merge forming larger planetary embryos with masses similar of Uranus and Neptune and also a extra ice giant producing more strong gravitational perturbation on the planetesimal disk.}
% \textbf{Gas drag is more efficient in damping the orbital eccentricities of planetesimals smaller than $100$ km in size. Larger planetesimals are also excited but generally remain on excited orbits sometimes even crossing the orbits of Jupiter and Saturn and being ejected.}

\begin{figure}
          \gridline{\fig{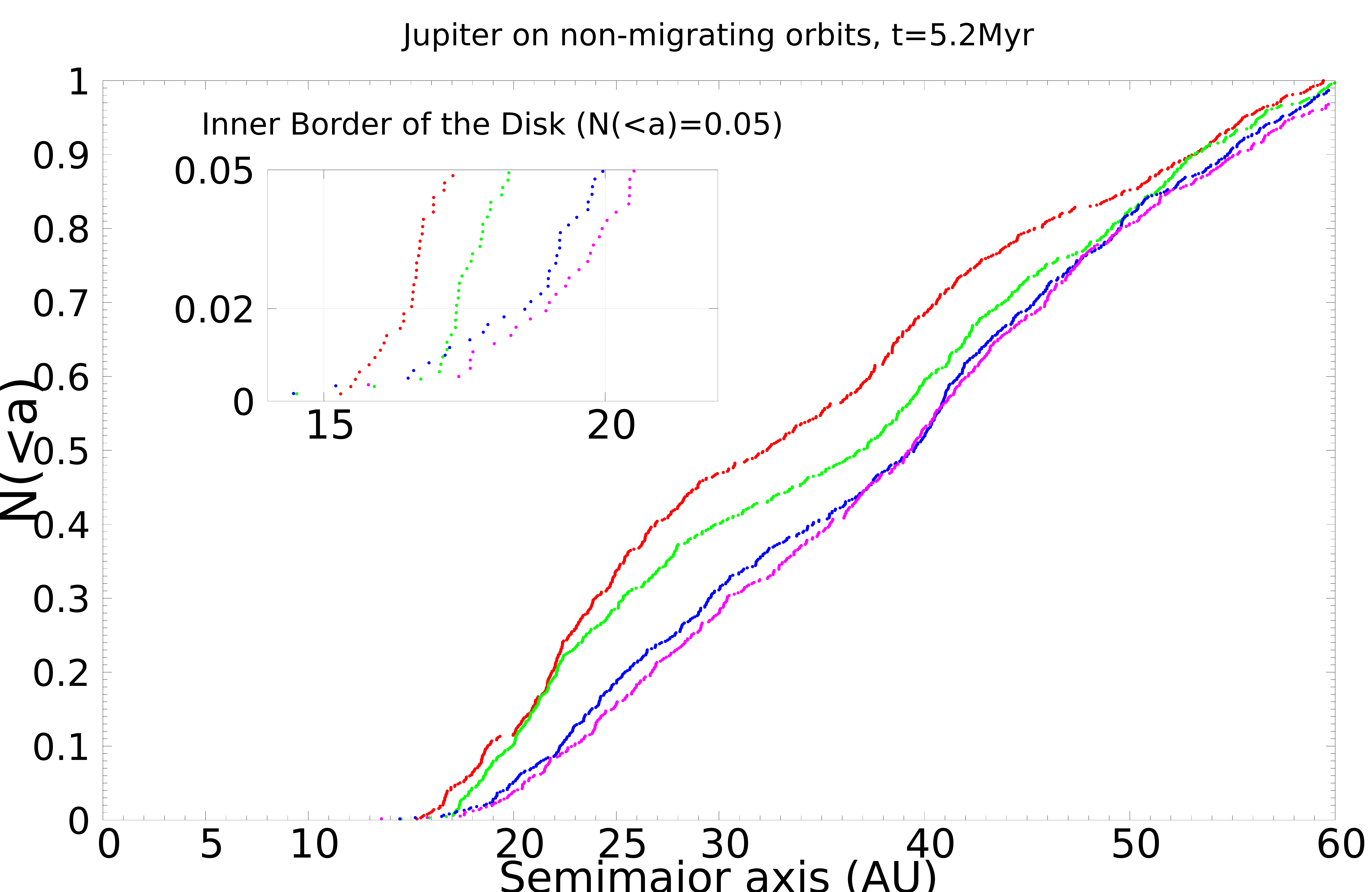}{0.52\textwidth}{(a)}
          \fig{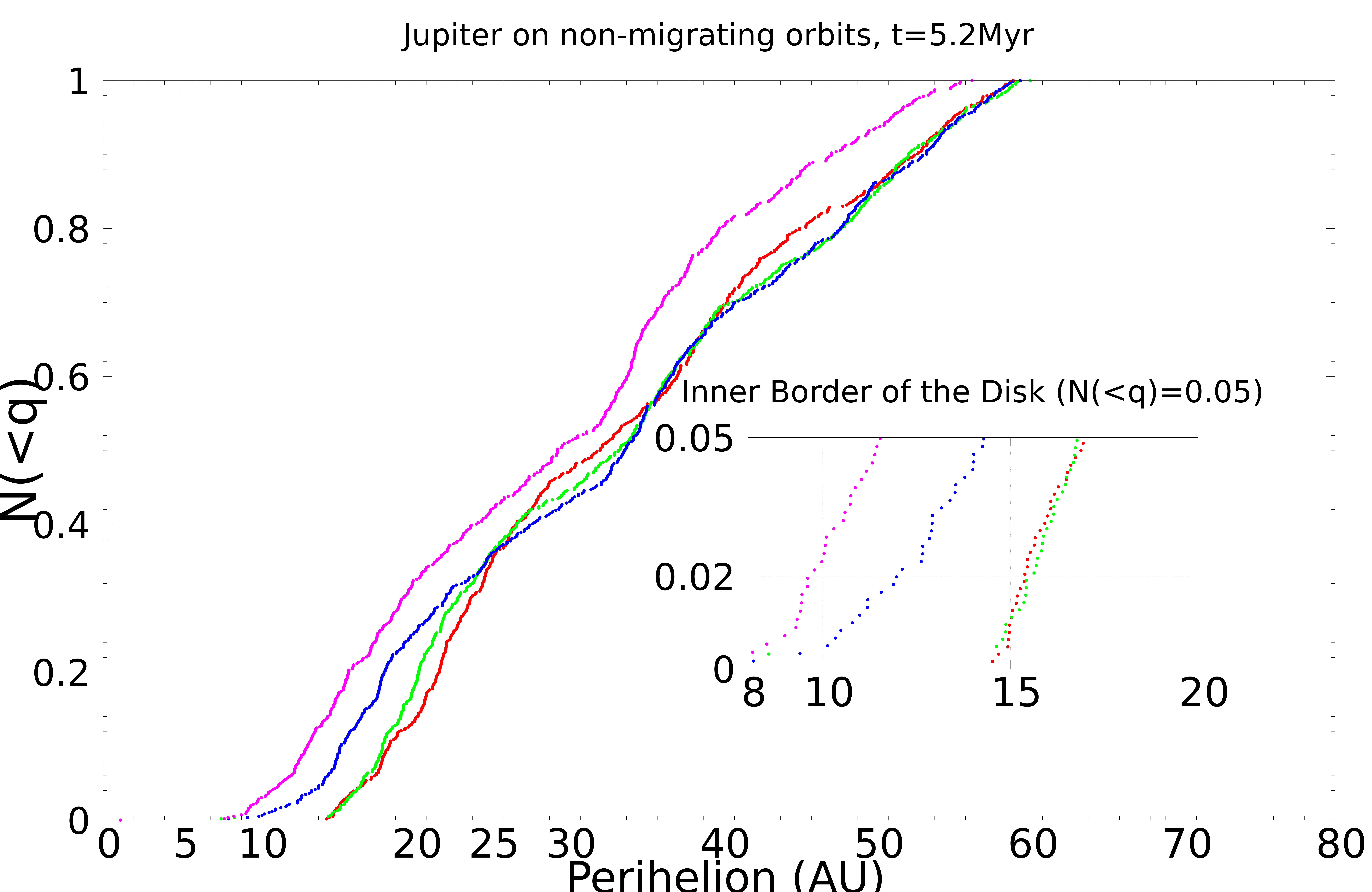}{0.52\textwidth}{(b)}
         }
%          \gridline{\fig{fig/Jup_static/c.pdf}{0.52\textwidth}{(c)}
%           \fig{fig/Jup_static/d.pdf}{0.52\textwidth}{(d)}
%          }
          
\caption{Cumulative normalized semi-major axis (panel (a)) and perihelion (panel (b)) distributions of the planetesimal disk at the dispersal of the gas in the simulation set \textbf{Jup\_static} (table \ref{tab:mathmode}).
The red, green, blue and magenta curve colors represent the cumulative distributions for different planetesimal's sizes, 1 km, 10 km, 100 km, 1000 km respectively.
We defined the border of the planetesimal disk in the end of our simulation as the value such their cumulative normalized distributions have values of $N(<a)=0.05$ and $N(<q)=0.05$ (smaller panels (a) and (b)).
We used these cumulative distributions to calculate the distance between the inner border of the disk and Neptune. 
Note: Neptune is defined as the outer most ice giant planet with semi-major axis of 12.58 AU and perihelion of 12.46 AU.}
\label{fig:comulativeJC}
\end{figure}

In the end of our simulation we defined the inner
edge of the planetesimal disk in semi major axis ($a_{edge}$) and perihelion distance ($q_{edge}$)
such that their cumulative normalized $a$ and $q$ distributions have values of $N(a<a_{edge})=0.05$ and $N(q<q_{edge})=0.05$ respectively (rather arbitrary but it is done so that rogue planetesimals with small $a$ or $q$ do not define the inner disk's edge; in other words, we accept that 5\% of the planetesimals are outliers, inwards of the defined disk's edge) (observe in Fig. \ref{fig:comulativeJC} (a) and (b)).
Neptune is defined as the outer most planet and at the end of the simulation it has semi-major axis of 12.58 AU 
and perihelion of 12.46 AU. 

We see that the inner edge of the disk in the end is quite close to Neptune's orbit 
and is planetesimal size dependent. Although the proto-planets were initially distributed up to $\sim40$ AU, 
they migrated out of the $20-40$ AU region rather quickly. Therefore they could dynamically excite the 
planetesimal population in that region, but not deplete it significantly. 
Moreover, gas drag partially damped the planetesimals' eccentricities and inclinations once the proto-planets left their natal region. 
Thus the final separation in semi-major axis between Neptune and the disk is $5.09$ AU for the km-size planetesimals and $8.16$ AU for the $1,000$ km-size planetesimals. 
The separation is smaller for small planetesimals because gas-drag tends to circularize 
the excited objects near their perihelion distance. If considered in $q$ (perihelion distance) space, however, 
the separation shrinks to $5.07$ AU and $0.55$ AU respectively. 
The big difference between $a_{edge}$ and $q_{edge}$ for the 1,000 km-planetesimals is due to the non-zero eccentricity of the latter.

The radial distance (measured with semimajor axes) between Neptune and the planetesimal disk is less important
when the planetesimals are on eccentric orbits and have lower perihelion distances. In this case, Neptune is in 
closer contact with the planetesimal disk. \citet{Deiennoetal2017} showed that even though there is a resolution dependence between the distance of Neptune and inner edge of the disk, the distance in the range that 
we found here would lead to an instability before 400 Myr, and \citet{Quarles_Kaib_2019} found even earlier instability times. 
We foresee that such a small separation will lead to an early instability, as we will test in section~\ref{instime}.

\subsubsection{Jupiter migrating outward}
\label{JO}

\begin{figure}
     \gridline{\fig{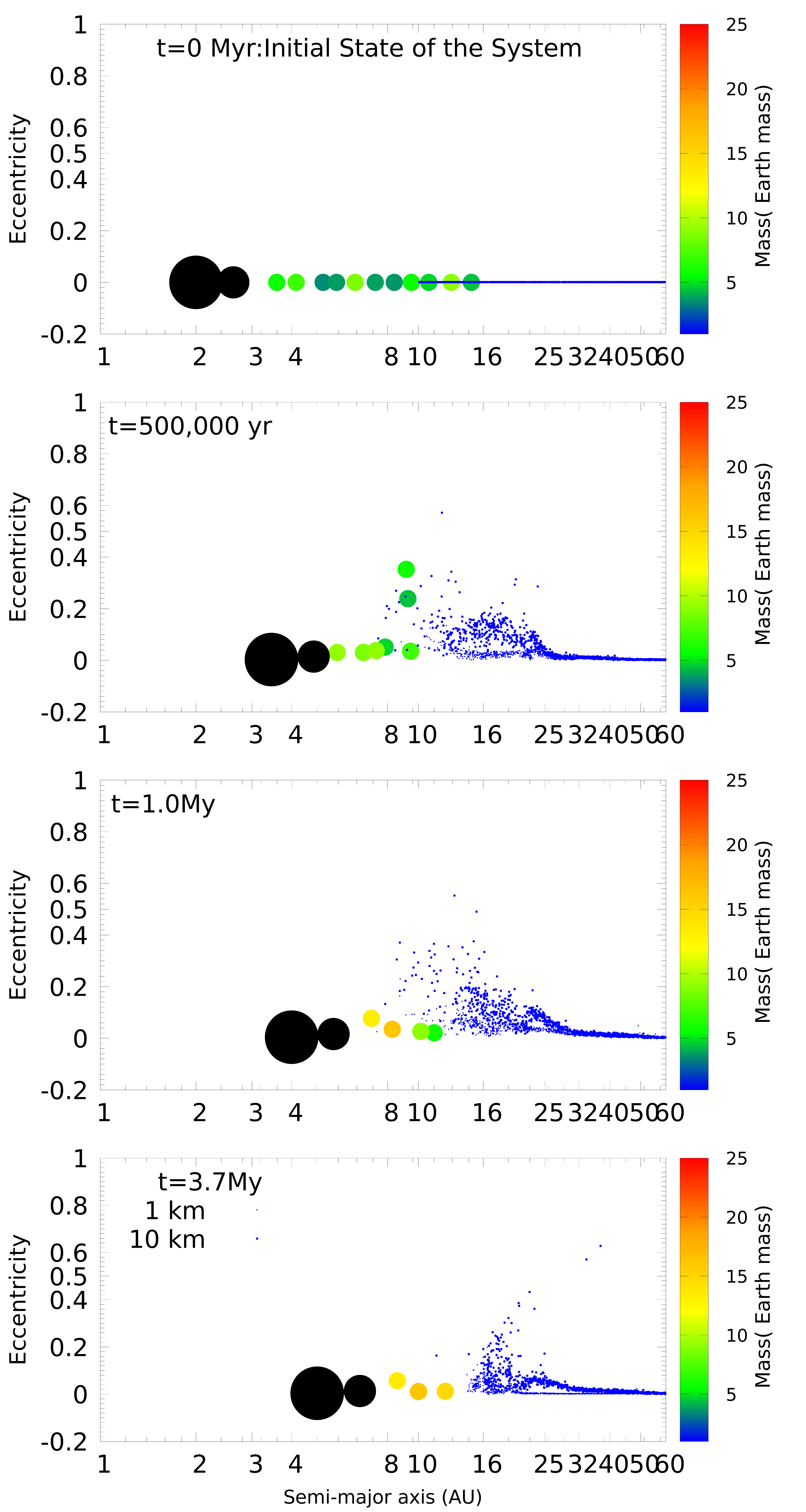}{0.5\textwidth}{(a)}
          \fig{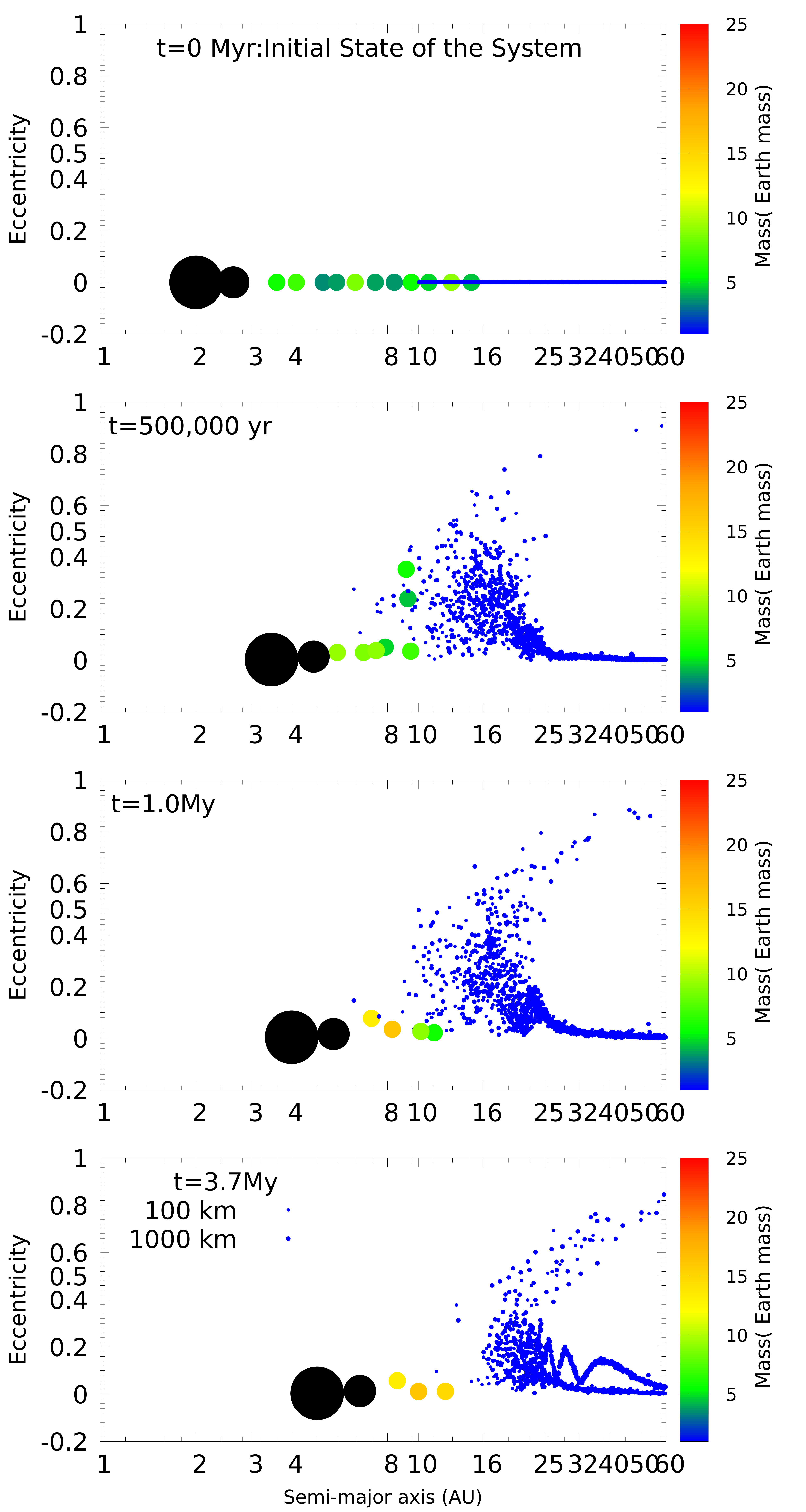}{0.5\textwidth}{(b)}
          }

\caption{The same as Fig. \ref{fig:nonmigrat1jc}, but for the simulation set \textbf{Jup\_outward} (table \ref{tab:mathmode}) where Jupiter is assumed to migrate from 2 to 5 AU. The total duration of the simulation is 3.7 My.}
\label{fig:migrateoutward2}
\end{figure}

\begin{figure}
     \gridline{\fig{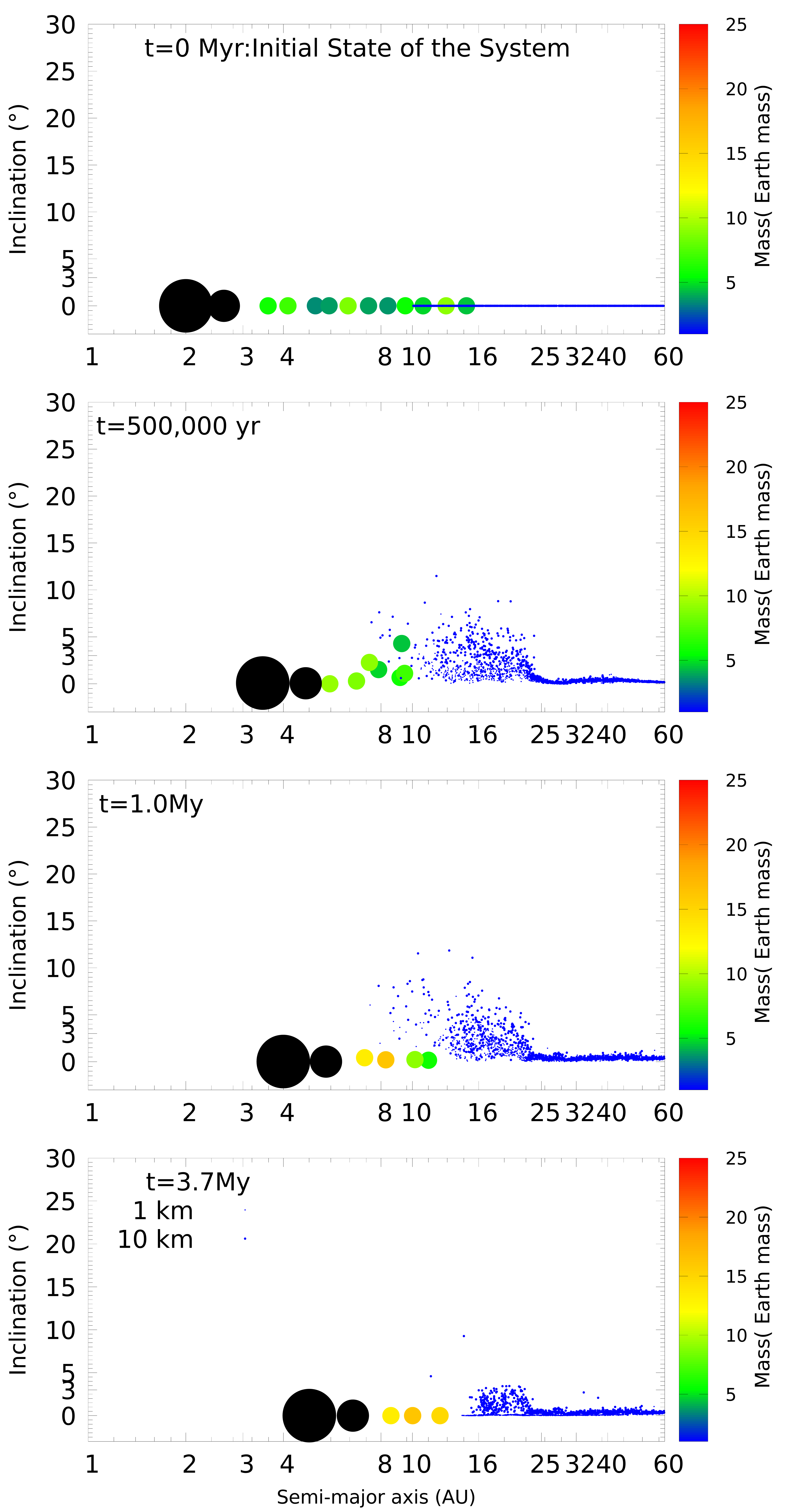}{0.5\textwidth}{(a)}
          \fig{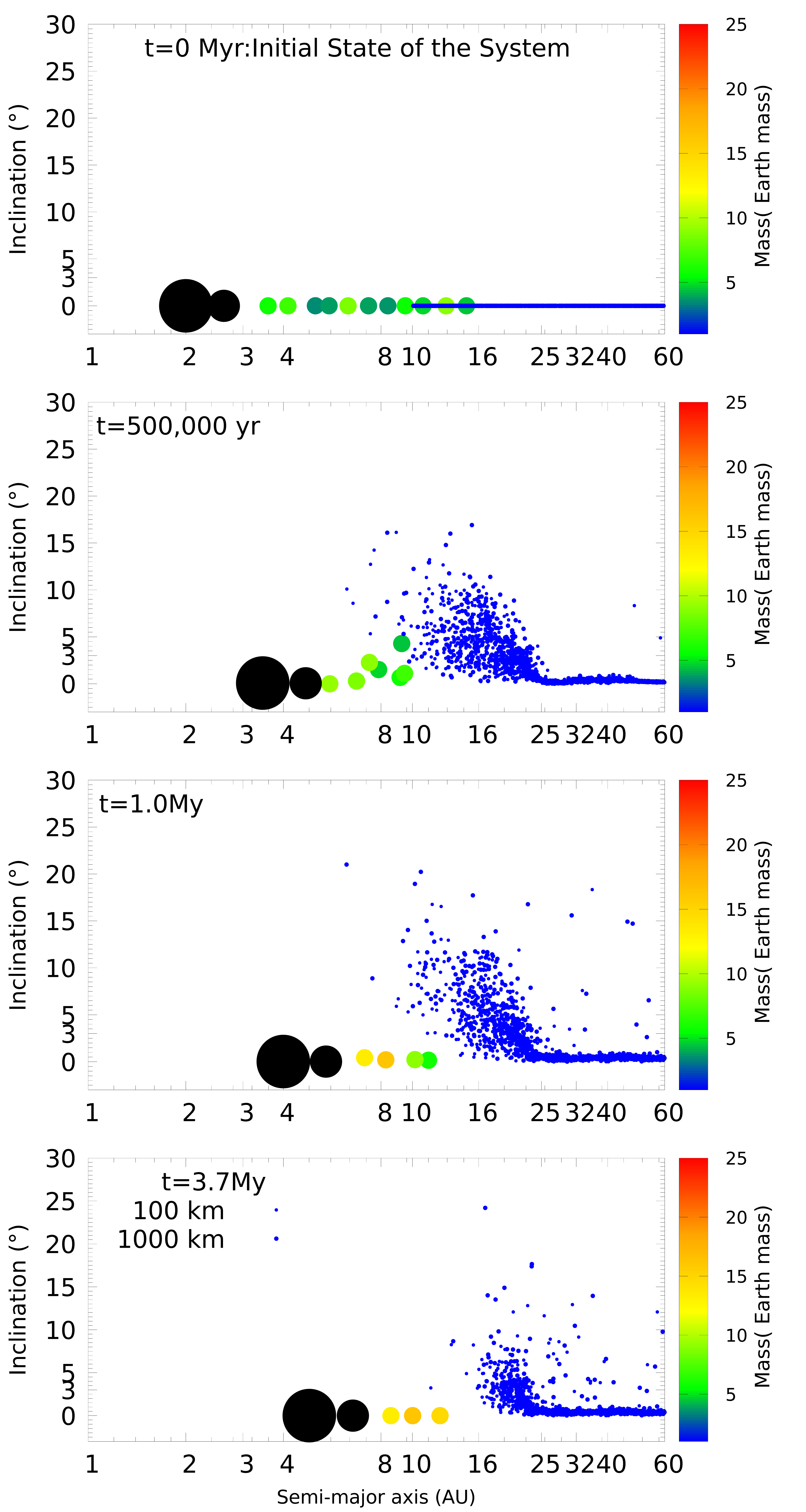}{0.5\textwidth}{(b)}
          }

\caption{The same as Fig. \ref{fig:nonmigrat1jcincli}, but for the simulation set \textbf{Jup\_outward} (table \ref{tab:mathmode}) where Jupiter is assumed to migrate from 2 to 5 AU. The total duration of the simulation is 3.7 My.}
\label{fig:migrateoutwardinc2}
\end{figure}

\begin{figure}
          \gridline{\fig{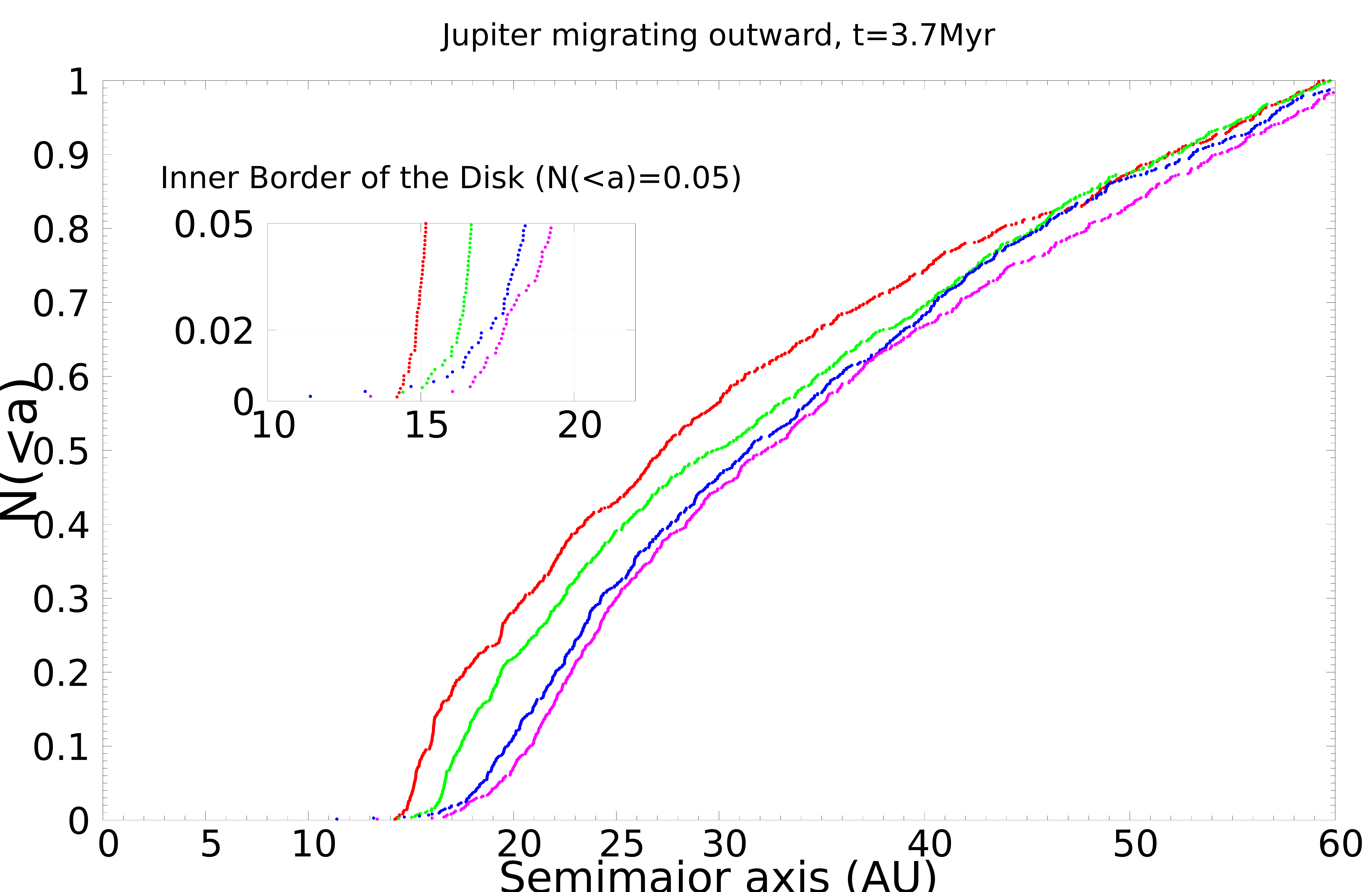}{0.52\textwidth}{(a)}
          \fig{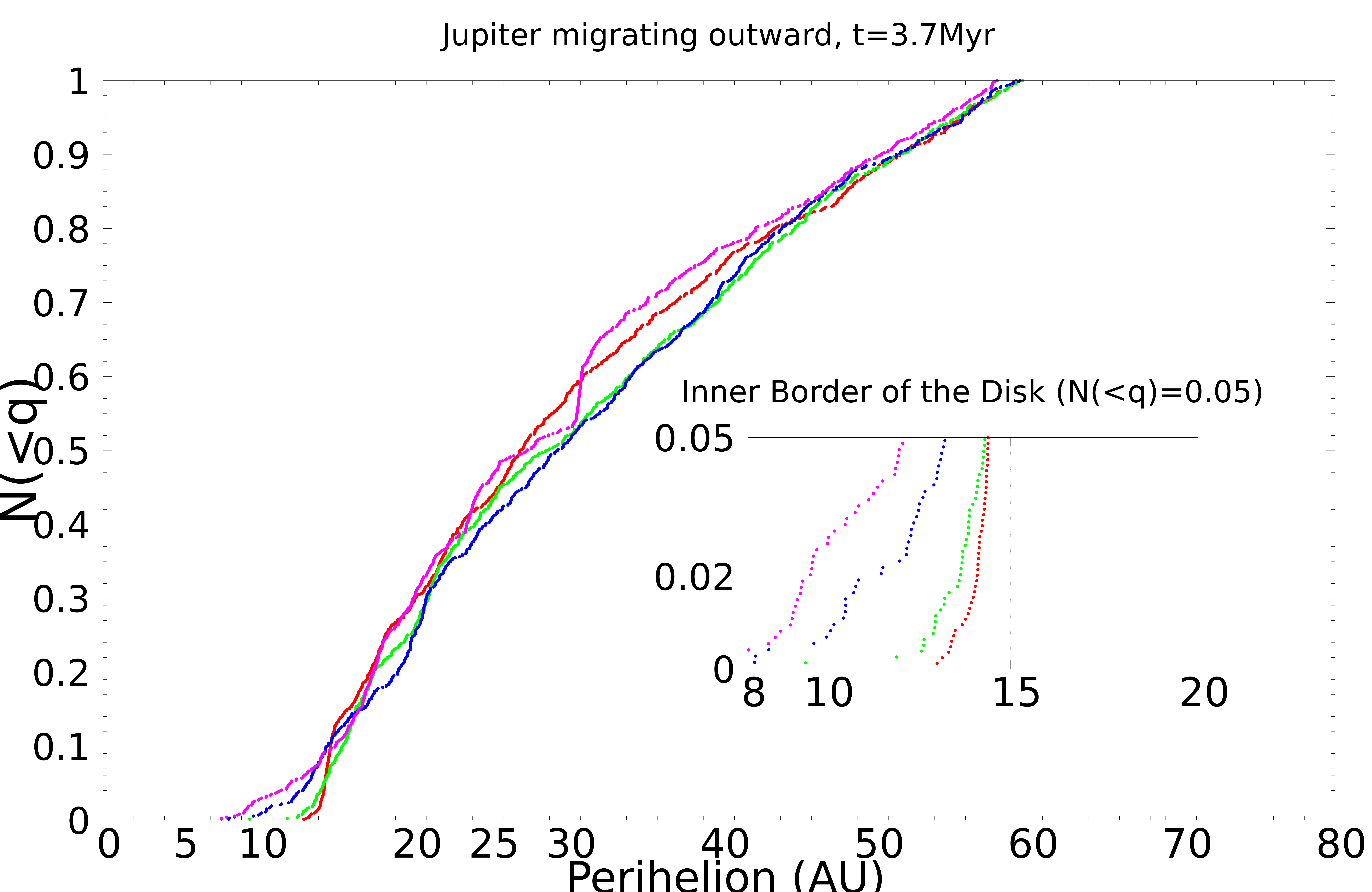}{0.52\textwidth}{(b)}
         }
%          \gridline{\fig{fig/Jup_outward/c.pdf}{0.52\textwidth}{(c)}
%           \fig{fig/Jup_outward/d.pdf}{0.52\textwidth}{(d)}
%          }
          
\caption{ The same as Fig. \ref{fig:comulativeJC}, but for the simulation set \textbf{Jup\_outward} (table \ref{tab:mathmode}).}
\label{fig:nonmigrat1jo}
\end{figure}

Figures \ref{fig:migrateoutward2} and \ref{fig:migrateoutwardinc2} show the eccentricity, inclination and semi-major axis evolution for the simulation set \textbf{Jup\_outward}. 
In this simulation, Jupiter is migrating outward from $2$ to $5$ AU and the simulation begins with a planetesimal disk extended from $10$ to $60$ AU (blue points). 
The gas lifetime is  $3.7$ Myr. The separation between Neptune and the inner edge of the planetesimal disk is even smaller than before. In fact, this separation is $2.63$ AU for the km-size planetesimals and $6.91$ AU for the $1,000$ km-size planetesimals in semi-major axis (Fig. \ref{fig:nonmigrat1jo} (a)). 
The separation in pericenter is $1.99$ AU for the km-size planetesimals and $0.46$ AU for the $1,000$ km-size planetesimals (Fig. \ref{fig:nonmigrat1jo} (b)). This is due to the fact that 
the planets moved towards the planetesimal disk and the average distance from the growing planets was larger than the average distance in the previous simulation. Another important effect in 
this simulation is the resonant shepherding, which transports planetesimals along with migrating planets in a size-dependent way ~\citep[e.g.][]{FoggNelson2005,raymond06}.

\clearpage
\subsubsection{Jupiter migrating inward}
\label{JI}
\begin{figure}
          \gridline{
          \fig{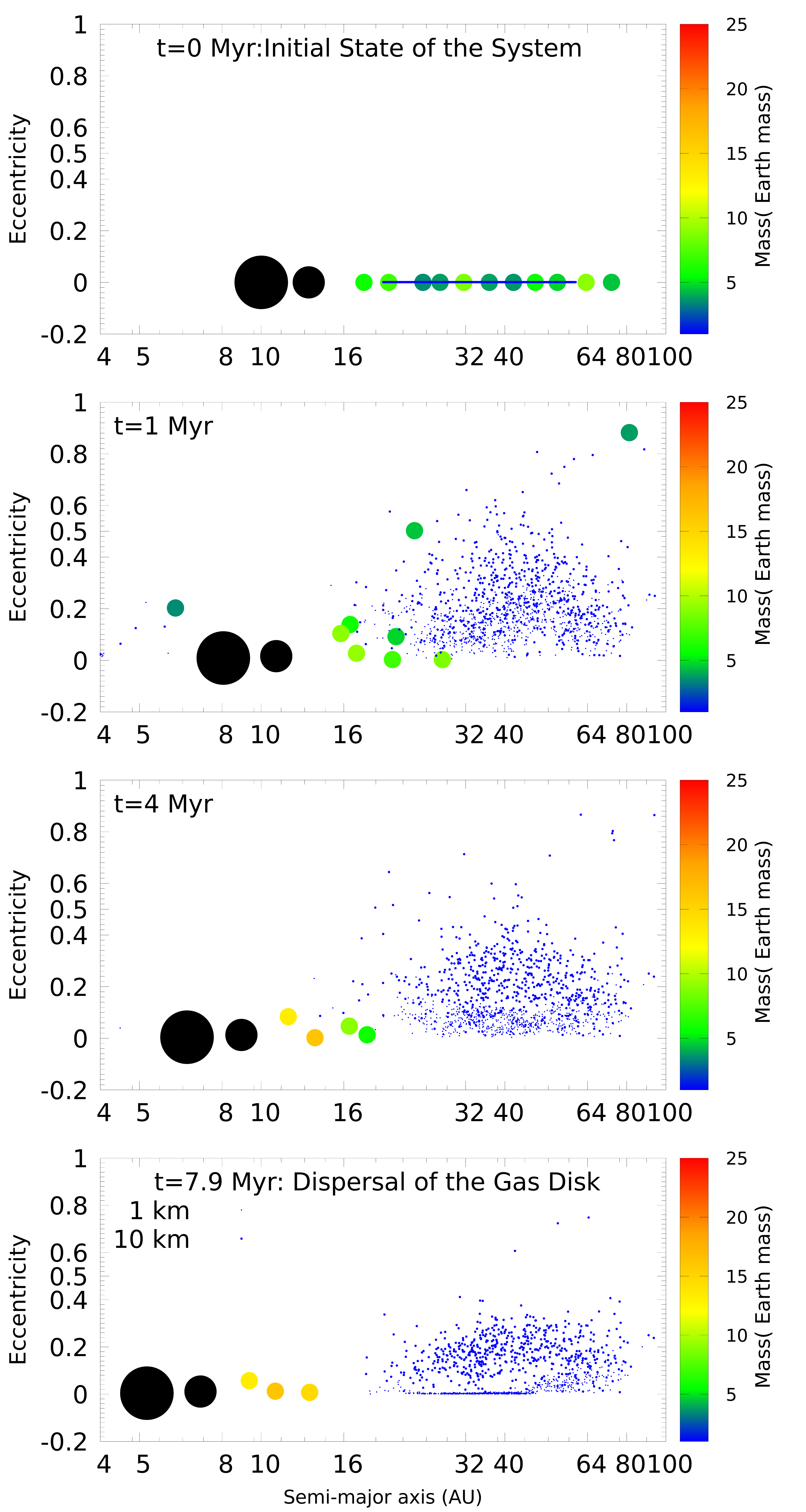}{0.5\textwidth}{(a)}
          \fig{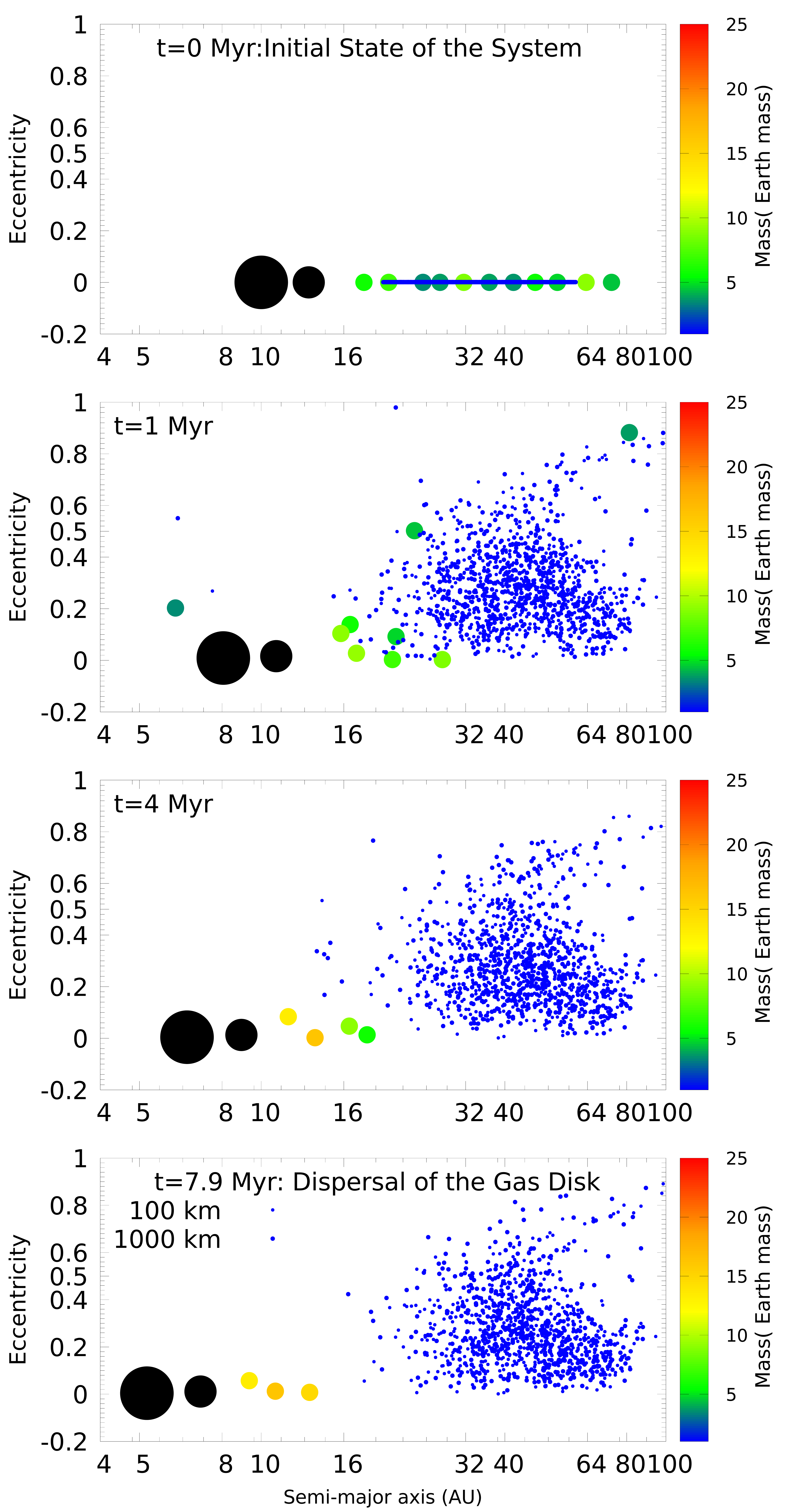}{0.5\textwidth}{(a)}}

\caption{The same as Fig. \ref{fig:nonmigrat1jc}, but for the simulation set \textbf{Jup\_10AU\_in} (table \ref{tab:mathmode}) where Jupiter is assumed to migrate from 10 to 5 AU. The total duration of the simulation is 7.9 My.}

\label{fig:JI1}
\end{figure}

\begin{figure}
          \gridline{
          \fig{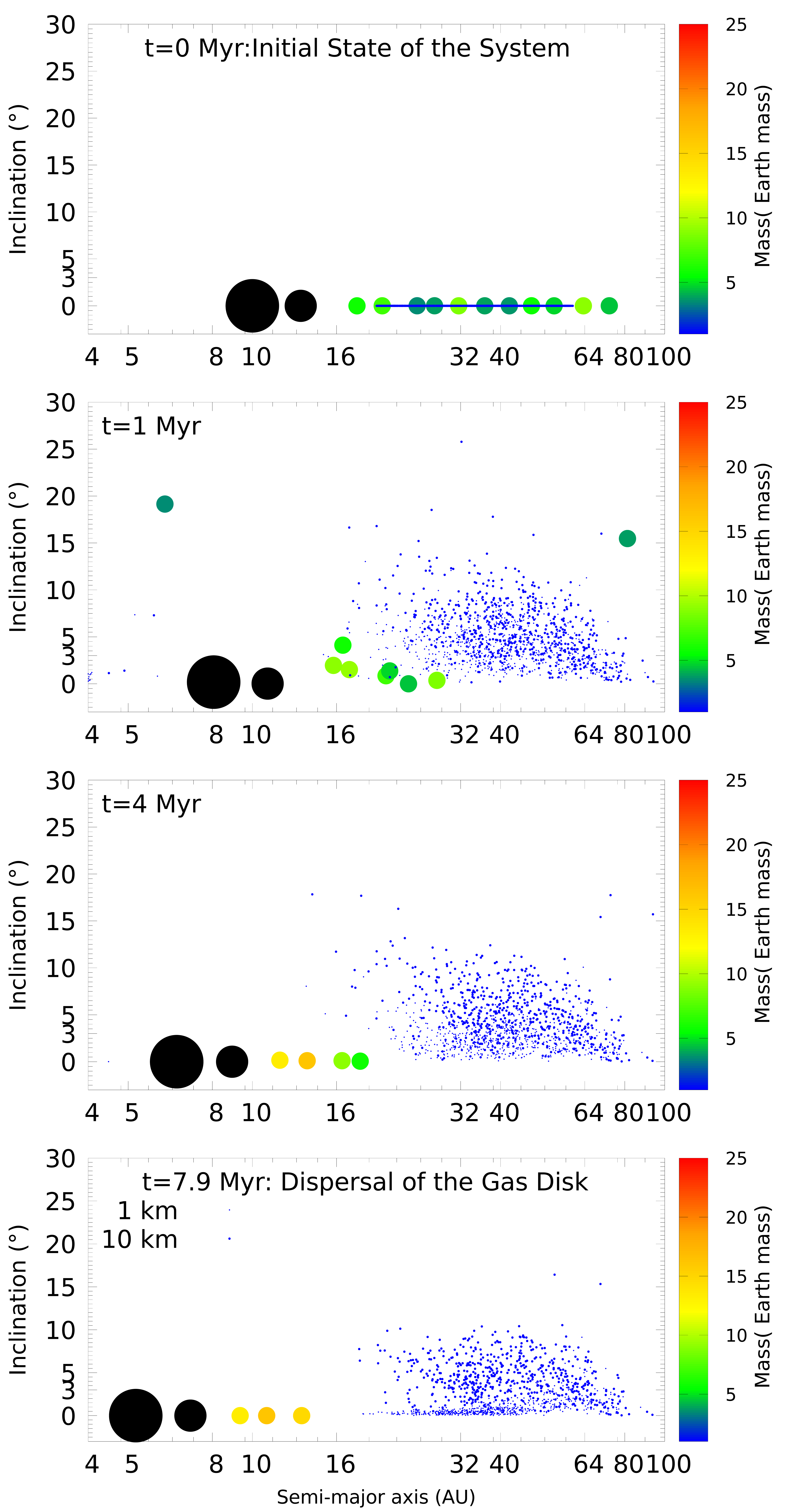}{0.5\textwidth}{(a)}
          \fig{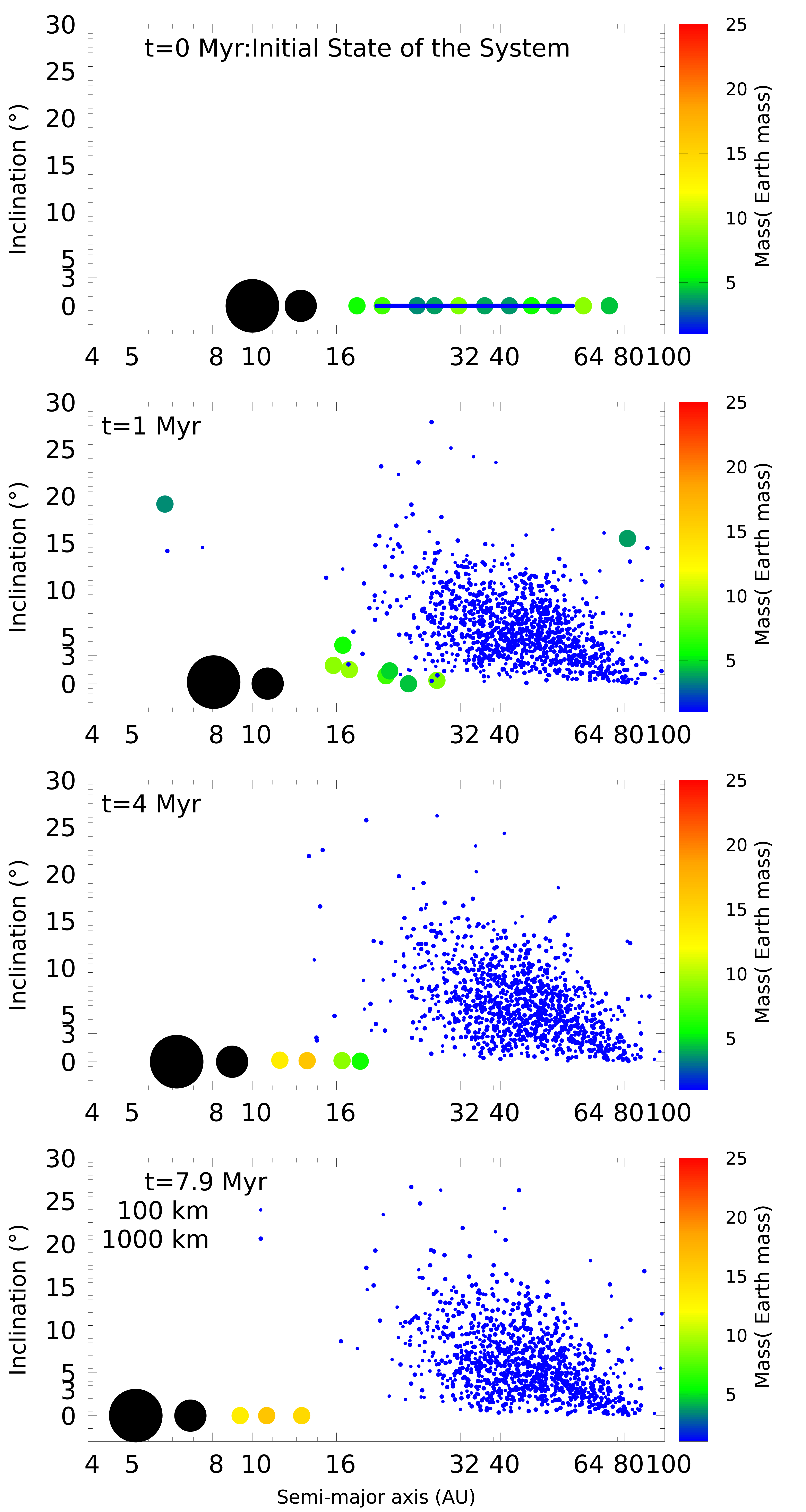}{0.5\textwidth}{(a)}}

\caption{The same as Fig. \ref{fig:nonmigrat1jcincli}, but for the simulation set \textbf{Jup\_10AU\_in} (table \ref{tab:mathmode}) where Jupiter is assumed to migrate from 10 to 5 AU. The total duration of the simulation is 7.9 My.}

\label{fig:JI11}
\end{figure}

\begin{figure}
          \gridline{\fig{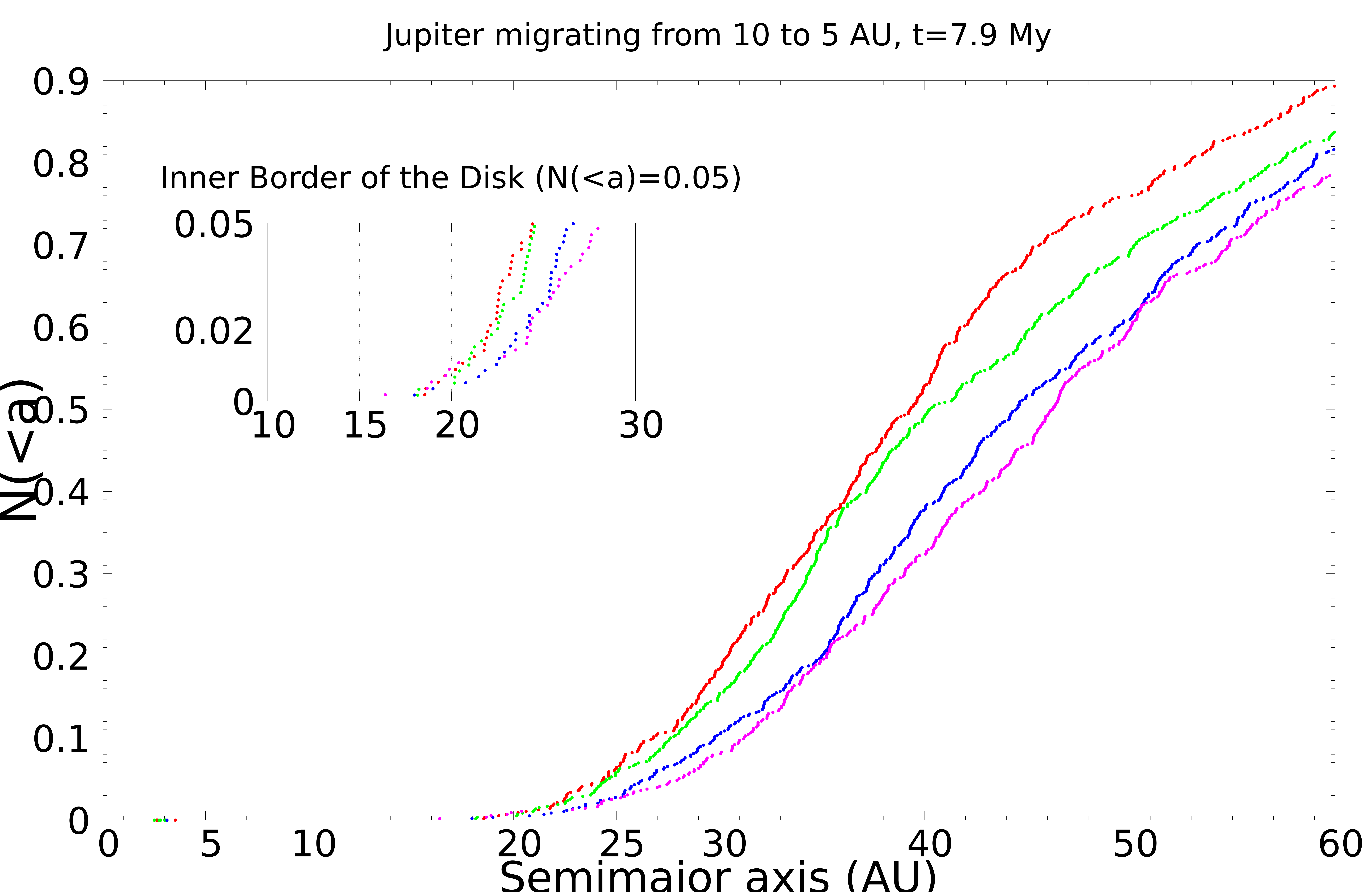}{0.52\textwidth}{(a)}
          \fig{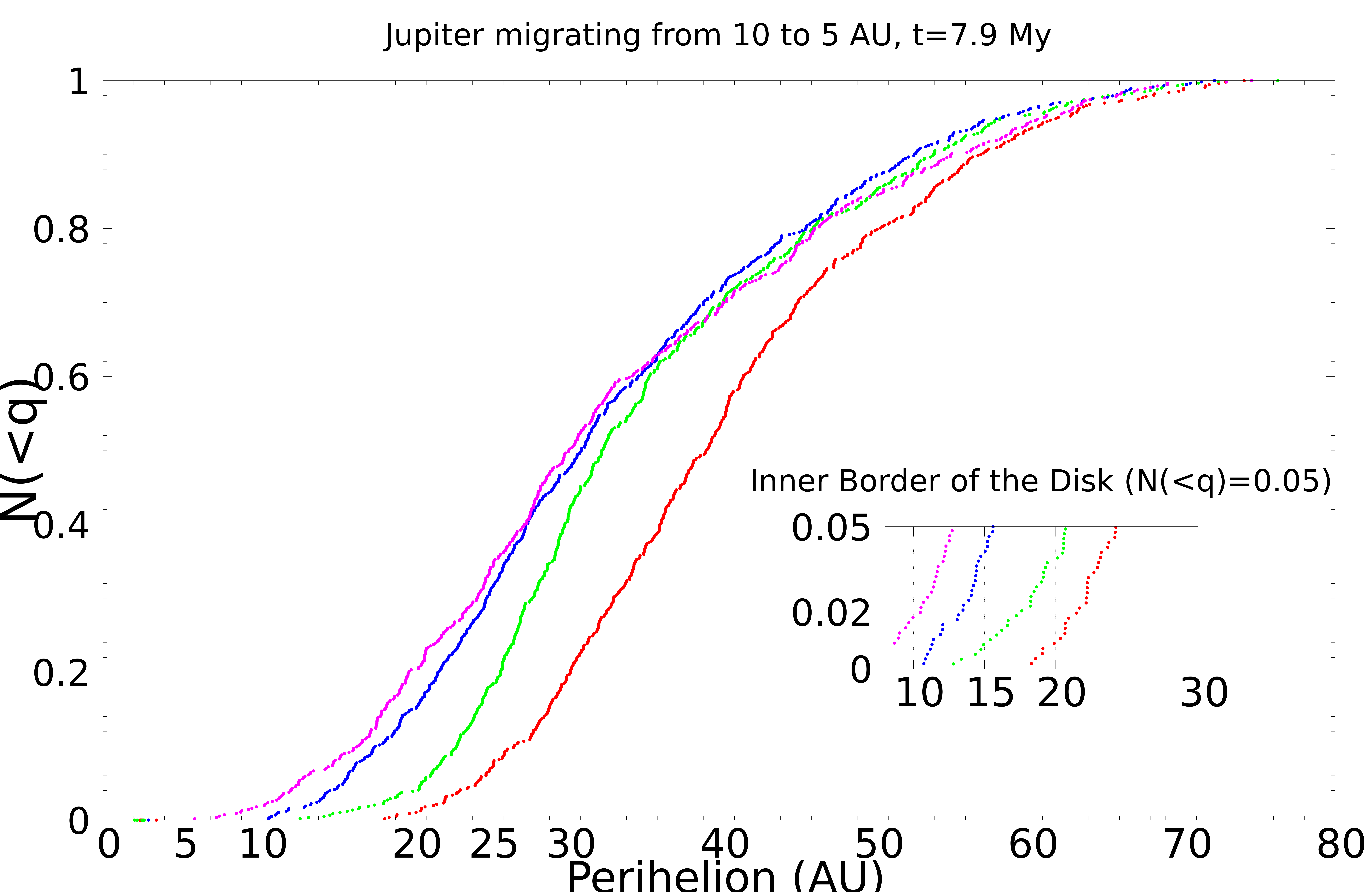}{0.52\textwidth}{(b)}
         }
%          \gridline{\fig{fig/Jup_10AU_in/c.pdf}{0.52\textwidth}{(c)}
%           \fig{fig/Jup_10AU_in/d.pdf}{0.52\textwidth}{(d)}
%          }
   \caption{The same as Fig. \ref{fig:comulativeJC}, but for the simulation set \textbf{Jup\_10AU\_in} (table \ref{tab:mathmode}).}
\label{fig:cumJI1}
\end{figure}

\begin{figure}
          \gridline{\fig{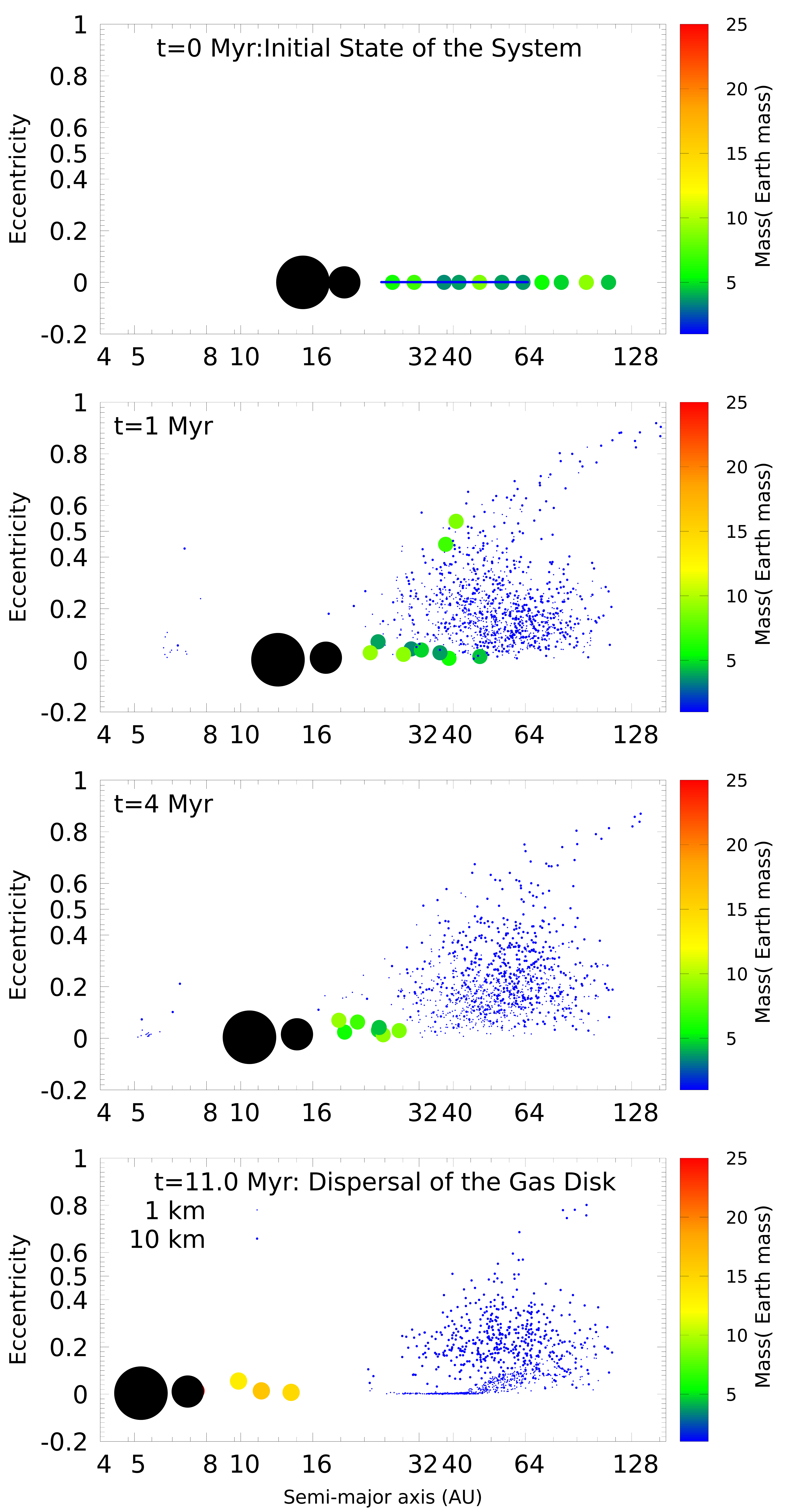}{0.5\textwidth}{(a)}
          \fig{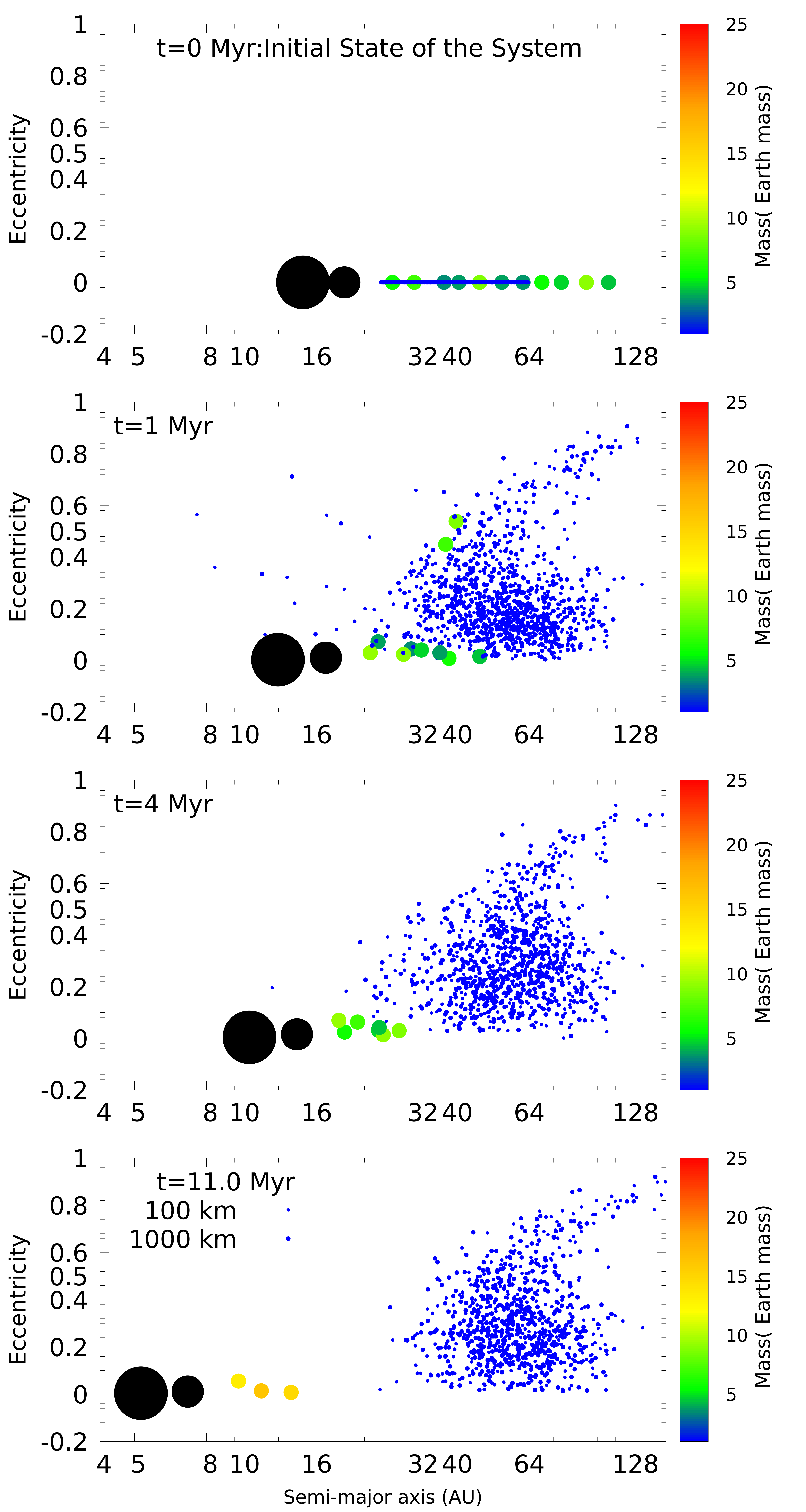}{0.5\textwidth}{(a)}
                    }

\caption{The same as Fig. \ref{fig:nonmigrat1jc}, but for the simulation set \textbf{Jup\_15AU\_in} (table \ref{tab:mathmode}) where Jupiter is assumed to migrate from 15 to 5 AU. The total duration of the simulation is 11 My.}
\label{fig:JI2}
\end{figure}

\begin{figure}
          \gridline{\fig{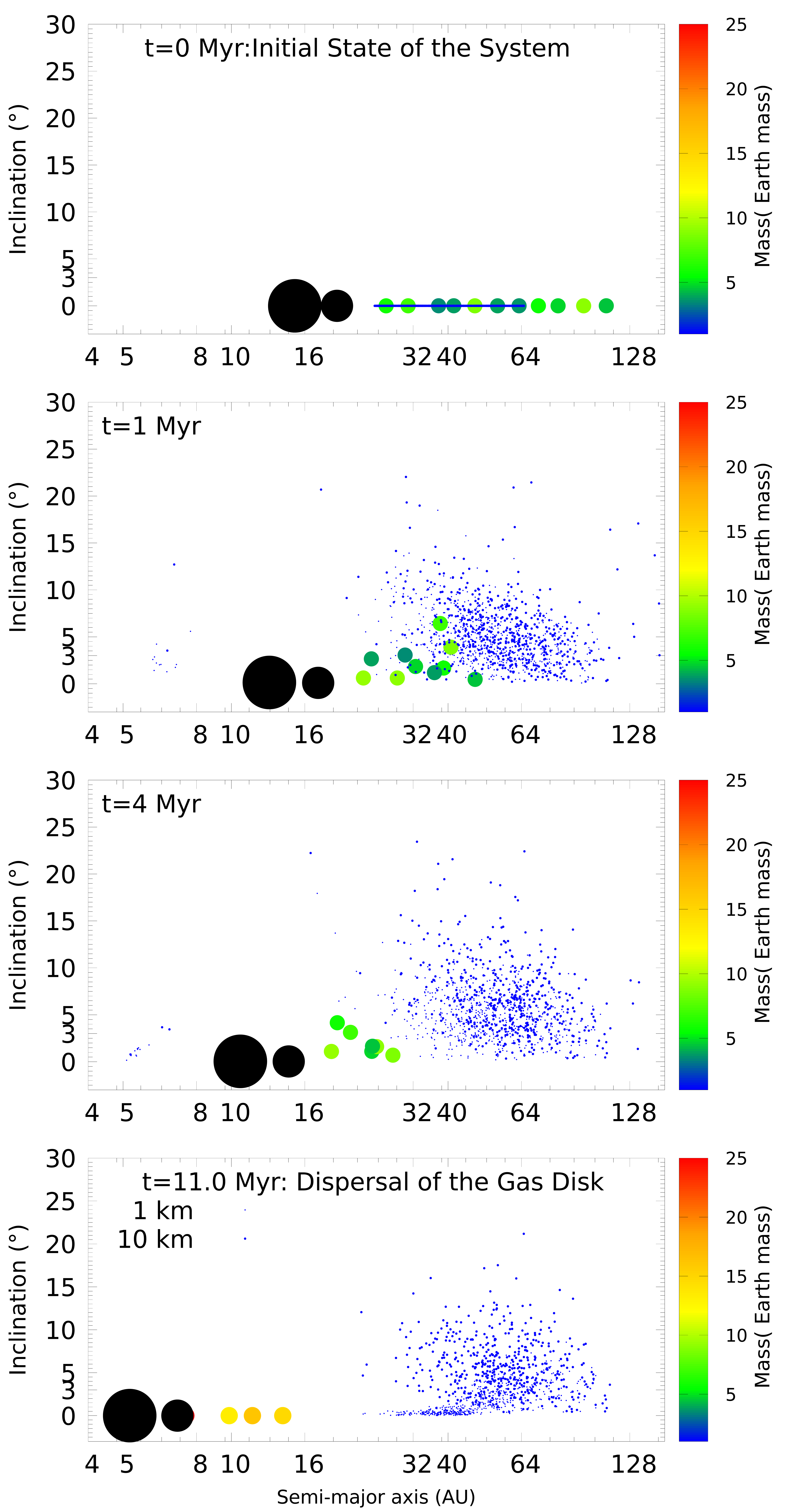}{0.5\textwidth}{(a)}
          \fig{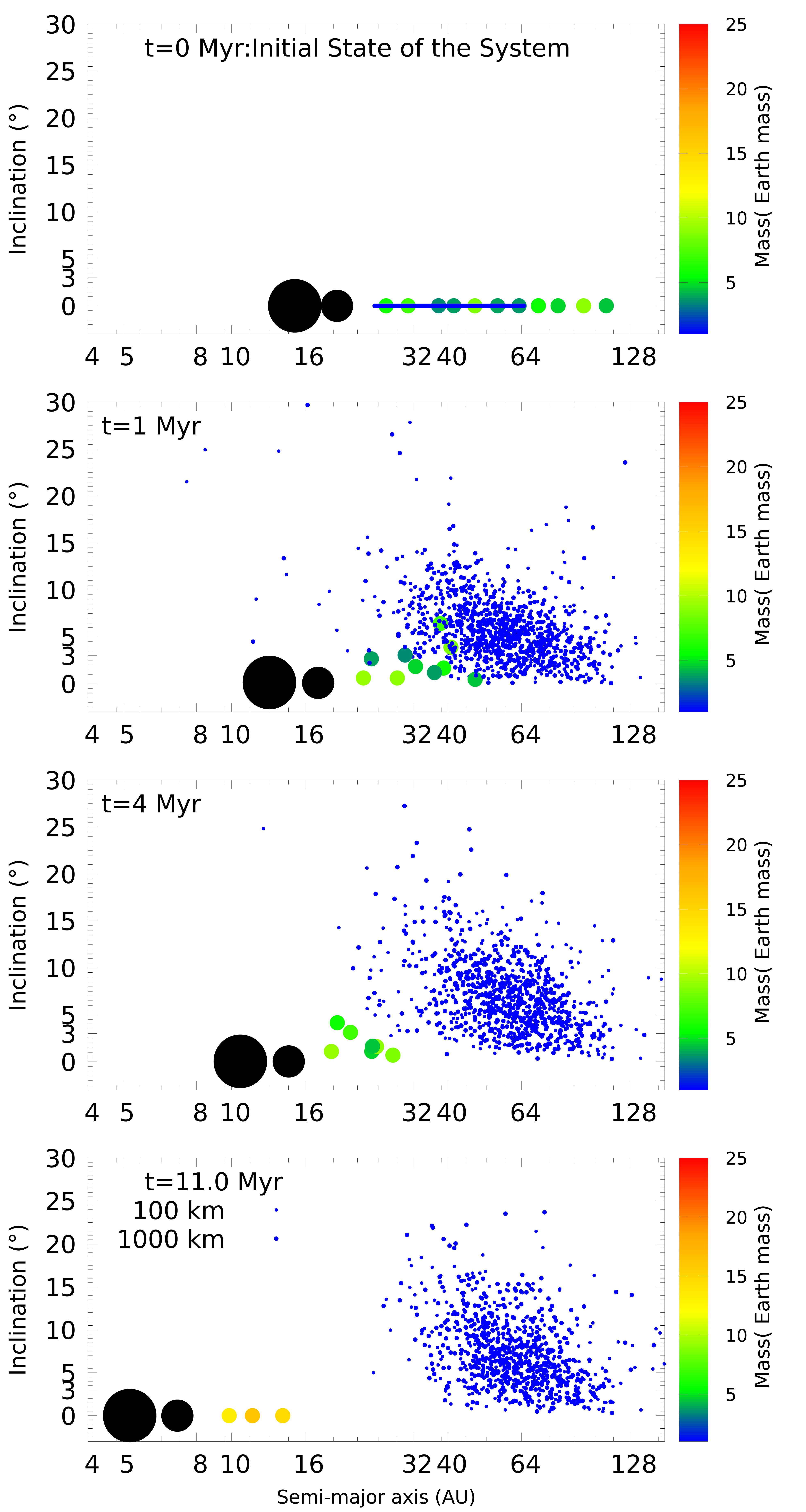}{0.5\textwidth}{(a)}
                    }

\caption{The same as Fig. \ref{fig:nonmigrat1jcincli}, but for the simulation set \textbf{Jup\_15AU\_in} (table \ref{tab:mathmode}) where Jupiter is assumed to migrate from 15 to 5 AU. The total duration of the simulation is 11 My.}
\label{fig:JI22}
\end{figure}

\begin{figure}
          \gridline{\fig{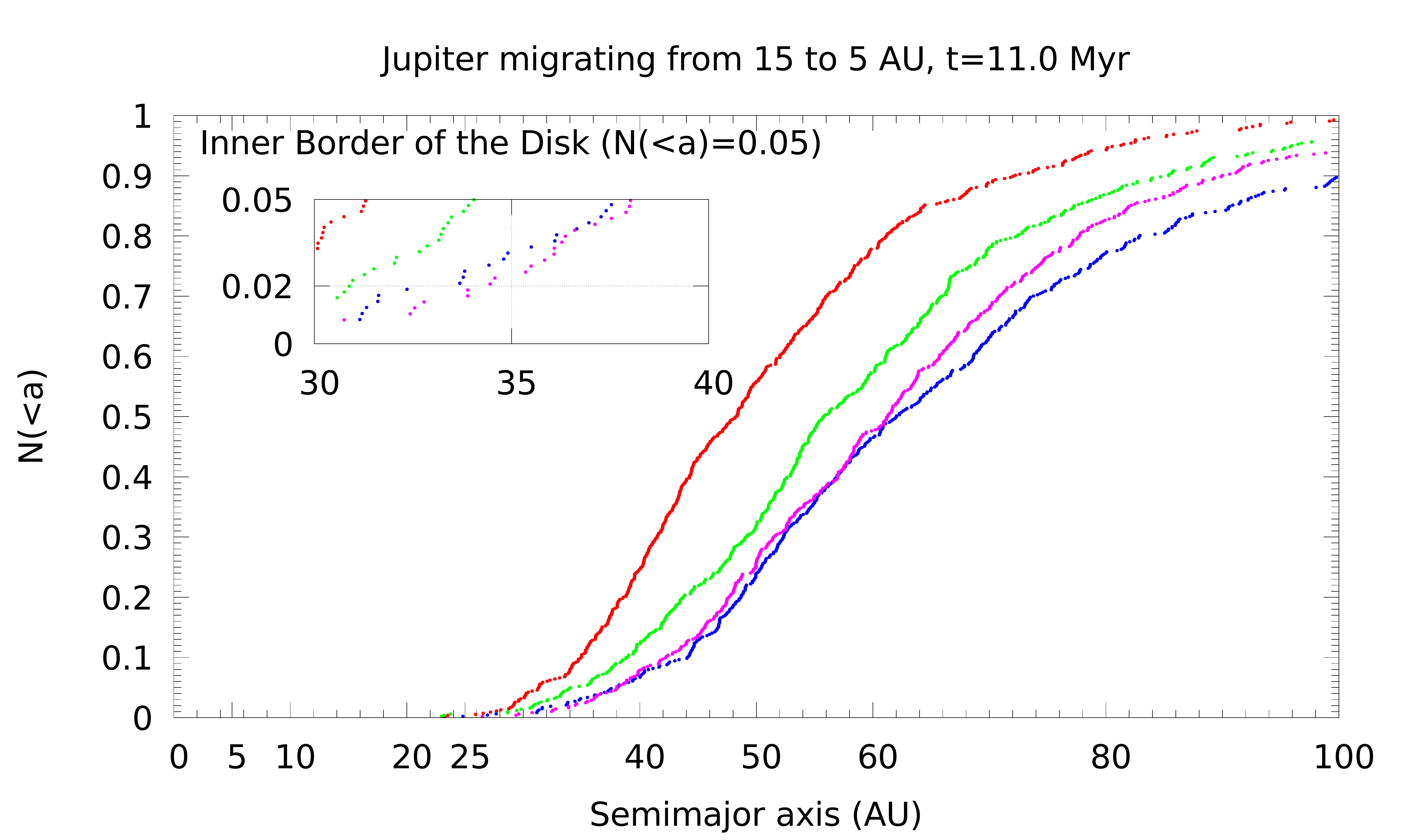}{0.52\textwidth}{(a)}
          \fig{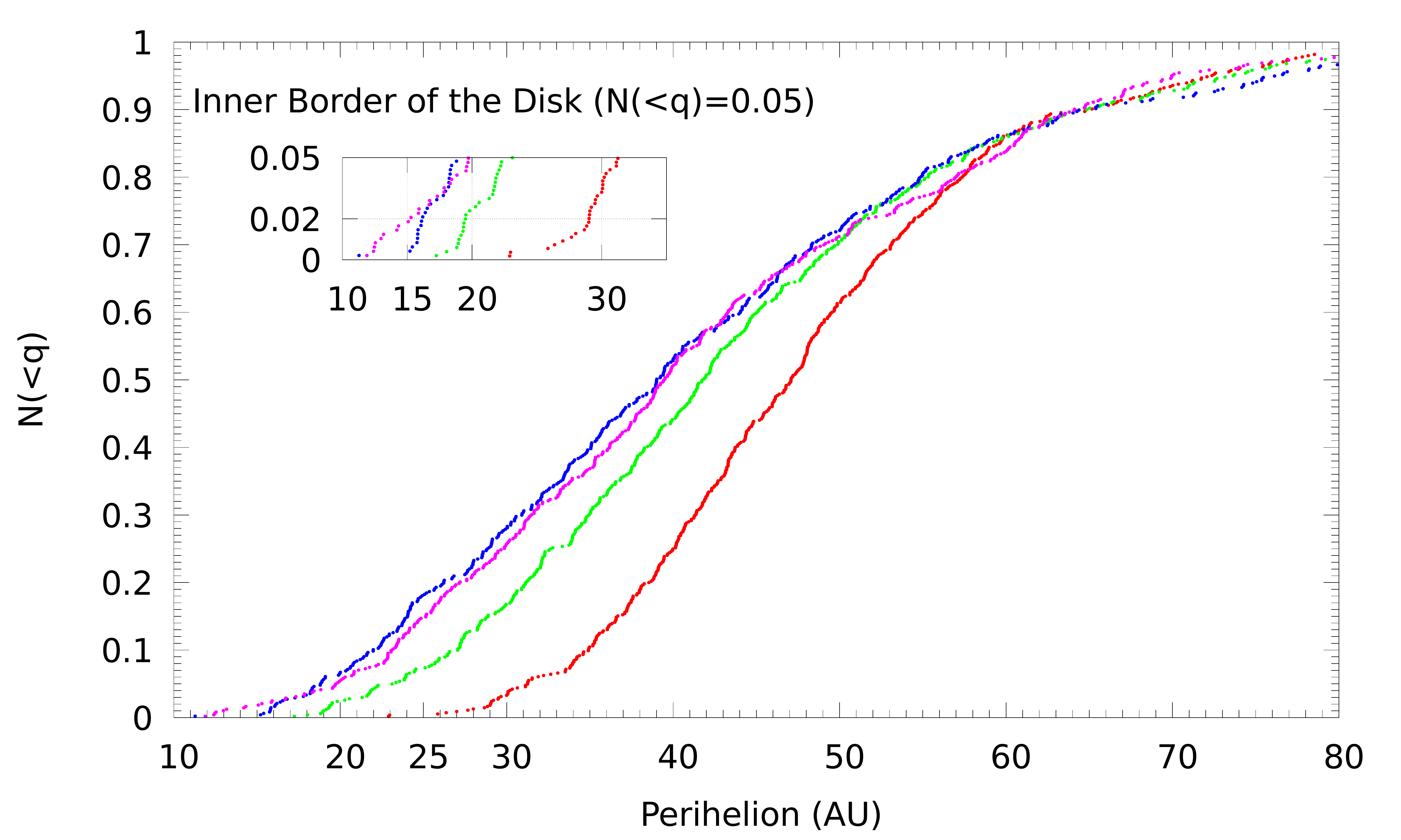}{0.52\textwidth}{(b)}
         }
%          \gridline{\fig{fig/Jup_15AU_in/c.pdf}{0.52\textwidth}{(c)}
%           \fig{fig/Jup_15AU_in/d.pdf}{0.52\textwidth}{(d)}
%          }
          
\caption{The same as Fig. \ref{fig:comulativeJC}, but for the simulation set \textbf{Jup\_15AU\_in} (table \ref{tab:mathmode}).}
\label{fig:cumJI2}
\end{figure}

\begin{figure}
          \gridline{
          \fig{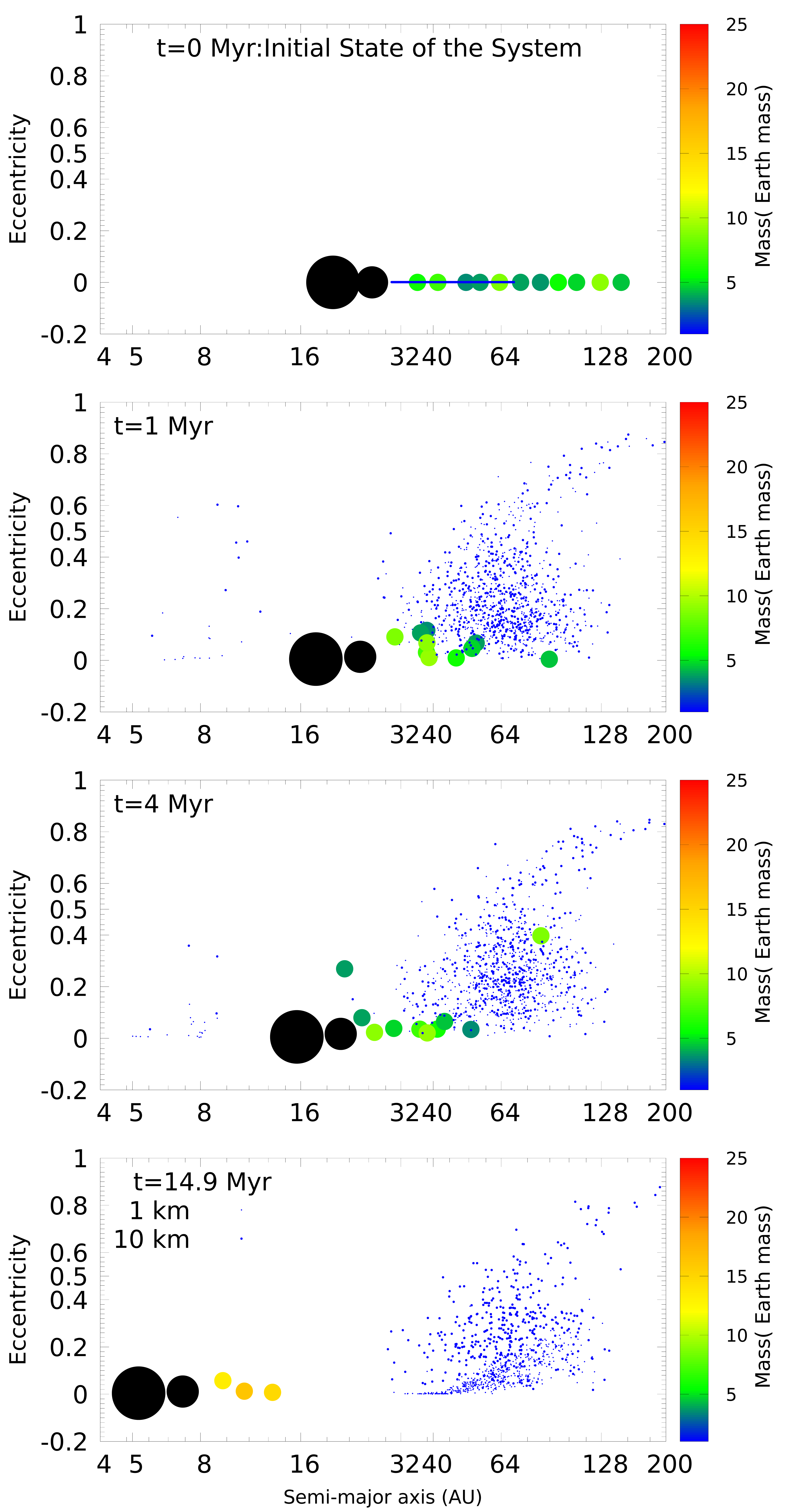}{0.5\textwidth}{(a)}
             \fig{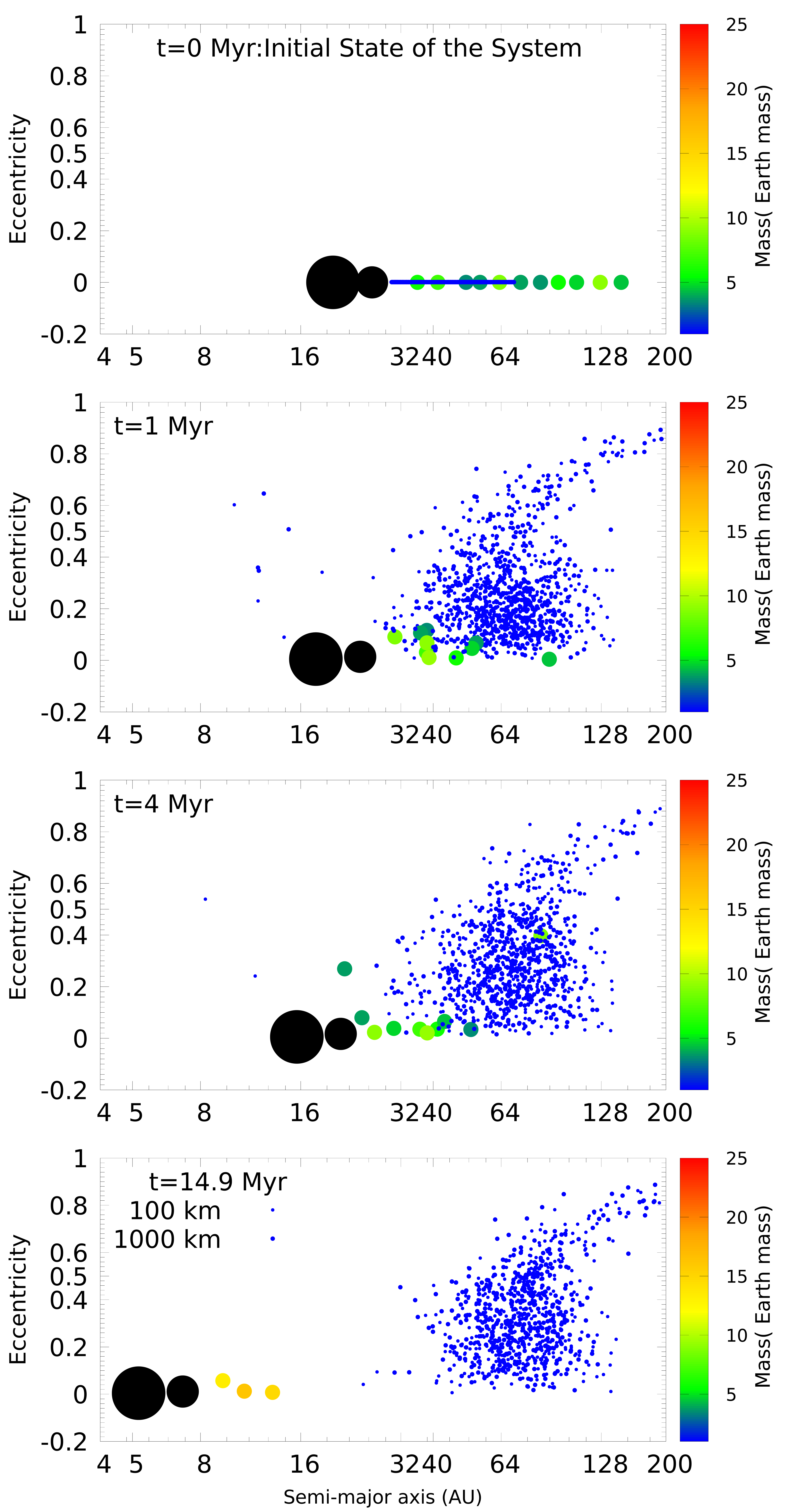}{0.5\textwidth}{(b)} }
         
\caption{The same as Fig. \ref{fig:nonmigrat1jc}, but for the simulation set \textbf{Jup\_20AU\_in} (table \ref{tab:mathmode}) where Jupiter is assumed to migrate from 20 to 5 AU. The total duration of the simulation is 14.9 My.}
\label{fig:JI3}
\end{figure}

\begin{figure}
          \gridline{
          \fig{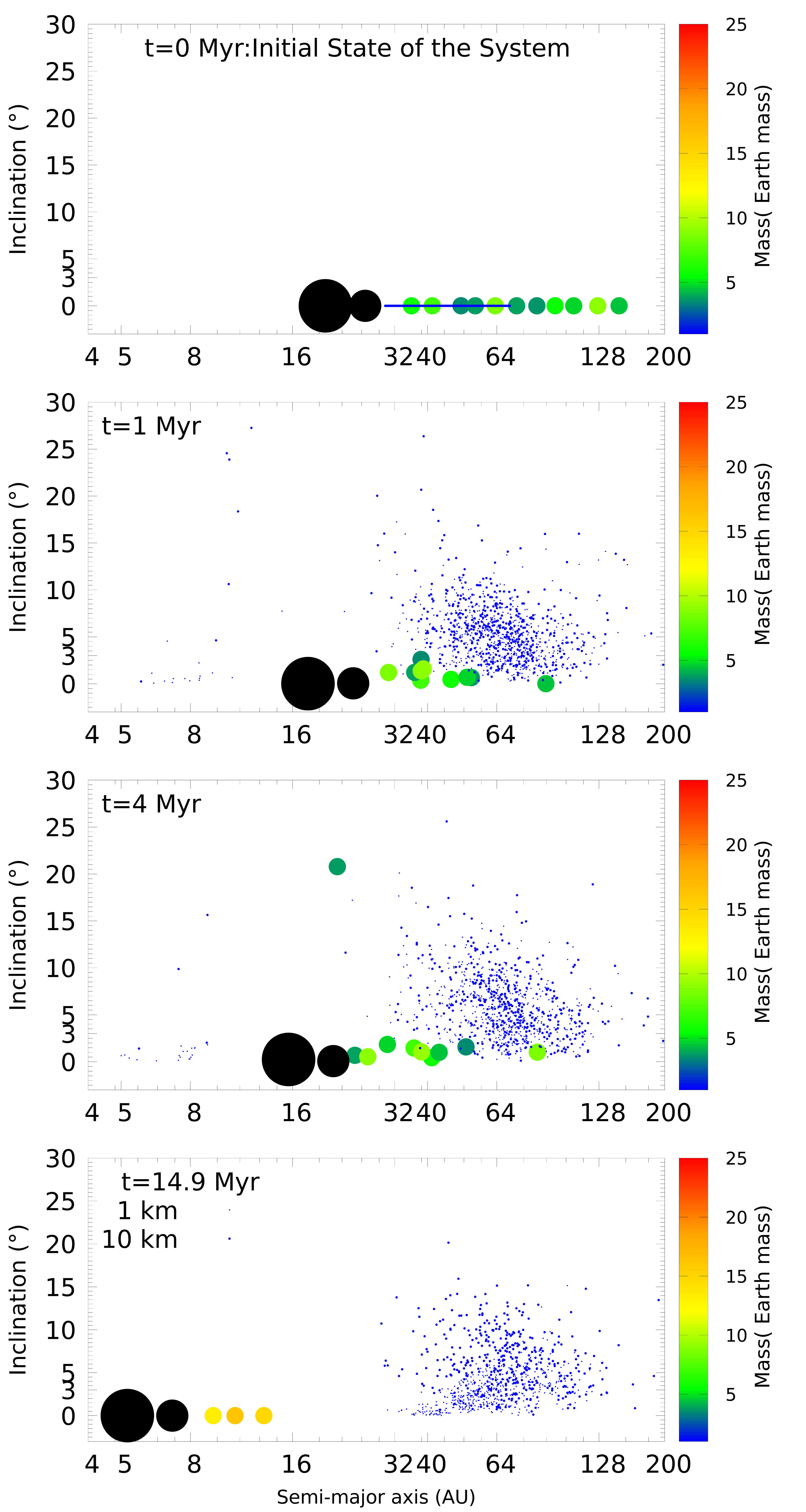}{0.5\textwidth}{(a)}
             \fig{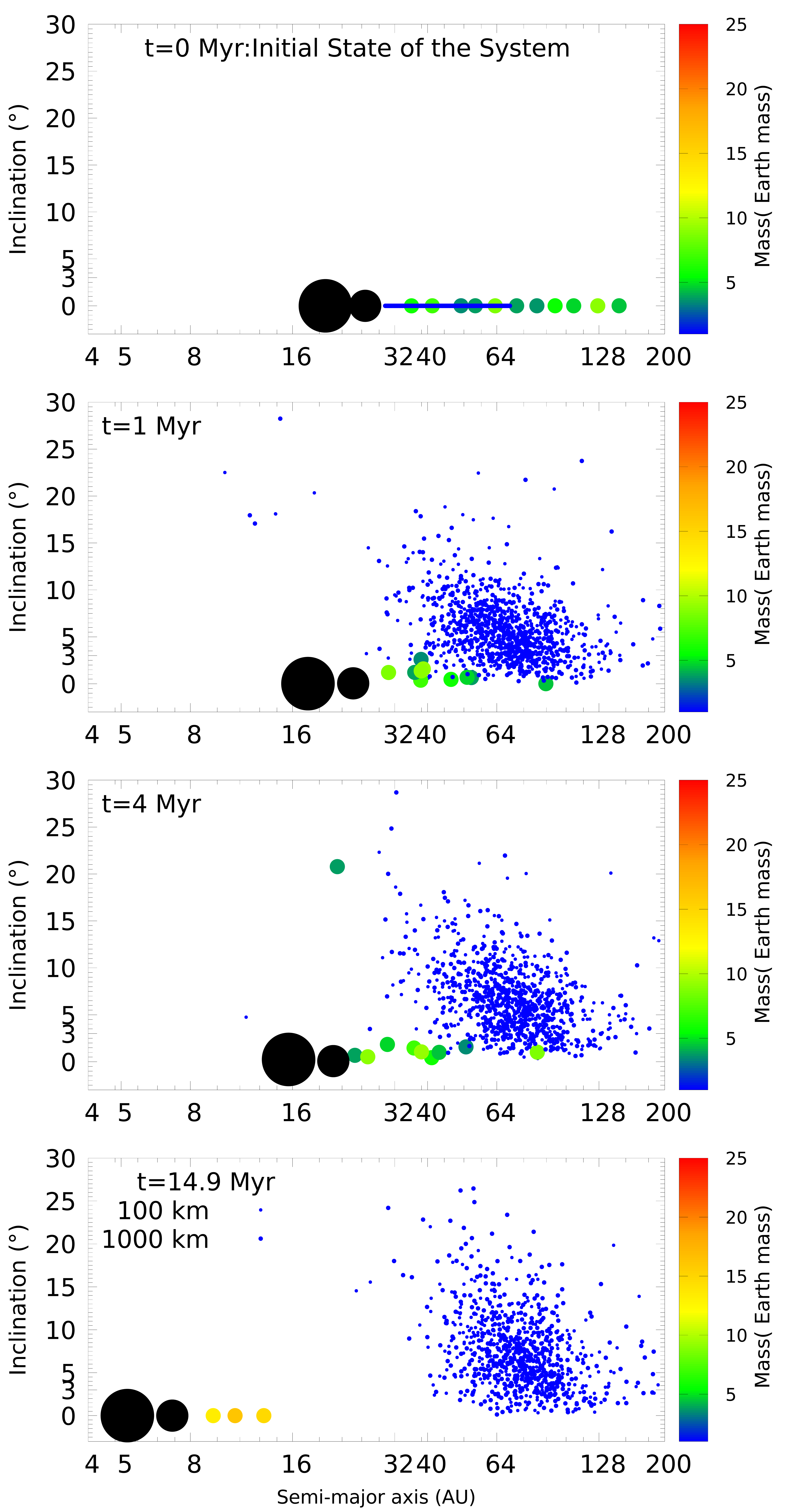}{0.5\textwidth}{(b)} }
         
\caption{The same as Fig.\ref{fig:nonmigrat1jcincli}, but for the simulation set \textbf{Jup\_20AU\_in} (table \ref{tab:mathmode}) where Jupiter is assumed to migrate from 20 to 5 AU. The total duration of the simulation is 14.9 My.}
\label{fig:JI33}
\end{figure}

\begin{figure}
          \gridline{\fig{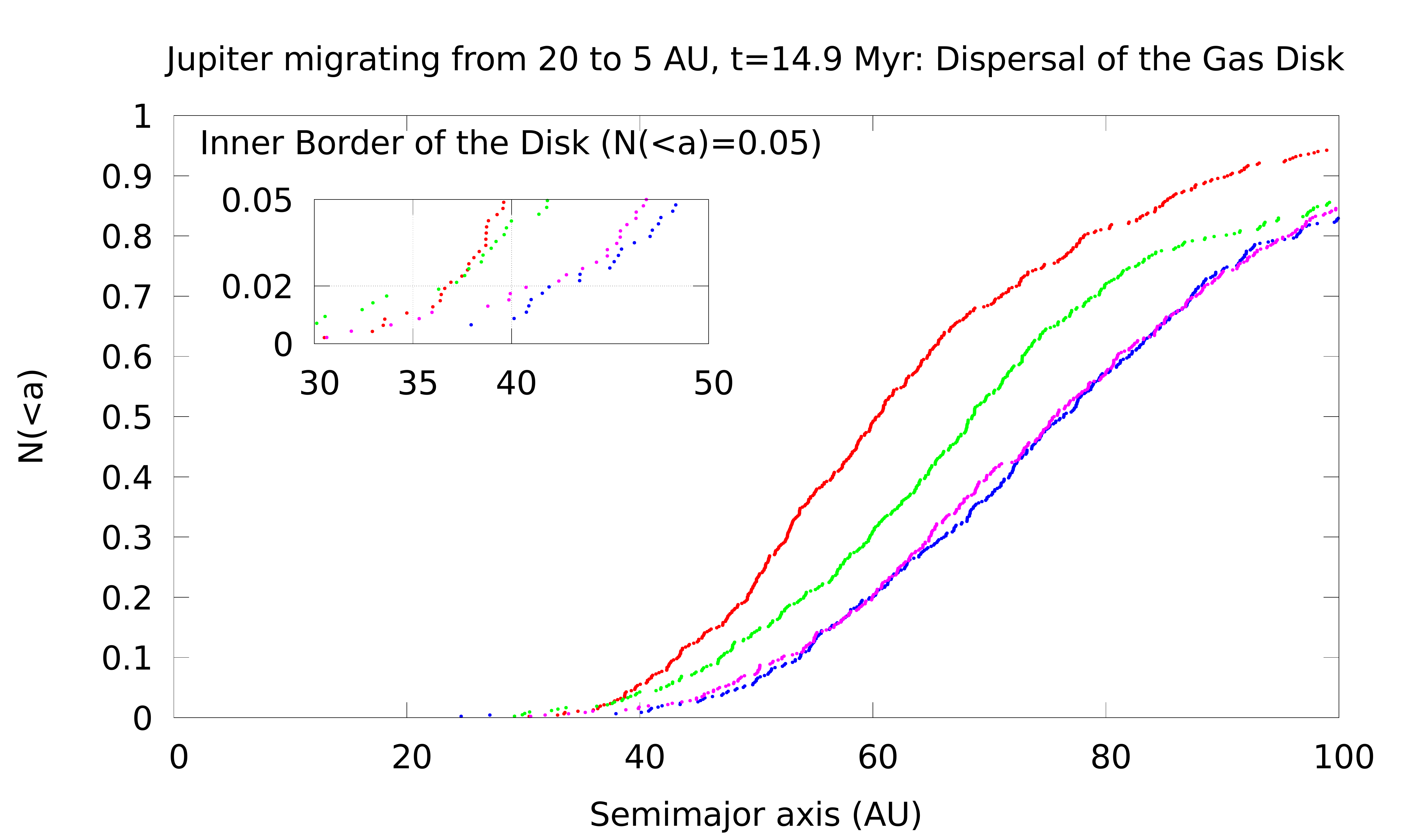}{0.52\textwidth}{(a)}
          \fig{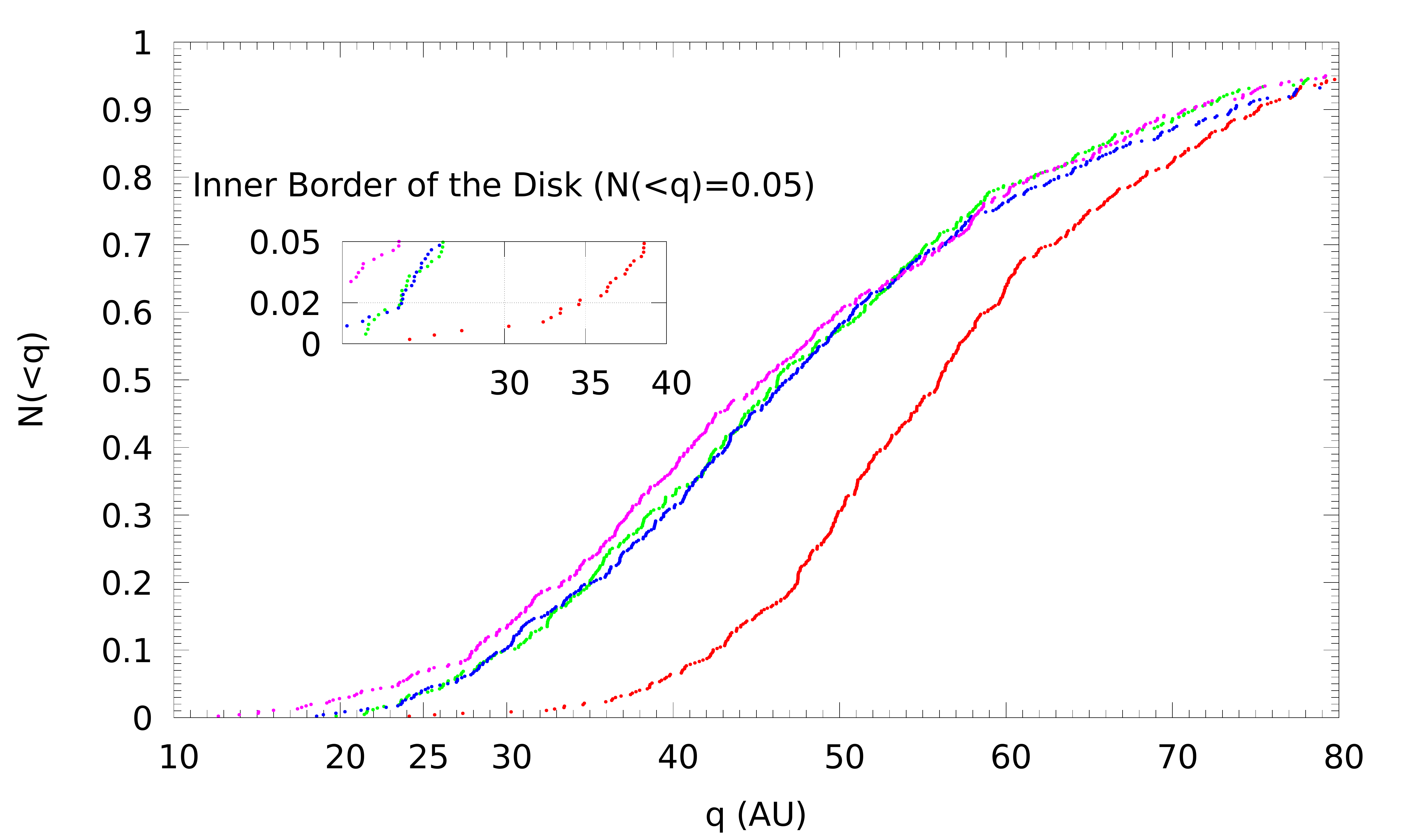}{0.52\textwidth}{(b)}
         }
%          \gridline{\fig{fig/Jup_20AU_in/c.pdf}{0.52\textwidth}{(c)}
%           \fig{fig/Jup_20AU_in/d.pdf}{0.52\textwidth}{(d)}
%          }
          
\caption{The same as Fig. \ref{fig:comulativeJC}, but for the simulation set \textbf{Jup\_20AU\_in} (table \ref{tab:mathmode}).}
\label{fig:cumJI3}
\end{figure}

The cases of inward migration of Jupiter, \textbf{Jup\_10AU\_in, Jup\_15AU\_in and Jup\_20AU\_in}, are presented in Figures \ref{fig:JI1}, \ref{fig:JI11}, \ref{fig:cumJI1}, \ref{fig:JI2}, \ref{fig:JI22}, \ref{fig:cumJI2}, \ref{fig:JI3}, \ref{fig:JI33} and \ref{fig:cumJI3}, respectively.
The simulations begin with a planetesimal disk extended from $20$ to $60$ AU (blue points).
We see that the wider is the migration range of Jupiter, the larger is the final gap between the position of Neptune and 
the inner edge of the disk. For instance, for $1,000$ km-size planetesimals, the separation in $q$ increases from $0.66$ AU, for a $5$ AU inward migration (Case \textbf{Jup\_10AU\_in}, see Fig. \ref{fig:cumJI1}), to $11.95$ AU for a $15$ AU inward migration (Case \textbf{Jup\_20AU\_in}, see Fig. \ref{fig:cumJI3}).
    The reason is that Saturn and the proto-planets starting farther out in the disk can remove more efficiently the $20-30$ AU population of planetesimals by scattering or accretion--concerning to planetesimal's accretion, 
the masses of the planets and proto-planets do not increase due to planetesimal accretion thus, we just removed the planetesimais which accreted with these objects.
Therefore, for large inward migration of Jupiter there may be 
the possibility that the planet instability occurs after a long time. We will check this in section ~\ref{instime}. Note however that, if Jupiter migrated from 10 AU or beyond, the final disk's inclination excitation beyond 
$40$ AU for $100$ km and larger objects exceeds significantly that ones observed in the cold Kuiper belt population,
as discussed next. 
% This is because proto-planets resided in that region for quite long time, before being removed by migration; so they could stir up 
% the local population significantly. On the other hand, the weak gas-drag suffered by large planetesimals is unable to bring the objects back to small-inclination orbits once all proto-planets have left.  
% These considerations suggest that it is unlikely that Jupiter migrated inwards by more than $5$ AU.  

\clearpage

\subsection{Which planetesimal disks are consistent with the primordial Kuiper belt?}
\label{primordialkuiperbelt}
\begin{figure}
          \gridline{\fig{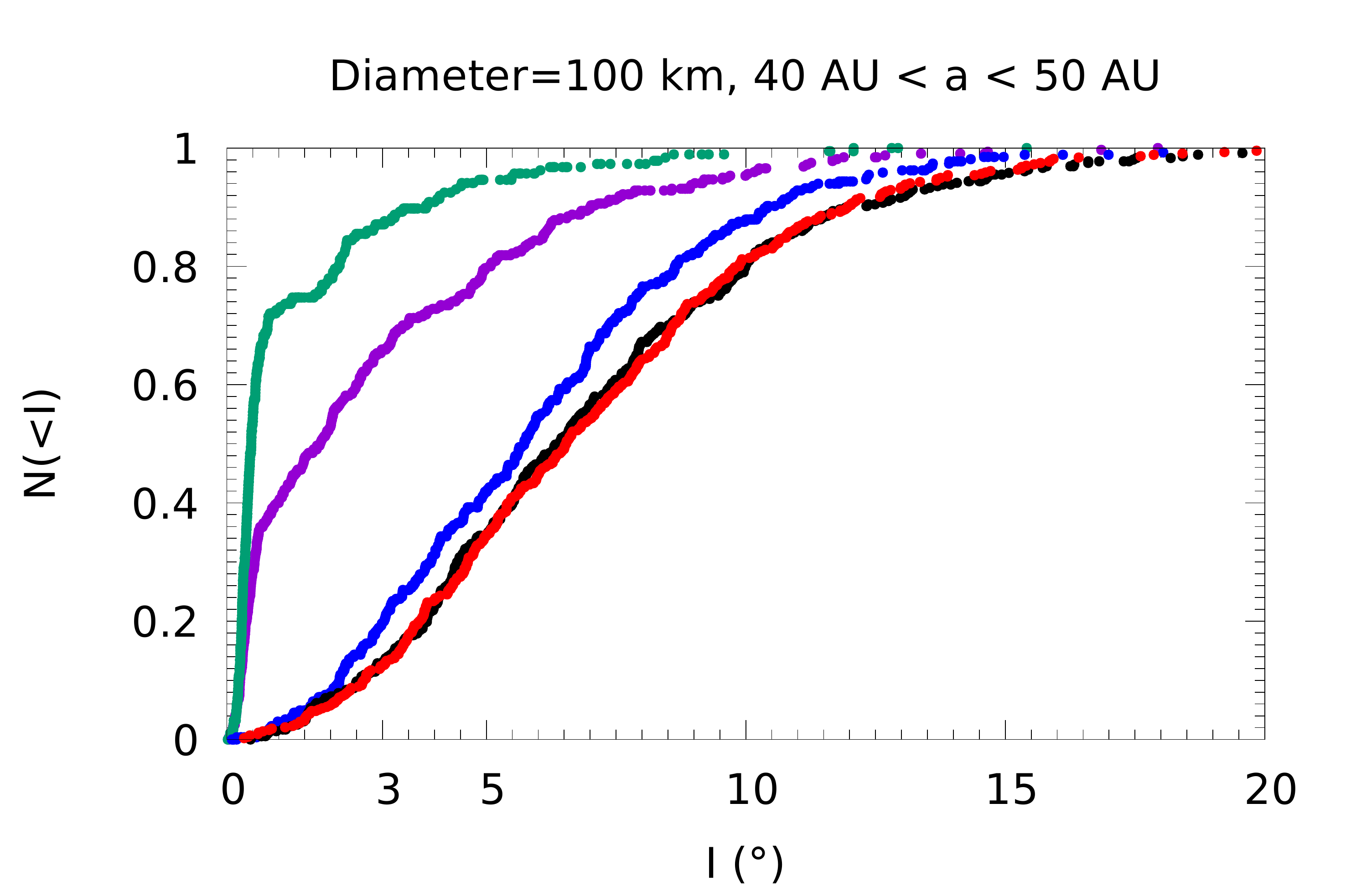}{0.6\textwidth}{}
         }
%          \gridline{\fig{/media/rafanw72/hda/Pesquisa/Morby/project_Morby/interpolation_gas_nbody/program_interpol/version10_migration/fix_time_larger/cumulative_distribution/Migration_Inward2/c.pdf}{0.52\textwidth}{(c)}
%           \fig{/media/rafanw72/hda/Pesquisa/Morby/project_Morby/interpolation_gas_nbody/program_interpol/version10_migration/fix_time_larger/cumulative_distribution/Migration_Inward2/d.pdf}{0.52\textwidth}{(d)}
%          }
          
\caption{Cumulative normalized inclination distribution ($N(<I)$) for 100 km objects presented in the local cold population $40-50$ AU at the end of our five simulations (\textbf{Jup\_static (purple), Jup\_outward (green), Jup\_10AU\_in (blue), Jup\_15AU\_in (black) and Jup\_20AU\_in (red)}, table \ref{tab:mathmode}). 
Note the cases with Jupiter migration from 10 or beyond the local $40-50$ AU population have a final inclination dispersion of $10$ degrees. Whereas, the others cases have a final inclination dispersion smaller than $5$ degrees.}
\label{fig:inclicum}
\end{figure}

The cold Kuiper belt population is very low in eccentricity and inclination and is notoriously confined with $4$ degrees from the invariant plane.  We can use this as a constraint on our planetesimal disks, as there is no clear mechanism to damp planetesimals' orbits after the dissipation of the gaseous disk.  
Figure \ref{fig:inclicum} shows the cumulative inclination distributions for 100 km planetesimals in all of our planetesimal disks.   Planetesimal disks in which Jupiter migrated from $10$ AU or beyond tend to leave the local $40-50$ AU population too excited to be compatible with the cold Kuiper Belt. 
For instance, in the simulation with Jupiter migrating from $10$ AU, the final inclination dispersion in the $40-50$ AU region is $10$ degrees (for $100$ km objects) ( blue curve in Fig. \ref{fig:inclicum}). 
 This is because the planetary embryos precursors of Uranus and Neptune are located beyond Jupiter and Saturn and therefore, 
 if Jupiter is originally beyond $10$ AU, some of the planetary embryos are initially resident in the Kuiper Belt region. 
 The embryos migrate away from their original location, but in doing so they excite the local planetesimal population. 
 Gas drag damping is weak for objects of $100$ km in size or larger, and therefore this excitation remains until the gas disk fade out, 
 in contrast with small inclination excitation observed for the cold Kuiper Belt population.
 We notice that the simulations with Jupiter on a fixed orbit, or migrating outward, give an inclination excitation in the $40-50$ AU region (green and purple curves in Fig. \ref{fig:inclicum}) that does not exceed the excitation of the cold Kuiper  Belt.

\section{Constraining the timing of the giant planet instability}
\label{instime}

We now address the timing of the giant planet instability. For statistical reasons, we make use of all simulations from \citet{Izidoroetal2015} that produced good matches to the outer Solar System, meaning that they formed at 
least two roughly equal-mass ice giants.  This constitutes a large enough sample that we can make statistical arguments.  We first test the stability of the giant planet systems alone to test for self-triggered instabilities (\S \ref{selftriggered}).  Next we 
test the stability of the same systems while including our self-consistently generated planetesimal disks (\S \ref{plantinstdisk}). 

Here we define time zero as the end of the gaseous disk phase, i.e. the end of the simulations presented in Section \ref{results}.  We measure the time of the giant planet instability relative to this.
The instability is defined as the beginning of the close-encounter phase among the planets. It happens when the giant planets break their resonant chain configuration. 

\subsection{Self-triggered Instability}
\label{selftriggered}

%%RAFAEL -- FILL IN THIS SECTION WITH YOUR NEW RESULTS.  I RECOMMEND A MULTI-PANEL peri/aphelion vs time plot showing each of Andre's good sims going unstable.  No need to show resonant angles evolving, just the orbital elements
% 
%     \begin{figure*}
%           \gridline{
%           \fig{/media/rafanw72/hda/Pesquisa/Morby/project_Morby/Sean_complementary/All_cases/dataprocess/good_cases/fig0.png}{0.5\textwidth}{(a)}
%           \fig{/media/rafanw72/hda/Pesquisa/Morby/project_Morby/Sean_complementary/All_cases/dataprocess/good_cases/fig1.png}{0.5\textwidth}{(b)}
%           }
%           
% \caption{Panel (a) shows the evolution of perihelion, semimajor axis, aphelion of Jupiter (red curve), Saturn (green curve), Ice1 (blue curve), Ice2 (pink curve) and Ice3 (gray curve) from the simulation of Izidoro et al. simulation without
% any planetesimal disk.
% \label{fig:izidoro1}}
% \end{figure*}

\citet{Izidoroetal2015} never tested whether their final systems of gas giants and ice giants remained stable after the gas disk phase. Many of such systems become unstable very quickly.  
% 
% For example, Figure~\ref{fig:izidoro1} shows that the giant planet system that we used in Section 2 goes unstable after 2 Myr. 
This happens when the planets, in particular the ice giants, are too eccentric for long-term stability or in a too compact resonant chain.  Indeed, it is well-known that resonant chains may go unstable ~\citep[e.g.][]{Matsumoto2012,Izidoro2017}.
% 
% 
%     \begin{figure*}
%           \gridline{
%           \fig{fig/izidoro/fig1.png}{0.5\textwidth}{}
% %           \fig{fig/izidoro/fig2.png}{0.5\textwidth}{(b)}
%           }
%           
% \caption{Panel shows the evolution of perihelion, semimajor axis, aphelion of Jupiter (red curve), Saturn (green curve), Ice1 (blue curve), Ice2 (pink curve) and Ice3 (gray curve) from the simulation of Izidoro et al. simulation without
% any planetesimal disk.
% \label{fig:izidoro1}}
% \end{figure*}
We statistically studied the stability of 27 giant planet systems from \citet{Izidoroetal2015}.  Each of these systems contained at least two roughly equal-mass ice giants similar to Uranus and Neptune, and many contained one or more additional surviving small planets (ice giant-mass or less).
For each of these systems we scaled the semi major axis of all planets to place Jupiter at 5 AU (see section \ref{interp}). Then we integrated each system for up to 1 Gyr to determine whether the system remained stable or underwent a self-triggered instability.  

Our initial conditions were the orbital elements of the planets at the end-time of the gas phase from \citet{Izidoroetal2015}. Since instabilities are so sensitive to initial conditions, we performed 10 simulations of each system to 
generate a distribution of outcomes. 
 For each simulation we did slight changes on the planet initial conditions: a random phase chose in the interval (-0.003,0.003) degrees is added to each orbital angles of the planets, 
including the mutual inclinations. These changes are small enough not to modify the stability conditions of the original planetary systems. Indeed, all closens of stable systems remain stable 
(in absence of planetesimal perturbations). We integrated each system for 1 Gyr using a time step of 0.5 years. 

\begin{figure}
          \gridline{
       \fig{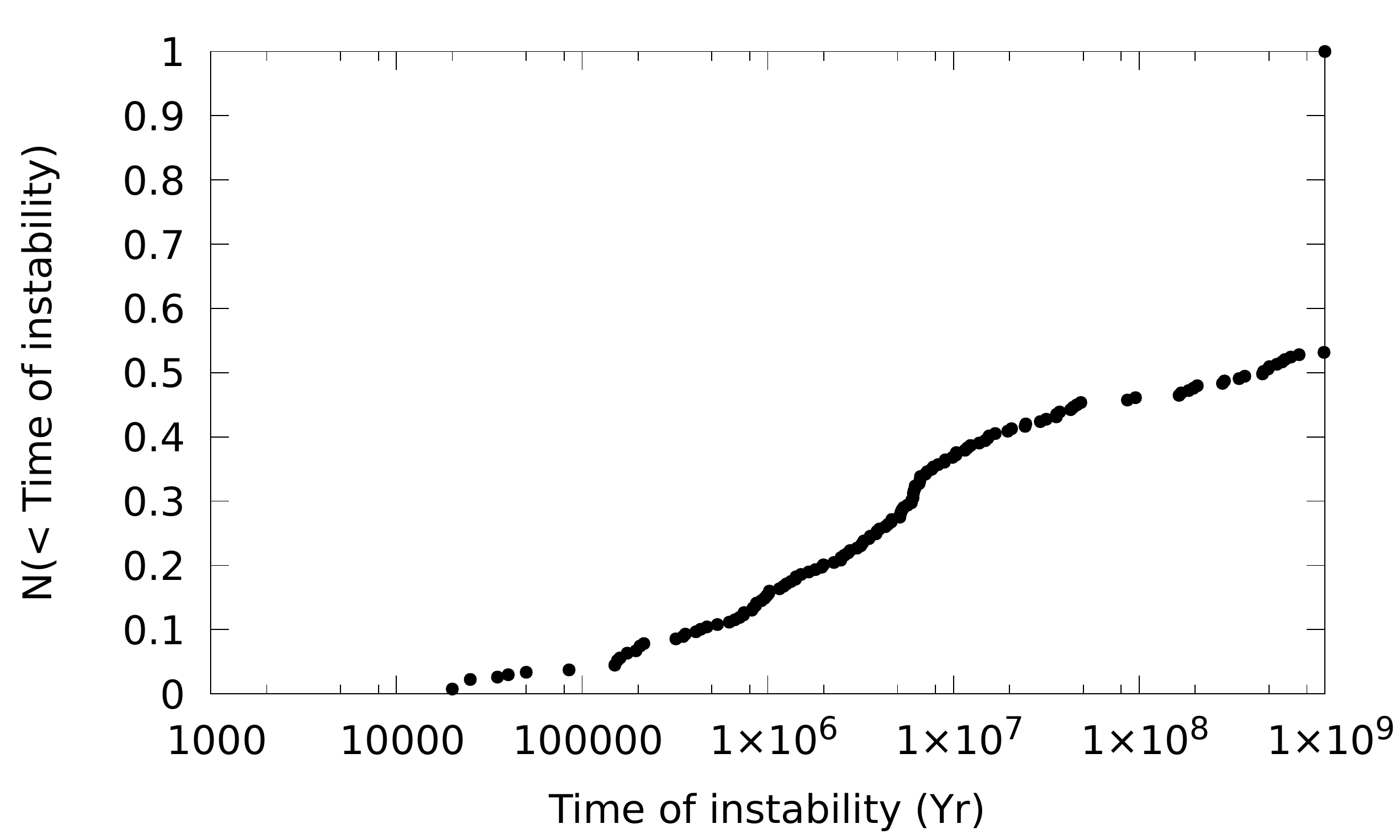}{0.5\textwidth}{}
%           \fig{/media/rafanw72/hda/Pesquisa/Morby/project_Morby/Sean_complementary/All_cases/dataprocess/good_cases_pc4/all_cases/traject.pdf}{0.5\textwidth}{(b)}
          }
%             \gridline{
%           \fig{/media/rafanw72/hda/Pesquisa/Morby/project_Morby/Sean_complementary/All_cases/dataprocess/good_cases/fig1.pdf}{0.5\textwidth}{(b)}
% %           \fig{/media/rafanw72/hda/Pesquisa/Morby/project_Morby/Sean_complementary/All_cases/dataprocess/good_cases/fig3.pdf}{0.5\textwidth}{(b)}
%           }

\caption{ Panel shows the cumulative normalized distribution of the time of instability calculated during our simulations after the gas dispersal and without any planetesimal disk.
\label{fig:siteizido}}
\end{figure}

Fig. \ref{fig:siteizido} shows the outcome of these simulations. Some systems go unstable quickly and others remain 
stable for 1 Gyr. Just under half (48\%) systems remain stable for 1 billion of years. 
  The vast majority (80\%) of unstable systems had instabilities within 10 Myr. The reason why many of \citet{Izidoroetal2015}'s planetary systems are self-unstable compared 
to the systems built as initial conditions of the Nice model \citep{Morbidellietal2007} is that in \citet{Izidoroetal2015}'s case (which we used in this work) the planetary
systems is continuously unstable until Uranus and Neptune are built and at that point there is not much gas to damp the orbits of the planets, weheras \citet{Morbidellietal2007} 
considered 4 full formed planets that were captured in mean motion resonance in sequence.

%Surprisingly, only systems with 4 and 6 giant planets (Jupiter, Saturn and 2 or 4 ice giants) were fully stable (Fig. \ref{fig:siteizido} (c)). 
The presence of an outer planetesimal disk may have a significant effect on the stability of a system of giant planets, 
in particular when the masses of the outermost planets are comparable to the disk mass~\citep{raymond10}. 
Thus, in the next section (\S \ref{plantinstdisk}) we test how the stability of giant planet systems changes when we 
include the planetesimal disks that we found in section \ref{results}. 

%We observed that the planetary systems with five giant planets got instabilities at the limit below of 282 My in our simulations. 
%Those systems are too eccentric for long-term stability, as we show in section \ref{plantinstdisk} that the system could become stable if the Izidoro et al. (2015a)'s simulations the gas disk was presumably not too thick. 
%In the framework of the Nice model, planetary systems which satisfied the criteria from many features of the Solar System with such high success are the five giant planet systems (Nesvorny and Morbidelli, 2012).
%However, four planets systems have had success as well to explain those features. In our simulations, four planets systems are stable for time enough to produce the Late Heavy Bombardment in an absence of a planetesimal disk. 
%Thus, we show in section \ref{plantinstdisk} how this picture change including the perturbation of the planetesimal disks that we found in our results of the section \ref{results}. 

\subsection{Planetesimal disk-triggered Instability}
%%RAFAEL -- UPDATE THIS SUBSECTION WITH YOUR NEW RESULTS
\label{plantinstdisk}

We now perform simulations to determine the timing of a planet instability triggered by the interaction with the 
planetesimal disk, making use of the planetesimal disks produced by Section \ref{results}.
The simulations presented in the Section \ref{results} treated planetesimals as test particles. We adopted 4 sizes for the computation of the gas-drag effects and hence the final orbital distributions. 
Now, if we want to investigate the effects of the planetesimal disk onto the planets, we need to combine the orbital distribution of these 4 categories of particles and assign a mass to them. We explain how we do this in the 
subsection \ref{massdiks} below. 
 
 \subsubsection{Mass of the Planetesimal disks}
 \label{massdiks}
The total mass of the primordial planetesimal disk is expected to be $20$ Earth's mass \citep{NesvornyMorby2012,Nesvorny2016a}. 
Its original size distribution was reconstructed by \citet{Nesvorny2016a} using many constraints. The first constraint is that Neptune's migration 
should have been grainy to explain why the fraction of the Kuiper Belt population in resonances is relatively small. 
This grainy migration requires close encounters of Neptune with massive Pluto-class planetesimals. 
Therefore, \citet{Nesvorny2016a} argued that the planetesimal disk contained $1000-4000$ Pluto-size objects.  For the intermediate sizes ($10<D<500$ km) 
\citet{Nesvorny2016a} adopted the \citet{Fraseretal2014} size distribution from observations of the Kuiper Belt and Jupiter Trojans.
This distribution shows a knee at $D\sim 100$ km. Other constraints come from the comet's size distribution and the requirement that the overall mass of the disk is finite. 
In summary, \citet{Nesvorny2016a} model the cumulative distribution $N(>D)$ as a piece-wise power law: 
\begin{equation}
\label{eq:discummm}
N(>D) \propto D^{-q},
\end{equation}

% \begin{equation}
% dN(>D) \propto D^{-q}dD,
% \end{equation}

with $q=1$ for $D>500$ km, $q=5$ for $100<D<500$km, $q=2$ for $10<D<100$km, $q=3$ for $1<D<10$km and $q=2$ for $D<1$ km. 
We consider that $ N (> 1000 km) $, $ N (> 500 km) $, $ N (> 100 km) $, $ N (> 10 km) $, $ N (> 1 km) $ and $ N (> 0.1 km) $ are the number of objects with diameter larger than 1000 km, 500 km, 
100 km and 0.1 km, respectively. The number of objects larger than 500 km  ($ N (> 500 km) $) is a constant to be determined by the total mass of the Planetesimal disk. 
Thus, we used the cumulative distribution (Eq. \ref{eq:discummm}) to relate the number of objects larger than a size D ($N>D$) with the respective exponent $q$:
\begin{equation}
\label{eq:eq500}
N (>500 km)= N (> 1000 km) \left(\frac{1000 km}{500 km}\right)^{q}, q=1
\end{equation}

\begin{equation}
N (>100 km)= N (> 500 km) \left(\frac{500 km}{100 km}\right)^{q}, q=5,
\end{equation}

\begin{equation}
N (>10 km)= N (> 100 km) \left(\frac{100 km}{10 km}\right)^{q}, q=2,
\end{equation}

\begin{equation}
N (>1 km)= N (> 10 km) \left(\frac{10 km}{1 km}\right)^{q}, q=3,
\end{equation}

\begin{equation}
\label{eq:eq01}
N (>0.1 km)= N (> 1 km) \left(\frac{1 km}{0.1 km}\right)^{q}, q=2.
\end{equation}

We used Equations \ref{eq:eq500} to \ref{eq:eq01} to calculate the proportional constant ($\gamma$) of the differential equation $dN(>D)$ defined as:

\begin{equation}
\label{eq:discummm2}
dN (> D) = \gamma D ^ {- q-1} dD,
\end{equation}

We can find $\gamma$ for each region of diameters larger than $D_ {1} $ and $D_ {2}$,
calculating the integral:
\begin{equation}
N (> D_1) - N (> D_2) = \int_ {D_1} ^ {D_2} \gamma D ^ {- q-1} dD.
\end{equation}

We considered that $N_0$ represents the objects larger than 500 km. Thus, 
\begin{equation}
\label{eq:integral1}
N_0= N(> 500km) =   \int_{500}^{+ \infty} \gamma_0 D^{-q-1}dD, q=1
\end{equation}

Solving the integral (Eq. \ref{eq:integral1}) in function of $\gamma_0$, we write: 
\begin{equation}
\gamma_0=  500 N_0
\end{equation}

We can calculate the other proportionality constants, be $\gamma_1$ the proportionality constant of the region between 100 and 500 km, with $q=5$:
\begin{equation}
N (> 100) - N (> 500) = \int_ {100} ^ {500} \gamma_1 D ^ {- q-1} dD.
\end{equation}

We can calculate the quantity of mass ($M_{D1,D2}$) between two different sizes $D_1$ e $D_2$. To do this, we considered spherical planetesimals with the bulk of density $\rho$. Thus,
\begin{equation}
\label{massatotal}
 M_{D1,D2}= \int_{D_1}^{D_2} \frac{4}{3} \pi \rho \left(\frac{D}{2}\right)^{3} dN(>D) = \int_{D_1}^{D_2} \frac{4}{3} \pi \rho \left(\frac{D}{2}\right)^{3} \gamma D^{-q-1} dD.
\end{equation}

We assume that our test particles with $D=1,000$ km represent all planetesimals with size larger than $500$ km. The particles whose distribution was computed 
assuming $D=100$ km represent the planetesimals between $30$ and $500$ km, those with $D=10$ km represent planetesimals between $3$ and $30$ km 
and those with $D=1$ km represent particles with $D<3$ km. Defining $N_o$ the number of planetesimals with $D>500$ km, the size distribution
reported above defines the number of planetesimals in each size interval. 
The cumulative distribution is then converted into an incremental distribution and the total mass is computed assuming a bulk density of 3 $g/cm^3$. With this procedure, 
the mass that we find is proportional to $N_0$, the number of bodies larger than $500$ km. The value of $N_0$ is then found imposing a total of $24$ Earth masses for the disk.
The resulting numbers for $N_0$, $N(>100km)$, $N(>10km)$ and $N(>1km)$ are: $6200$, $1.5872\times10^{7}$, $1.5872\times10^{9}$, $1.5872\times10^{12}$, in good agreement with Fig. 15 of \citet{Nesvorny2016a}. 
The total number of particles that we have at each size at the end of the simulations reported in the previous section is of course much smaller than the real number of planetesimals in the disk.
In one case for example, we have 564, 605, 629, 637 particles with a diameter of 1000, 100, 10 and 1 km respectively at the end of the gas phase.
Thus, we created ``super-particles'', with a mass equal to the total mass in the considered size interval, divided by the number of particles that survived in the end of our simulations.

With the mass of the planetesimal disk in hands, the next step is to perform the simulations of the evolution of the planetary system
under the effects of the different planetesimal disks showed in the section~\ref{results}.
However, for a smooth transition from the previous simulation (that featured the planets and test particles) we grow the masses of the particles 
from 0 to their final mass ($M_f$) with a function 
\begin{equation}
M(t)=M_f(1-\exp(-t/3 My)). 
\end{equation}
In this way, planetesimals are growing their mass slowly and smoothly until reach the final Super-Particles mass after $3$ Myr of simulation. This procedure was followed to avoid an abrupt transition from a massless disk to a 
massive disk, which could cause artificial instabilities.

\begin{figure}
          \gridline{\fig{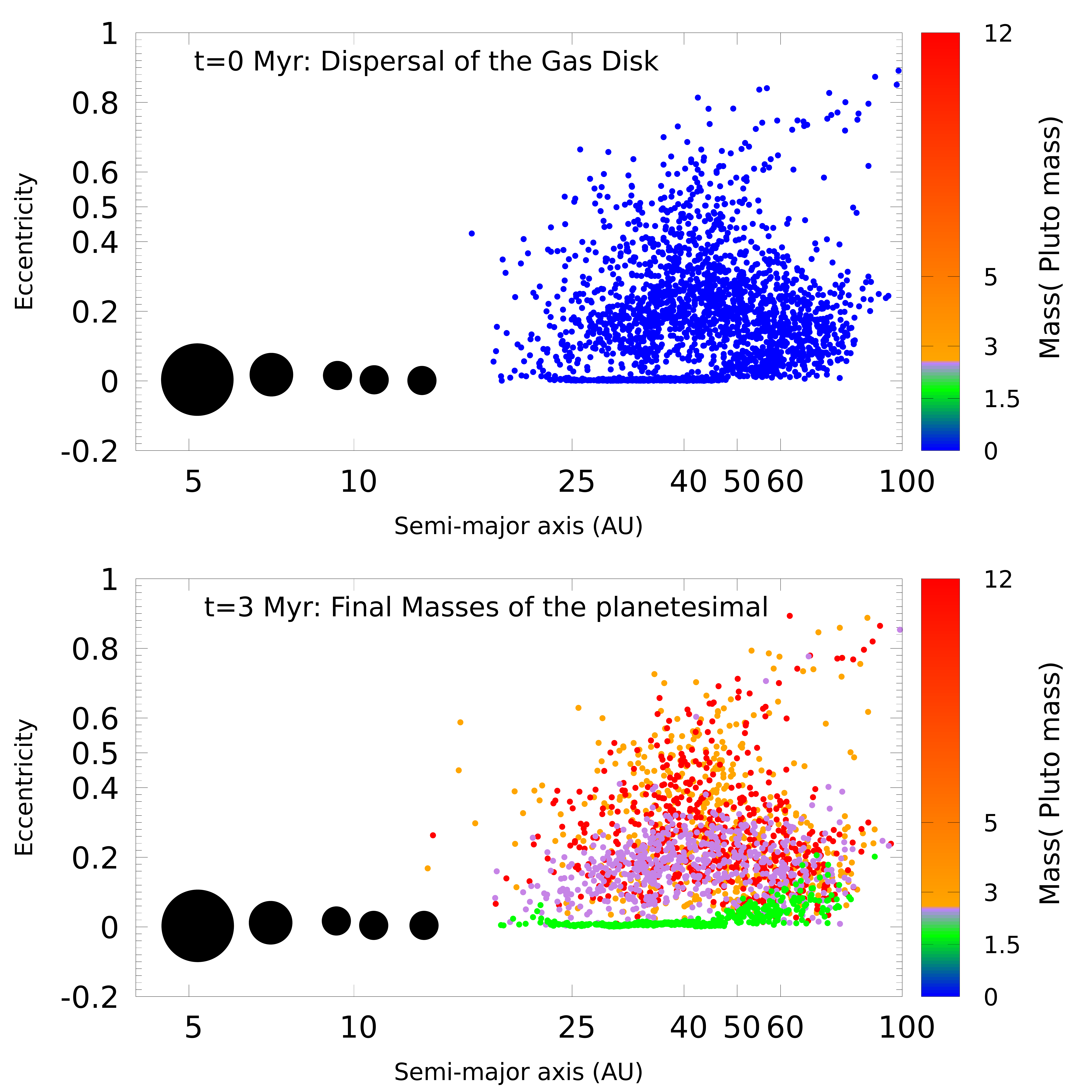}{0.5\textwidth}{}}
          
\caption{Two snapshots of our simulations with giant planets interacting with the planetesimal disk, starting from the end-state of the simulation with Jupiter migrating from 10 to 5 AU (Fig. \ref{fig:JI1}) 
The color scale represents the super-particles' mass in units of Pluto's mass.}
\label{fig:massdistribution}
\end{figure}

In Figure \ref{fig:massdistribution}, we show two snapshots of one of our simulations (the one starting from the endstate of the simulation with Jupiter migrating from $10$ to $5$ AU)
where the planets interact with the planetesimal disk. The color box represents the super-particles' mass in Pluto's masses. At time $3$ My (Fig. \ref{fig:massdistribution} b) the particles have their final mass.
Note that the most massive super-particles (red) correspond to the test particles with $D=100$ km in the simulations of the previous section because 
most of the mass in the \citet{Nesvorny2016a} distribution falls in the 30$<D<$500 km range, that is represented by our D=100 km Super-Particles.
The group of particles most dynamically excited (orange) correspond to super-particles with masses of 3 Pluto's mass and D=1000 km. 
The green super-particles correspond instead to Super particles with D=1 km, strongly damped by gas drag during the planet formation-migration phase. The purple super-particles
represent the super-particles with D=10 km with mass around of 2.4 Pluto's mass.

 Because our super-particles are quite massive (several Pluto masses), when they encounter the planets they can force the latter to have spuriously large orbital jumps, that favor the rapid onset of instability. 
For this reason, following  \citet{Gomesetal2005} when super-Particles come close enough to Neptune's orbit (i.e. their perihelion distance is within 2.85 Neptune's Hill radii from Neptune's aphelion distance), 
they are cloned 18, 30 or 150 times depending on the size of the super-Particles. The clones have all initially the same position and slightly different velocities. In this way, Neptune encounters only particles individually with 
10 percent of the mass of Pluto, avoiding artificially large orbital jumps.

\begin{figure}
          \gridline{
           \fig{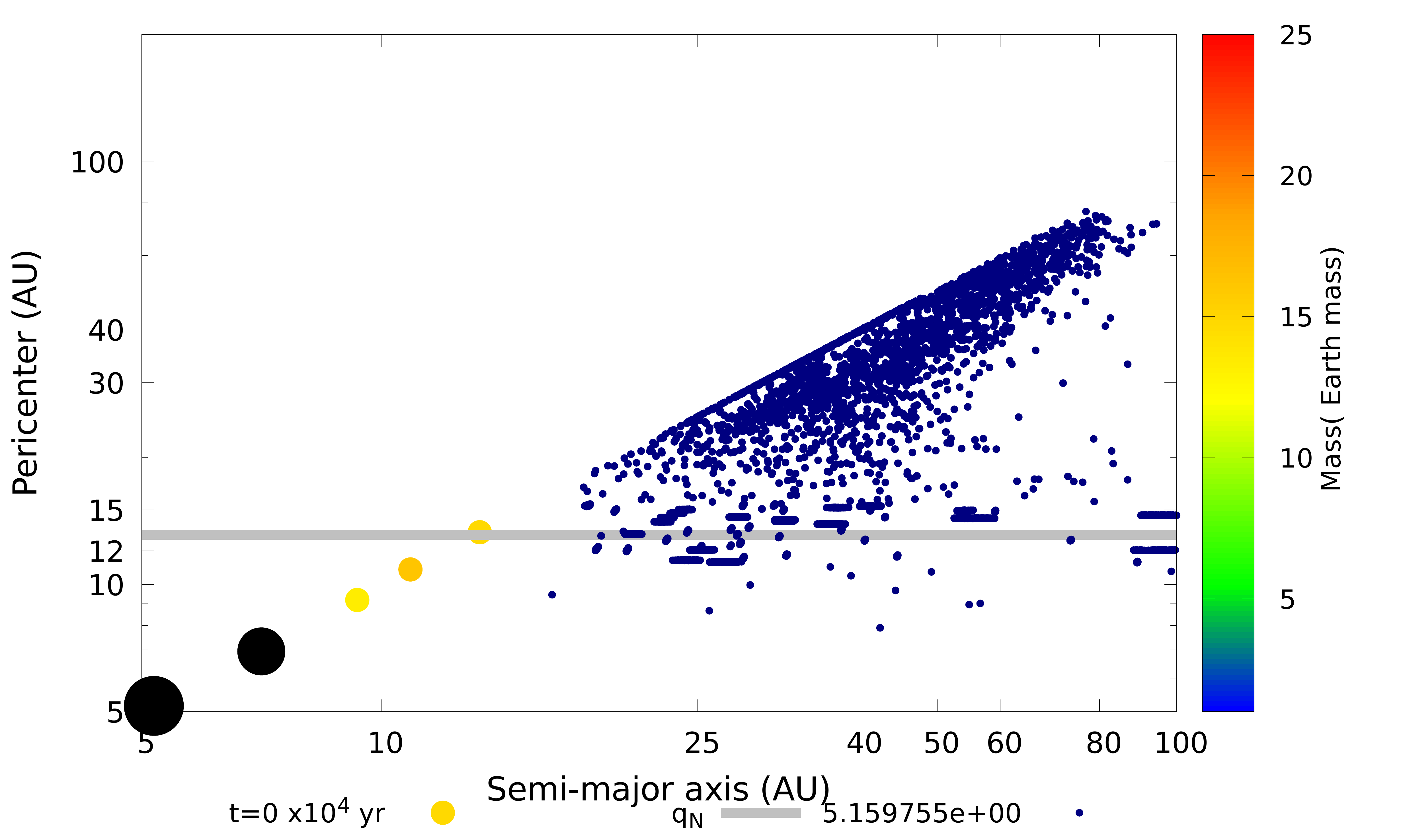}{0.5\textwidth}{(a)}
          \fig{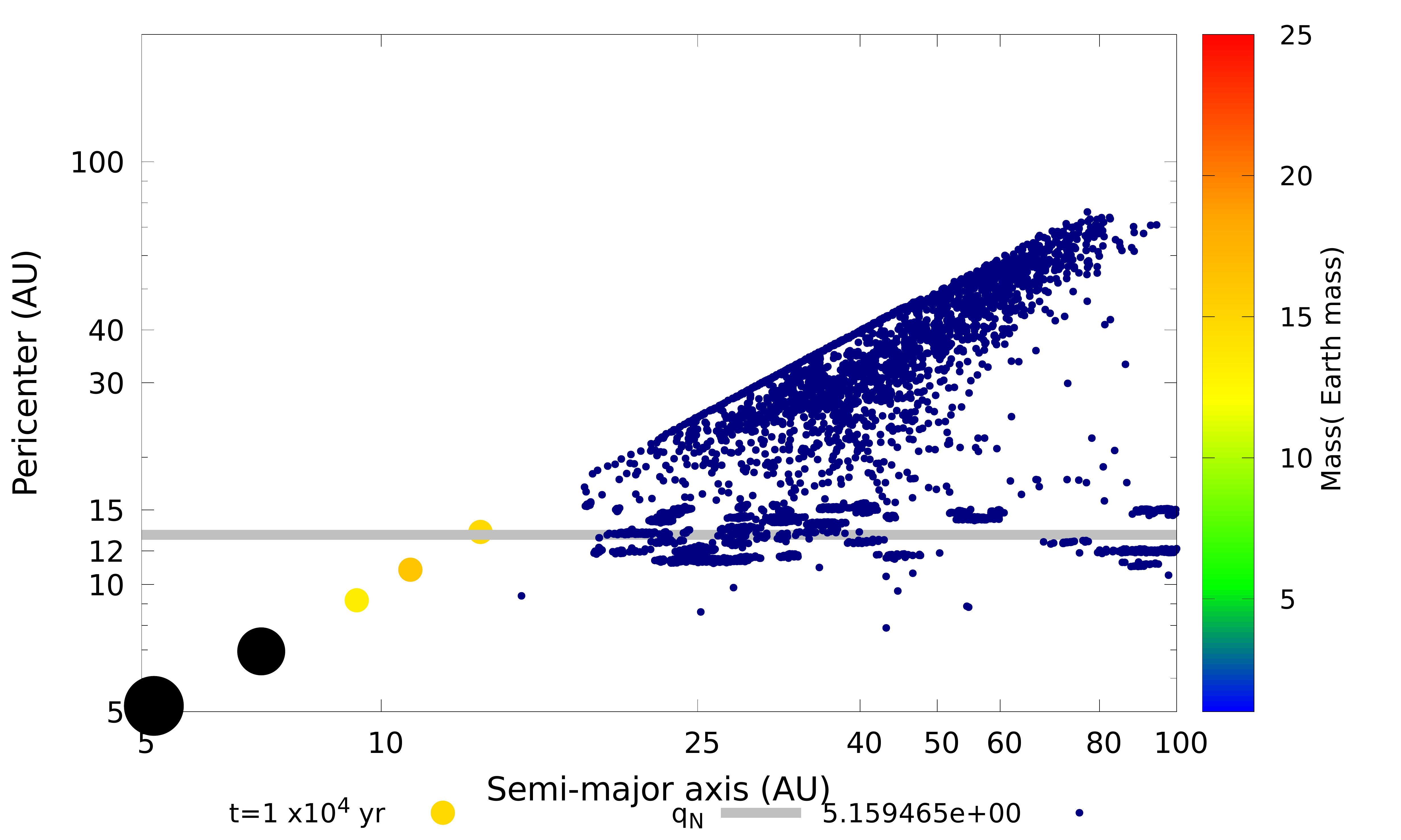}{0.5\textwidth}{}
          }
             \gridline{
           \fig{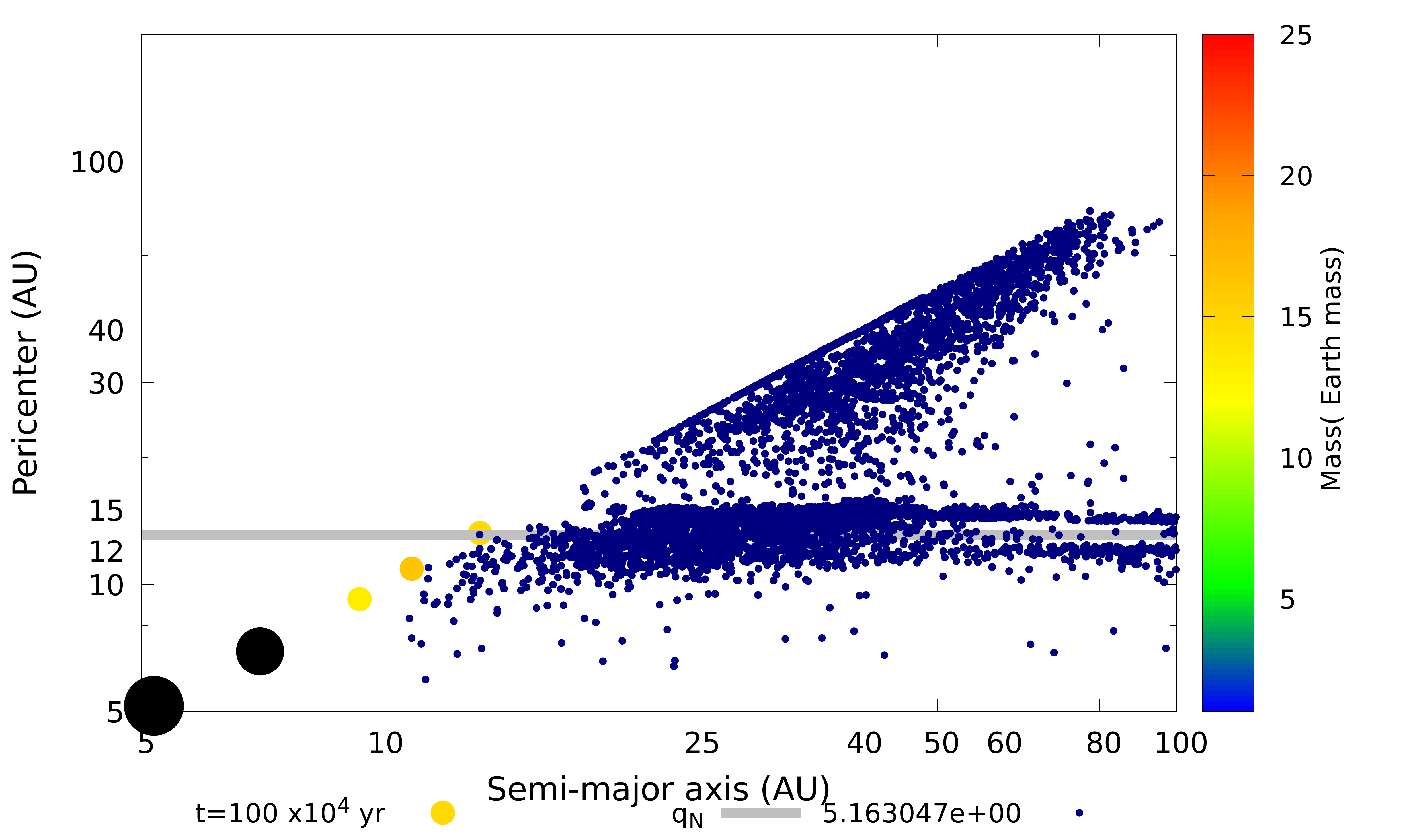}{0.5\textwidth}{(b)}
           \fig{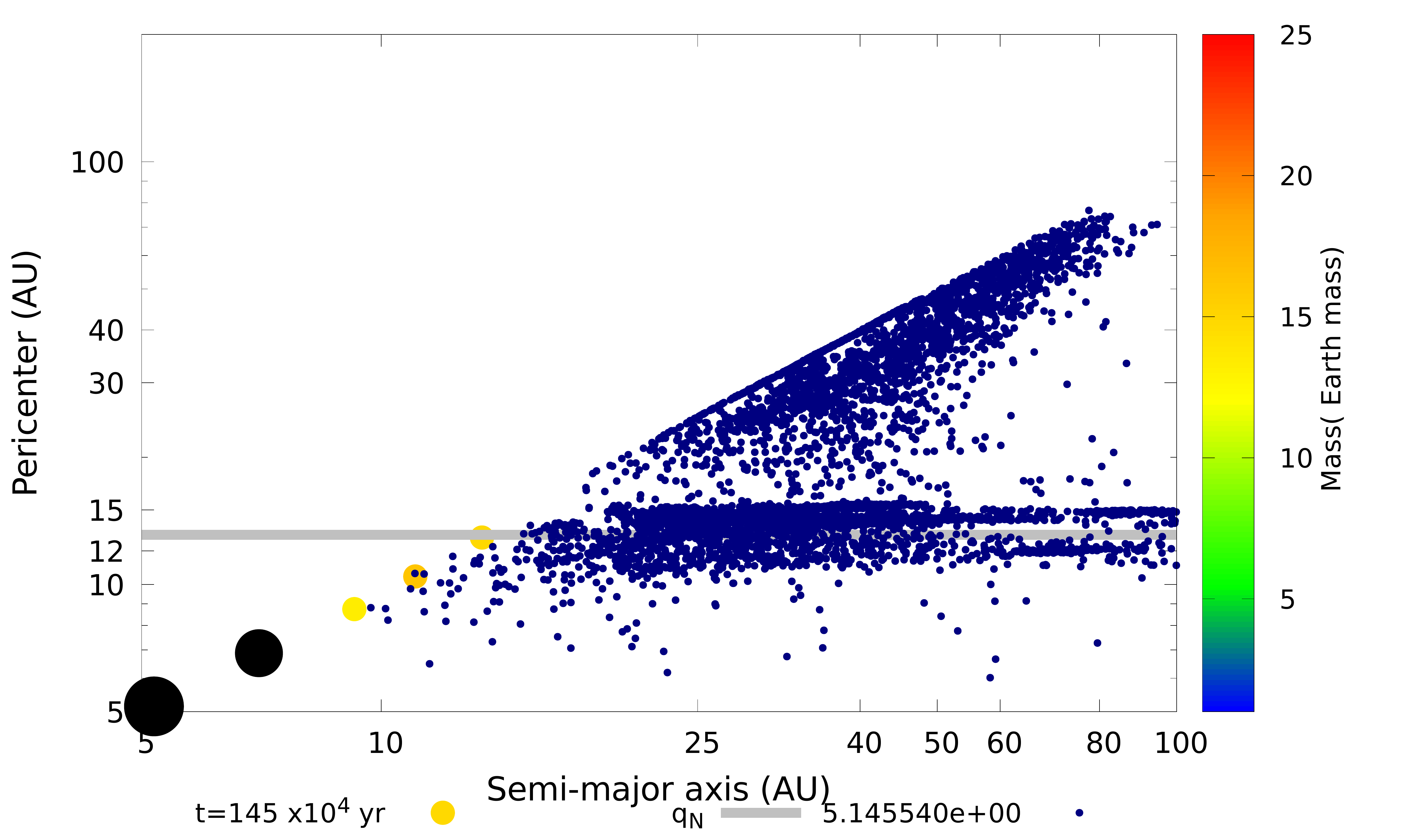}{0.5\textwidth}{(b)}
           }

\caption{Evolution of the pericenter of the planets and planetesimals as a function of the semimajor axis. Clone production takes place around Neptune's pericenter line (gray line). 
Panel (a) represents the initial moment of the simulation and the pericenter of the planets and planetesimals in 10,000, 1 million years and 1.45 million years
are showed in panels (b), (c) and (d) respectively.\label{fig:clones}}
\end{figure}

 Neptune is always defined as the outermost ice giant in our planetary systems. Let $q_ {N}$ be the pericenter of Neptune and $q_ {p}$ the pericenter of a planetesimal. 
Let $\delta =\frac{5}{30}q_{Neptune}$, we clone the planetesimals when $q_ {N} - q_{p} <= \delta $. In this condition, a new planetesimal is created with small deviations in the velocity of the particle to be 
cloned. The positions of the new planetesimals were kept fixed. In Fig. \ref{fig:clones}, 
we show the production of clones in four snapshots of one of our simulations with the planetesimal disc. 
These snapshots represent the pericenter of the planets and planetesimals as a function of the semimajor axis. At the start, the planetesimal that are already close 
to Neptune's orbit are immediately cloned. The new planetesimals are clustered around Neptune's pericente liner (gray line).

 In this way, Neptune only undergoes close encounters with particles with masses of 10 percent the mass of Pluto, avoiding artificial jumps. 
The goal of this procedure was to avoid numerically inducing early instabilities. If instabilities 
take place in our simulated systems with 0.1 Pluto-mass planetesimals, then they should 
certainly have taken place in real systems with Pluto-mass planetesimals (e.g., Quarles and Kaib 2019).

\subsubsection{Statistical analyses for Planetesimal disk-triggered Instability }
Once again, we statistically studied the stability of the same 27 giant planet system from \citet{Izidoroetal2015} 
which we used in Section \ref{selftriggered} but now making use of the planetesimal disks produced by the cases 
\textbf{Jup\_static}, \textbf{Jup\_outward}, \textbf{Jup\_10AU\_in} and \textbf{Jup\_15AU\_in} 
\footnote{We did not perform simulations with the case \textbf{Jup\_20AU\_in} because of computational resources and also it is inconsistent with the primordial Kuiper Belt. (See in Section \ref{primordialkuiperbelt})}. 
It is important to say we used the same planetesimals' disks from the different migrition histories for Jupiter 
for all 27 planetary systems. However, all of the planetesimals' disks are produced from a single case of formation of 
Uranus and Neptune. To maintain the same semimajor axis relative to the outermost ice giant we scale the planetesimals' 
orbits for each planetary system.
To generate a distribution of outcomes, we performed 10 simulations of each of the 27 cases randomizing the orbital angles of the planetesimals' orbits. We have done a total of 1080 simulations. 
For the planets, the initial conditions were the orbital elements of the planets at the end-time of the gas phase from \citet{Izidoroetal2015}.

\begin{figure}
\gridline{\fig{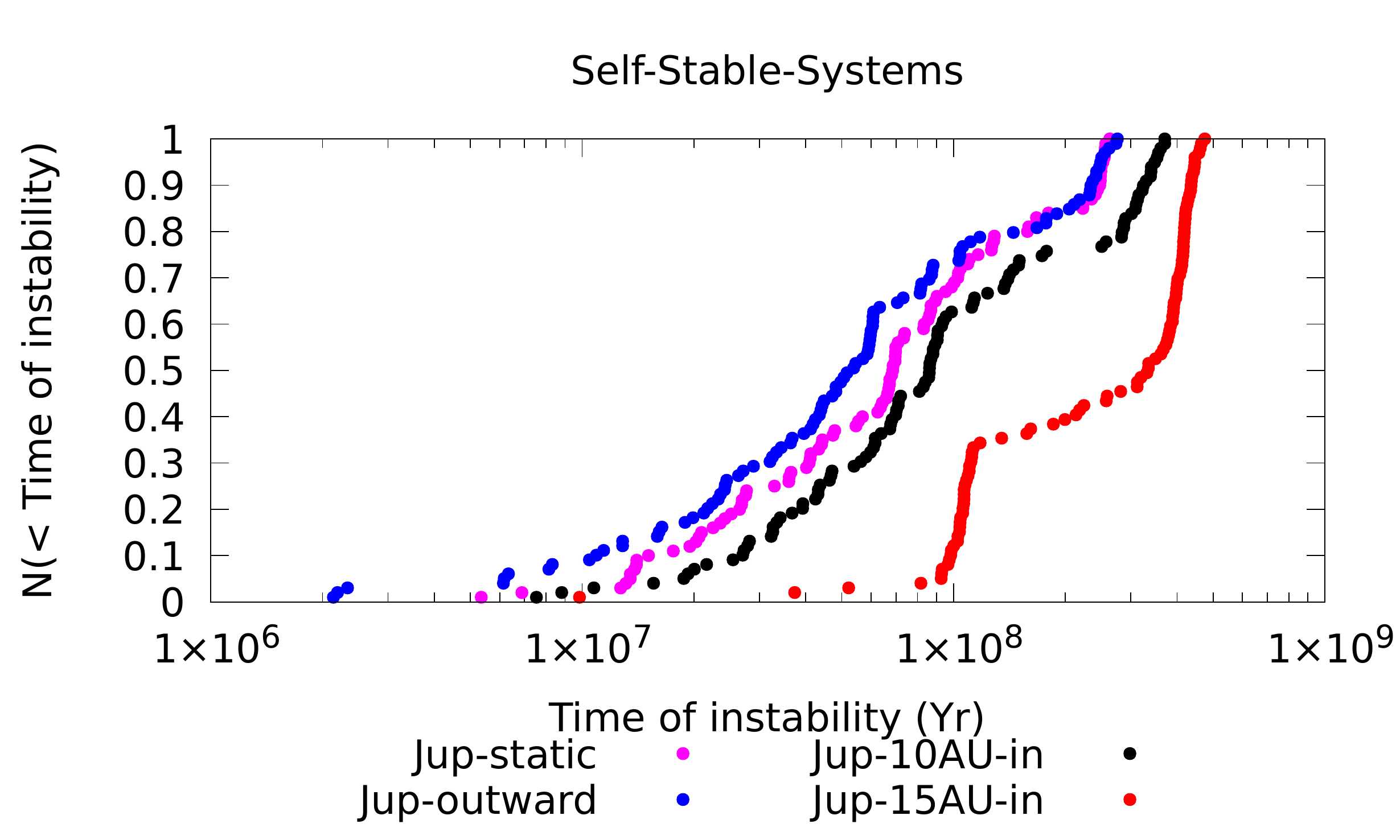} {0.5\textwidth}{(a)}
          \fig{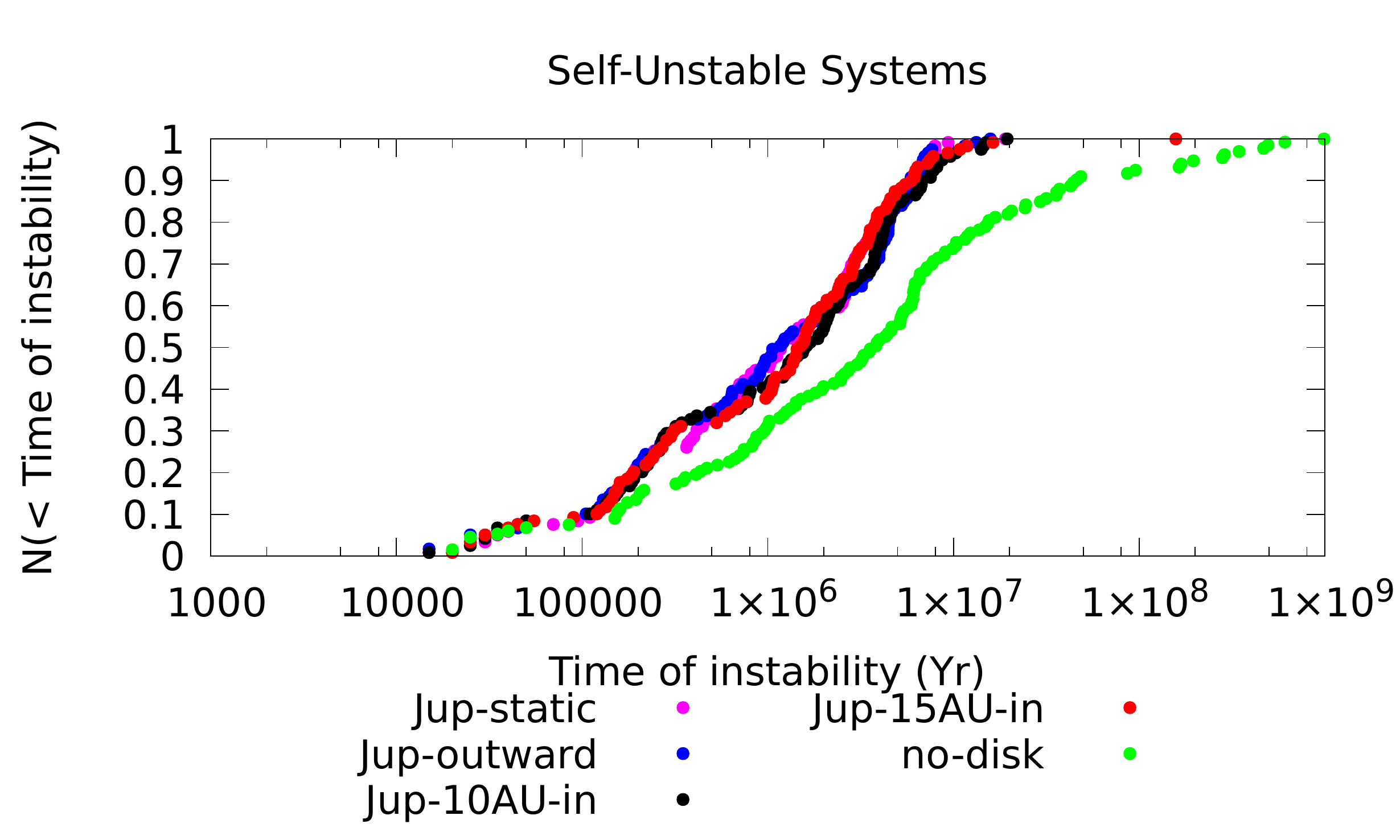} {0.5\textwidth}{(b)}
          }
\caption{Cumulative normalized distributions of the timing of instabilities for the cases calculated during our simulations after the gas dispersal using the planetesimal disks produced by the cases \textbf{Jup\_static},
\textbf{Jup\_outward}, \textbf{Jup\_10AU\_in} and \textbf{Jup\_15AU\_in}, in pink, blue, black and red respectively. In the Panel (a) we only considered self-stable systems (i.e systems that are stable for 1 Gy without any planetesimal
disk). In the Panel (b) we used the unself-stable systems (i.e systems that are unstable in less than 1 Gy without any planetesimal disk) to calculate the distribution of time of instabilities. The green points represent the distribution of unself-stable systems getting unstable without any planetesimal disk for a direct comparison. 
\label{fig:cumulativefinal}}
 \end{figure}

The outcome of these simulations is shown in Fig. \ref{fig:cumulativefinal}. The distribution of instability times is shown separately: first we used only the self-stable systems which  are stable for 1 Gy in case without any planetesimal disk (Fig. \ref{fig:cumulativefinal} (a)) and calculated the distribution times using the planetesimal disks; and second we used only the unself-stable system which are those systems unstable in less than 1 Gy in case of any planetesimal disk (Fig. \ref{fig:cumulativefinal} (b)) to calculate the distribution times now using the planetesimal disks. 
For comparison with the results of Section \ref{selftriggered} while 48\% of simulations without planetesimal disks were stable for 1 billion years without planetesimal disks, all of the simulations with disks were unstable within 500 Myr.

According with the Panel (a) of Fig. \ref{fig:cumulativefinal}, as we expected, the fraction of stable system depends of the initial distance between Neptune and the inner border of the disk. For example, the case with Jupiter on a non-migrating orbit at 5 AU results in a median instability time of 60 My, whereas the case where Jupiter migrates from 15 to 5 AU, which results in a much wider separation between Neptune and the inner edge of the disk, results in a median instability time of 134.70 My. These results were predicted by Deienno et al. (2017) based on the planetesimal disc results from sect. 2.2 
and are broadly consistent with the results of Quarles and Kaib (2019). Although we have seen in Fig. \ref{fig:inclicum} that the case with Jupiter migrating from 10 to 5 AU overexcites the cold Kuiper belt, Fig. \ref{fig:cumulativefinal} (a) shows that the distribution of instability times in that case is similar to that of no-Jupiter-migration case. So we retain this distribution as an upper bound of the instability times of acceptable short-range inward migration cases. Thus we conclude that the median instability times range in the interval 50.0-80.3My, with 75\% of the instabilities occurring within 190My.

For the unself-stable systems (Panel (b) of Fig.  \ref{fig:cumulativefinal}) the times of instabilities, in general, are not dependent on the way that Jupiter migrated. Thus, the distribution of the time of instabilities is not dependent on the gap between the disk and the planets.  However, the median of instability for the four cases with the planetesimal disk is 1.5 My with 99\% of the all unself-stable systems go unstable in less than 10 My.
Only 75\% of the unself-stable systems go unstable in less than 10 My without any planetesimal disk (green points in Panel (b) Fig.  Fig.  \ref{fig:cumulativefinal}). It shows that the planetesimal disk plays an important role to drive the instabilities to early times.

 \section{Discussion}
 \label{discussion}

Our results strongly favor an early giant planet instability. 
A natural outcome of successful ice giant formation simulations \citep{Izidoroetal2015} is a self-triggered 
instability within 10 Myr of the dissipation of the disk (\S \ref{selftriggered}). If instead the giant planets emerged from the disk 
on stable orbits, then interactions with the planetesimal disk should nonetheless have triggered the instability within 
500 Myr. If we discard the cases where Jupiter migrated inwards by 5 AU or more because they result in a cold Kuiper belt 
that is too dynamical excited (Fig. \ref{fig:cumulativefinal}), the instability times are within 200My, of which 70\% are within 70My.  
% From a dynamical point of view an early instability also bypasses the problem of the dynamical fragility of the terrestrial planets~\citep{brasser09,AgnorLin2011,KaibandChambers2016}. Instead, an early instability would have removed most of the material from Mars's formation region, explaining the small mass of Mars and its short accretion timescale~\citep{Clementetal2018,clement2019a}.  It also may have contributed to depleting and dynamically exciting the asteroids~\citep{Morbidellietal2010,Nesvornyetal2017,Deiennoetal2018,clement2019b}.
In this section we discuss how our result fits within a broader context. 
First we compare our dynamically-inferred timeframe with empirical constraints for the giant planet instability (\S \ref{empcons}). 
% Then we play devil's advocate and speculate as to what sequence of events may in principle have delayed the instability, 
% leaning heavily on processes not included in our simulations (\S 4.2). 
Finally, we present the limitations of our simulations (\S \ref{limit}) to motivate future work that might 
improve the results.

\subsection{Empirical Constraints}
\label{empcons}
The division between an ``early'' and a ``late'' instability remains fuzzy.  This is essential in understanding the timeline of Solar System evolution in the context of other landmarks.  For instance, 
the Sun's gaseous disk dissipated 3-5 Myr after CAIs (calcium-aluminium-rich inclusion, probably first solids to condensate in the sun's natal disk), judging from the timescale for the disappearance of disks around other young stars~\citep{haisch01,pascucci09} as well as the ages of the oldest chondrules~\citep[e.g.][]{krot05,bollard15}. Mars' growth was complete on a similar timescale of 5-10 Myr~\citep{nimmo07,dauphas11}, whereas the Moon-forming impact did not occur until roughly 50-100 Myr later~\citep[e.g., review by ][]{kleine09}. 

\citet{Morbidellietlal2018} showed that the lunar cratering record and the highly-siderophile element abundances of the 
Earth and Moon could be reconciled by an instability that happened in the first hundred million years of Solar System history, 
provided that the highly siderophile elements have been removed from the lunar mantle during a late crystallization of the 
lunar magma ocean.
Other re-analyses of the cratering record and the age distribution of Apollo samples also point to an early instability, 
but without firm timing constraints~\citep{boehnke16,zellner17,michael18}. 

Some dynamical analyses have derived more quantitative estimates.  An instability within 100 Myr after CAIs is 
required to explain the survival of the Patroclus-Menoetius Jupiter binary Trojan, assuming it was formed in the 
primordial Kuiper belt and dynamically captured during the instability~\citep{Nesvornyetal2018}. 
\citet{Nesvorny2015a} found that the instability could not have happened earlier than $\sim 10$ Myr by arguing that 
Neptune's slow, pre-instability migration is needed to excite the inclinations of Kuiper belt objects.
Morbidelli and Nesvorny (submitted, 2019) found that the current size-distribution of the Kuiper belt could be obtained starting 
from the size distribution coming from streaming-instability models \citep{simon2016} provided that the 
trans-Neptunian planetesimal disk was not dispersed before $\sim$ 50 My. 
% 
% 
% 
% 
% Of course, in our model the Kuiper belt may have been partially pre-excited by the progenitors of the ice giants; in some of our simulations the Kuiper belt is over-excited. We therefore do not consider this a firm timing constraint, although a careful analysis of the dynamical sculpting of the Kuiper belt during the gas disk phase is in order.

On the other hand, a very early instability occurring before the completion of the accretion of the terrestrial planets would bypasse 
the problem of the dynamical fragility of the terrestrial planet system \citep{brasser09,AgnorLin2011,KaibandChambers2016}.
Moreover, such an early instability could have removed most of the material from Mars's formation region, explaining the small mass of Mars and its short accretion timescale
\citep{clement2018b}. It could have also contributed to depleting and dynamically exciting the asteroids \citep{Morbidellietal2010,Nesvornyetal2017,Deiennoetal2018,clement2019}. 
These considerations seem to favor an instability within the first $\sim$10 My, possibly even less. 

However an instability before the Moon-forming impact -- generally thought to be the last giant impact in the inner Solar System -- should 
have left a chemical imprint on Earth's interior that is not observed. In fact, the isotopic signatures of atmospheric and mantle 
Xenon are distinct \citep{Caracausietal2016}. \citet{Martyetal2017} used the isotopic signature of comet 67P measured by the ROSETTA 
spacecraft to argue that $\sim 20\%$ of present-day atmospheric Xenon is of cometary origin. 
This naively suggests that the cometary bombardment (necessarily associated to the giant planet instability) 
occurred after the formation of the Earth's crust, which would imply a giant planet instability later than 
the Moon-forming impact, no earlier than 50-100 Myr after CAIs~\citep[e.g., ][]{kleine09}.  
% 
% First, at face value the data only require that some comets impacted Earth after Moon formation, not that cometary impacts were absent at 
% earlier times.  Indeed, the same comets that brought the atmosphere's Xenon would only have delivered $\sim 1\%$ of 
% Earth's water~\citep{marty16}, a tiny fraction of Earth's volatile budget. 
% While the bulk of Earth's late accretion was likely Enstatite chondrite-like~\citep{rubie15,dauphas17}, 
% it is hard to imagine that no comet-like objects impacted Earth after the Moon-forming impact, especially given that the asteroid 
% belt was contaminated with planetesimals that originated past 10 AU~\citep[and likely out to $\sim 20$ AU or beyond][]{raymond17} 
% during the giant planets' growth and migration.  
However, if the in-gassing of Xenon in the silicate magma was inefficient, cometary Xenon may not have penetrated into the terrestrial mantle during the magma-ocean phase that followed the giant impact. 
It is also known that a large fraction of the atmosphere of the proto-Earth might have survived the giant impact, particulalry if there was no surface ocean at that time \citep{Schlichting2018}. Thus, it may be possible that the cometary bombardment predated the Moon-forming event, although this requires further geochemical investigations. 

In conclusions, it is not possible to conclude from firm constraints when the giant planet instability occurred within the first 100My. Unfortunately, the present study, with a median instability time of 36.78-61.5 My 
and a 75\% instability time of 136 My when excluding the cases with long-range inward migration of Jupiter does not help in assessing what is the most probable time.

\subsection{Limitations of our work}
\label{limit}
As with any numerical study, our simulations are a simplified and idealized version of reality.
One main limitation is that our results are based on a small number of outcomes from \citet{Izidoroetal2015}.
Nevertheless, these simulations represent, to our knowledge, the only models that quantitatively explain the origin of Uranus and Neptune.
We performed N-body simulation of Jupiter, Saturn and a small number of planetary embryos from \citet{Izidoroetal2015} simulations. Thus, we artificially 
forced Jupiter to migrate and rescaled the semi-major axes of the others planets and embryos from the original integration to Jupiter's. The evolution of the other elements, for example,
eccentricities and inclinations is kept the same as in the original simulation. This procedure is not ideal. Even if all bodies were locked in mutual mean motion resonances, the global migration 
of the system would affect the orbital eccentricities. However, in all cases the eccentricities should be small (of order $h^2$, where h is the aspect ratio of the disk). We expect that 
the slightly different eccentricities considering different migration paths for Jupiter would not have dramatic consequences on the resulting dynamical structure of the planetesimal disk. 
Instead, the dynamical sculpting of the planetesimals disk depends mainly on how long the embryos stay in the disk and therefore on the migration of the whole chain (determined by the migration of Jupiter).
A more realistic model would require very computationally costly hydrodynamical simulations. The advantage of our procedure is that we can compare the results obtained imposing just different migration patterns for Jupiter, 
keeping all other parameters equal. If we had done different simulations for each migration pattern of Jupiter, too many aspects of the evolution would have changed (due to the fact that all evolutions are chaotic), making it
difficult to determine what was the cause of the different final results.

Our study of the stability of self-triggered and planetesimal disk-triggered simulations included a total of 1080 simulations whereas our 
self-consistent planetesimal disks were generated from a single simulation of Uranus and Neptune formation.
 Pebble accretion can explain the rapid growth of super-Earths and Ice giants in different parts of the gas-disk including the region
inside of the cold Kuiper Belt (or the current trans-Neptunian region \citep{LambrechtsJohansen2014,johansen17,I2019,B2019,Lambrechtsetal2019}).
Other formation histories are possible. For example, Jupiter could have formed far out, then Saturn and embryos form in sequence, in the same location once
the previous planet has migrated away. In that case, the cold Kuiper belt would be much less excited, even 
in the case of an initial distant location of Jupiter (e.g. 20 AU). Our result show actually how far out 
the formation of embryos could have ocurred before exciting the cold Kuiper belt too much, rather than how far out Jupiter 
could form.
Given that our simulations are built upon those of \citet{Izidoroetal2015}, we inherit the limitations of that study.  For example, as discussed above, our model for the underlying gaseous disk is plausible but uncertain. 
In addition, we did not include potentially important processes such as collisional fragmentation~\citep{leinhardt12,genda12} and pebble accretion~\citep{johansen17}.  Our simulations also start with an already-formed 
Jupiter and Saturn.  If their growth dramatically re-shaped the outer Solar System then our initial conditions would have to be re-thought.

 \section{Conclusions}
 \label{conclusion}

In this paper we broadly constrained the timing of the giant planet instability using simulations of the early evolution of 
the outer Solar System. We started from the best simulations of \citet{Izidoroetal2015} of the growth of the ice giants during 
the gaseous disk phase via inward migration of $\sim 5 M_\oplus$ cores that are blocked by the already-formed Jupiter and Saturn. 
We generated plausible outer planetesimal disks that were dynamically sculpted in a self-consistent way during this formation process, 
taking into account both the size-dependence of aerodynamic gas drag and a range of possible migration histories for Jupiter and 
Saturn~\citep[see][]{pierens14}.
We determined that a large fraction ($\sim$ 50\%) of the giant planet configurations generated by \citet{Izidoroetal2015} 
became unstable within 10 Myr of disk dissipation. The leads to the possibility that the giant planet instability was self-triggered 
by the planets themselves, which is new in the context of the Solar System's history but a well-known process in a more 
general context~\citep[e.g.][]{chambers96,marzari02,ford08}.

When we introduce the different outer planetesimal disks, the giant planet configurations that would have 
remained stable if they had been alone went unstable within 500 Myr, and generally much faster.
The median instability timescale is 36.78-61.5 My. If we exclude the disks sculpted during long-range inward migrations of Jupiter, which are inconsistent with 
the small dynamical excitaiton of the cold Kuiper belt, the instability time is within 136My in 75\% of the cases, 
which is consistent with the conclusions of \citet{Nesvornyetal2018} on the survival of the Trojan Patroclos as 
primordial binary from the Kuiper belt.  
Unfortunately, given our statistics and the available constraints, it is difficult to conclude when the giant 
planet instability happened within the first $\sim$100My. In particular it is difficult to assess whether the giant planet 
instability pre-dates or post-dates the Moon-forming event.

\section*{Acknowledgments}
We acknowledge helpful discussions with Sarah Stewart on the origin of Earth's Xenon.
We also thank John E. Chambers and Rogerio Deienno for useful comments and suggestions on the submitted manuscript.
R.R.S acknowledges support provided by grants \#2017/09919-8, \#2016/24561-0 and \#2015/15588-9, S\~ao Paulo Research Foundation (FAPESP) and the good hospitality received from Observatoire de la C\^ote d'Azur.
A.I. thanks finnancial support from Fapesp via grants  \#2016/12686-2 and  \#2016/19556-7.  
S.~N.~R. thanks the Agence Nationale pour la Recherche for funding via grant ANR-13-BS05-0003-002 (grant MOJO).


\bibliography{bib}{}
\bibliographystyle{aasjournal}



\end{document}